\documentclass[12pt]{article}
\begin{document}
\input epsf
\def\be{\begin{equation}}
\def\bea{\begin{eqnarray}}
\def\ee{\end{equation}}
\def\eea{\end{eqnarray}}
\def\d{\partial}
\def\la{\lambda}
\def\eps{\epsilon}

\begin{titlepage}
\begin{center}
\hfill EFI-07-17\\

\vskip 1cm

{\LARGE {\bf Strings ending on branes from supergravity\\ \ \  \\
  }}

\vskip 1.5cm

{\large  Oleg Lunin }

\vskip 1cm

{\it Enrico Fermi Institute, University of Chicago,
Chicago, IL 60637
}

\vskip 3.5cm

\vspace{5mm}

\noindent

{\bf Abstract}

\end{center}

We study geometries produced by brane intersections preserving eight 
supercharges. Typical examples of such configurations are given by 
fundamental strings ending on Dp branes and we construct gravity solutions 
describing such intersections. The geometry is specified in terms of two 
functions obeying coupled differential equations and the boundary conditions 
are determined by distributions of D branes. We show that a consistency of 
type IIB supergravity constrains the allowed positions of the branes. The 
shapes of branes derived from gravity are found to be in a perfect agreement 
with profiles predicted by the DBI analysis. We also discuss related 
1/4--BPS systems in M theory.

\vskip 4.5cm

\end{titlepage}

\newpage

\setcounter{tocdepth}{2}

{\footnotesize\tableofcontents}

\setcounter{tocdepth}{0}

\newpage


\section{Introduction}
\renewcommand{\theequation}{1.\arabic{equation}}
\setcounter{equation}{0}


Much of the progress in string theory over the last decade was based on the 
improvement in our understanding 
of nonperturbative objects such as D branes. Originally branes appeared 
independently from the open string
analysis \cite{LeighPolch} and from solving equations for closed strings 
\cite{HorowStrom} and latter it 
was realized that these two approaches gave complimentary descriptions of 
the same objects \cite{Polch}.  
The idea of duality between open-- and closed--string pictures culminated in 
the discovery of AdS/CFT correspondence \cite{malda} which was formulated as 
an equivalence between a field theory described by open 
strings and a theory of closed strings on a geometry produced by branes. D 
branes have also been crucial for 
improving our understanding of black holes \cite{StromVafa}. Most of these 
developments 
emerged from a progress in studying flat branes both in the open string 
picture and in supergravity.

Unfortunately curved branes are not understood as well as their flat 
counterparts.
One of the reasons for this gap is the fact that flat branes preserve $16$ 
supercharges, while the 
objects with curved worldvolume preserve at most half of this 
amount\footnote{Here 
we are discussing the situation in asymptotically--flat space, in 
$AdS_5\times S^5$ one can have curved branes preserving sixteen 
supersymmetries and they have been studied both in the probe approximation 
\cite{giant,rey,Drukker} and in supergravity 
\cite{LLM,LinMaldGom,yama,myWils,GomRom,Gutprl}.}. While in the open string picture a 
dynamics of curved branes with fluxes has been studied in the past 
(a prototypical example of such computation was presented in \cite{CalMal}), 
gravity description of such objects is not well--developed. 
Extension of open/closed duality to the case of curved branes could 
potentially lead to new decoupling limits
and to discovery of interesting examples of gauge/gravity pairs with lower 
supersymmetry.  

Another motivation for finding geometries with lower supersymmetry comes 
from a desire to
classify intersecting branes\footnote{See \cite{gauntRev,smithRev} for a 
review of progress in this classification.}. Such intersections can be used 
to gain information about 
physics of black holes (the classical example is D1--D5--P intersection used 
in state counting of \cite{StromVafa}) or 
about dynamics of gauge theories at strong coupling \cite{HanWitten}. It 
turns out that the brane intersections are closely related to curved branes 
with fluxes, for example in \cite{CalMal} it was demonstrated that a curved 
D brane with electric flux on its worldvolume mimics behavior of fundamental 
strings ending on a brane. As we will see below, on the gravity side the 
descriptions of the intersections and curved branes are also unified. 

We will mostly be interested in branes ending on other branes and the rules 
for 
such intersections can be derived using quantization conditions for various 
charges \cite{StrmTwns}. 
If the number of branes is small, their low--energy dynamics is 
well--described by the DBI action in flat space and in the past this action 
has been used to study various intersections. However, as the number of 
branes increases, their effect on metric cannot be neglected, and one needs 
to find the geometries produced by the branes. For the parallel  stacks of 
flat branes this task has been accomplished in \cite{HorowStrom}, but for a 
generic brane intersection the relevant metrics are not known. The goal of 
this paper is to derive the geometries produced by 1/4--BPS intersections. 
In contrast to the traditional approach where the positions of the branes are 
specified from the beginning, we will only require 
a certain amount of supersymmetry to be preserved and solve the equations 
away from the branes. Then the brane profiles will be {\it derived} from the 
consistency conditions. Thus in the closed string 
picture we will view D branes as dynamical objects and this treatment is a 
direct counterpart 
of the DBI analysis which one performs for the open strings.

We will begin by looking at a bunch of fundamental strings ending on a single 
stack of D3 branes. The probe analysis for such configuration has been 
presented in \cite{CalMal}, in particular one finds that this setup is 
invariant under $SO(3)\times SO(5)$ transformations (we will review this in 
section 2). It turns out that this 
isometry is sufficiently restrictive to allow one to derive a gravity 
solution preserving $8$ supercharges. Of course, a 
generic $1/4$--BPS geometry is not expected to have an $SO(3)\times SO(5)$ 
isometry group, but based on the symmetric example, we managed to guess the 
relevant geometry for the general case 
and the result is presented in section \ref{SecGnrSln}. It turns out that in 
order to satisfy consistency conditions coming from 
supergravity, the brane sources cannot be introduced arbitrarily, but rather 
they should follow particular profiles, 
and we find these shapes to be in a perfect agreement with results coming 
from the open string analysis.  

This paper has the following organization. In section 2 we will review the 
basic ideas of \cite{CalMal} and extend their probe analysis to branes in 
nontrivial backgrounds. In particular, we will find the restrictions on the 
trajectories of the 
probes coming from the dynamics of DBI action. An M theory counterpart of 
the F1--D3 system contains  
membranes ending on M5 branes and we will find classical solutions of the PST 
action which are relevant for this case. Section 3 is devoted to geometric 
description of a stack of D3 branes with a single spike 
and due to an enhanced symmetry of this setup, we are able to derive the 
appropriate solution without making additional assumptions. In section 4 we 
propose a generalization of this geometry to a situation without bosonic 
symmetries (and check that supergravity equations are satisfied for this 
case as well) and, by requiring consistency of the equations in the presence 
of sources, we find the locations of D branes. These positions are shown to 
be in a perfect agreement with results 
of DBI analysis presented in section 2. 
In section 5 we use various duality chains to produce solutions of eleven 
dimensional supergravity and again an agreement between consistent boundary 
conditions and PST analysis on the probe side is found.


\section{Curved branes in the probe approximation}

\label{SectProbe}
\renewcommand{\theequation}{2.\arabic{equation}}
\setcounter{equation}{0}


The main goal of this paper is to construct geometries which describe 
fundamental strings ending on D3 or D5 
branes. If the number of branes (and strings) is small, the metric is 
well--approximated by the flat space everywhere except for the locations of 
the branes. Since D branes are dynamical objects, these locations cannot be 
arbitrary, but rather they should be determined by solving equations for 
various fields living the worldvolume of the branes. In this section we will 
recall the form of such solutions corresponding to strings ending on D 
branes and we will derive the expressions for the profiles of the brane 
probes. In section \ref{SbctMtchPrb} we will also perform a similar analysis 
for a membrane ending on M5 brane in eleven dimensions. 

Since we will be solving equations coming from the DBI action, the analysis 
of this section pertains to a description in terms of open strings. 
On the other hand, the remaining part of this paper is devoted to 
supergravity, which gives a picture from the 
point of view of closed strings. 
Once these two analyses are compared, we will find a nontrivial agreement 
which can be interpreted as open/closed string duality. In the decoupling 
limit this duality reduces to a standard AdS/CFT.


\subsection{Supersymmetric brane intersections.}
\label{SectSbScan}


We begin with recalling some general facts about intersecting branes in IIB 
string theory. 
In this theory supersymmetry transformations are parameterized by two 
Majorana--Weyl spinors 
which have the same chirality, and it is convenient 
to combine them into a 32--component real object $\eps$ which satisfies a 
chirality projection:
\bea
\eps=\left(\begin{array}{c} \eps_1\\ \eps_2\end{array}\right),\quad 
{\bf 1}_2\otimes \Gamma_{11}\eps=-\eps:\qquad \Gamma_{11}\eps_{1,2}=
-\eps_{1,2}.
\eea
Ten--dimensional flat space preserves $32$ supersymmetries corresponding to 
arbitrary constant values 
of $\eps_1$ and $\eps_2$ (modulo the chiral projection). By adding a brane 
to $R^{9,1}$ one breaks half of the supersymmetries and the appropriate 
projections are 
\cite{ScanRef} (see also \cite{smithRev} for a review):
\bea\label{BraneSUSY}
{\rm F1}:&\quad \Gamma=\sigma_3\otimes \Gamma_{(2)},\quad & 
\Gamma\eps=\eps,\nonumber\\
{\rm NS5}:&\quad \Gamma=\sigma_3\otimes \Gamma_{(6)},\quad &
\Gamma\eps=\eps,\\
{\rm D}(2p-1):&\quad \Gamma=i\sigma^p_3\sigma_2\otimes \Gamma_{(2p)},\quad &
\Gamma\eps=\eps.\nonumber
\eea
Here $\Gamma_{(2p)}$ is a product of gamma matrices with indices pointing 
along the worldvolume of the brane. Each of the branes appearing in 
(\ref{BraneSUSY}) preserves $16$ real supercharges and there are two other 
interesting objects which have the same amount of SUSY --- a plane wave and 
a KK monopole\footnote{To unify the description of KK monopoles in 
ten and eleven dimensions, we characterize the monopole by four nontrivial 
coordinates rather than by $5+1$ (or $6+1$) worldvolume directions.}: 
\bea\label{GeomBranes}
{\rm P}:&\quad \Gamma={\bf 1}_2\otimes \Gamma_{(2)},\quad & 
\Gamma\eps=\eps,\nonumber\\
{\rm KK}:&\quad \Gamma={\bf 1}_2\otimes \Gamma_{(4)},\quad &
\Gamma\eps=\eps.
\eea
These configurations have a pure geometric nature and they do not involve 
fluxes. 

Once the building blocks preserving half of the supersymmetries are 
specified, one can start combining them to 
produce configurations with lower SUSY. Such supersymmetric intersections 
are only possible if the projectors for the ingredients commute with each 
other. 
The main example studied in this paper involves fundamental strings ending 
on a D3 brane:
\bea\label{BranePreScan}
\begin{array}{c|ccccccccc}
&1&2&3&4&5&6&7&8&9\\
D3&\bullet&\bullet&\bullet&&&&&&\\
F1&&&&\bullet&&&&&\\
\end{array}
\eea 
Looking at spinors preserved by each object:
\bea
D3:&& i\sigma_2\otimes \Gamma_{0123}\eps_3=\eps_3,\nonumber\\
F1:&& {\bf 1}_2\otimes \Gamma_{04}\eps_1=\eps_1,\nonumber
\eea
we observe that two projectors can be diagonalized simultaneously and the 
entire configuration
preserves $8$ supercharges. In fact, one more object can be added to this 
system without breaking additional supersymmetry:
\bea
D5_{56789}:&& \sigma_1\otimes \Gamma_{056789}\eps_5=\eps_5,\nonumber
\eea
so it is useful to look at the following configuration:
\bea\label{BraneScan}
\begin{array}{c|ccccccccc}
&1&2&3&4&5&6&7&8&9\\
D3&\bullet&\bullet&\bullet&&&&&&\\
D5&&&&&\bullet&\bullet&\bullet&\bullet&\bullet\\
F1&&&&\bullet&&&&&\\
\end{array}
\eea
Performing a similar analysis, one can classify all brane intersections 
preserving $8$ 
supercharges\footnote{We omit the configurations which can be found by an 
application of S 
duality (e.g. $(D3_{123},\mbox{NS5}_{56789},D1_{4})$).}:
\bea\label{IntersBraneIIB}
&&(D3_{123},D5_{56789},F1_{4}),~
(D3_{123},D3_{145},KK_{2345}),~(D3_{123},D5_{12456},\mbox{NS5}_{12789}),
\nonumber\\
&&\quad\qquad(D5_{12345},D7_{1234678},\mbox{NS5}_{12349}),\quad
(D7_{1234567},F1_{8},D1_9)\\
&&\qquad(D5_{12345},D5_{16789},P_{1}),\quad
(D3_{123},D7_{1456789},P_1),\quad (D1_{1},P_1)\nonumber\\
&&\quad\qquad(D5_{12345},D1_{1},KK_{2345}),\quad
(D3_{123},D7_{1234567},KK_{4567}).\nonumber
\eea
To construct the geometries corresponding to intersections appearing in the 
last two lines, one needs to superpose harmonic functions and some of the 
resulting solutions are well--known
\cite{HorTseytl}.
The geometries describing localized intersections presented in the first two 
lines has not been written before, and our main goal is to find the 
appropriate metrics. It turns out that 
once the description of (\ref{BraneScan}) is known, the other configurations 
appearing in 
(\ref{IntersBraneIIB}) can be recovered by application of various dualities, 
so most of our discussion will be concentrated on (\ref{BraneScan}) and we 
will come back to other configurations in section \ref{SectSbOthIIB}.

Finally let us comment on branes in M theory. One still has geometric 
objects characterized by projections  
(\ref{GeomBranes}), and in addition there are M2 and M5 branes which 
preserve the following pieces of  
the 32--component real spinor $\eps$ \cite{11DScan,gueven,MDBI,PST}:
\bea
{\rm M2}:&\quad \Gamma=\Gamma_{(3)},\quad & \Gamma\eps=\eps,\nonumber\\
{\rm M5}:&\quad \Gamma=\Gamma_{(6)},\quad &
\Gamma\eps=\eps.\nonumber
\eea
The intersections preserving $8$ supersymmetries can be classified in this 
case as well:
\bea\label{MthryInter}
(M5_{12345},M5_{1789(10)},M2_{16}),~(M5_{12345},KK_{1234},P_5),
~(KK_{1234},M2_{12},M2_{34}),&&\\
(M5_{12345},M5_{12367},KK_{4567}),~(KK_{1234},KK_{5678},M2_{9(10)})&&
\nonumber\\
(KK_{1234},KK_{1256},KK_{3456}),~
(M5_{12345},KK_{6789}),~(M2_{12},P_1).&&\nonumber
\eea
and we will discuss the corresponding geometries in section \ref{SectMsoln}. 

However before we start constructing supergravity solutions, it is useful to 
perform a brane probe analysis.
We will see that some intersections are described by curved branes with 
fluxes on their worldvolumes and we will find the shapes of such branes. 
This analysis will be presented both 
in type IIB string theory (using D3--D5--F1 system as an example) and in M 
theory (where we discuss M5--M5--M2 intersection). 

\subsection{Bions in flat space}
\label{SectSbBion}

\begin{figure}[tb]
\begin{center}
\epsfysize=2.2in \epsffile{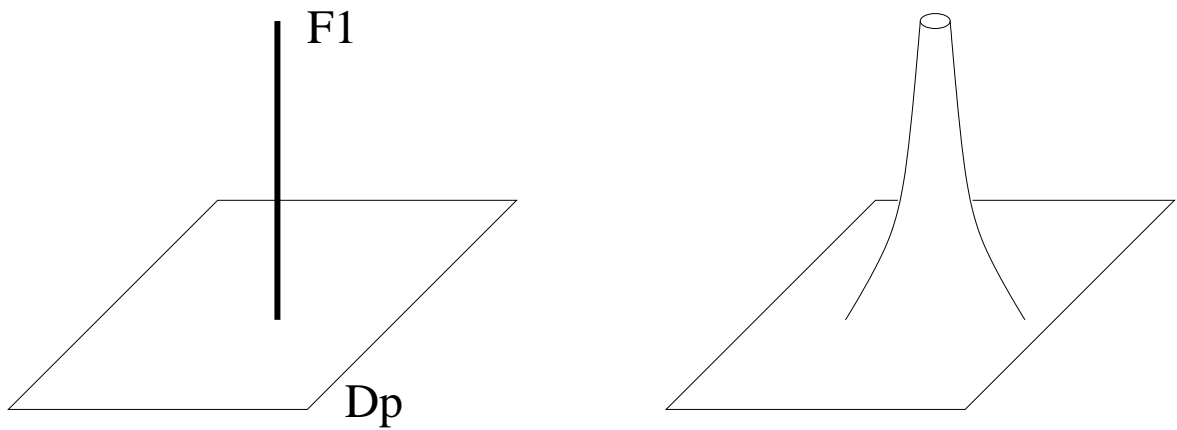}
\end{center}
\caption{Two different pictures for fundamental string ending on a Dp brane: 
the naive 
configuration (a) and the description in terms of spike introduced in 
\cite{CalMal} (b). 
} \label{FigSpike}
\end{figure}

We begin by recalling the solution found by Callan and Maldacena
\cite{CalMal}. The basic idea of that work can be summarized in the 
following way. Suppose one wants to 
describe a fundamental string ending on Dp brane (as depicted in figure 
\ref{FigSpike}a).  This configuration is 
expected to preserve eight real supercharges. 
An observer living  on the D brane sees a pointlike charge, so an electric 
field should be excited on the worldvolume of the brane. This modifies 
the shape of 
the D brane, and the correct physical picture is given by figure 
\ref{FigSpike}b
rather than \ref{FigSpike}a: the fundamental string is replaced by a curved 
brane with flux. 
Let us review this construction in more detail.


The starting point for the analysis of \cite{CalMal} was a special 
embedding of Dp brane into the ten dimensional space, so that the brane was 
stretched along the directions $X^0,\dots X^p$, while it was also allowed to 
have a nontrivial profile in one of the transverse coordinate 
$X\equiv X^{p+1}$. To have interesting dynamics, one also allows for a 
non--vanishing electric field on the worldvolume of the brane. In the static 
gauge 
($X^0=\xi^0$,\dots,$X^p=\xi^p$) the DBI action for Dp brane becomes
\bea\label{CMact}
S_{Dp}&=&-T_{p}\int d^{p+1}\xi\sqrt{-\mbox{det}(G+2\pi\alpha' F)}
\nonumber\\
&=&
-T_{p}\int d^{p+1}\xi\sqrt{(1-E^2)(1+(\nabla X)^2)+({\bf E}\nabla X)^2}.
\eea
Here electric field is defined as $E_i=2\pi\alpha' F_{i0}$.
Since we are looking for a static solution, it is convenient to choose a 
gauge ${\bf E}=\nabla A$, this leads to the equations of motion for two 
scalars $X,A$: 
\bea
&&\nabla_i\left[\frac{(1-(\nabla A)^2)\nabla_i X+
(\nabla A\nabla X)\nabla_i A}{
\sqrt{(1-(\nabla A)^2)(1+(\nabla X)^2)+(\nabla A\nabla X)^2}}\right]=0,
\nonumber\\
&&\nabla_i\left[\frac{(1+(\nabla X)^2)\nabla_i A+
(\nabla A\nabla X)\nabla_i X}{
\sqrt{(1-(\nabla A)^2)(1+(\nabla X)^2)+(\nabla A\nabla X)^2}}\right]=0.
\nonumber
\eea
In \cite{CalMal} it was pointed out that these equations linearize it we take 
$A=X$. Moreover, in this case the solution saturates the BPS bound since it 
has a very simple energy density:
\bea
{\cal E}={\bf E}\frac{\delta S_{Dp}}{\delta {\bf E}}-{\cal L}=
T_p\left(1+(\nabla X)^2\right).
\eea

To summarize, the construction of \cite{CalMal} gives a family of 1/4--BPS 
configurations which are 
parameterized by one harmonic function $X$:
\bea\label{ProbeHarmonic}
\nabla^2 X=0,\quad F_{0i}=\frac{1}{2\pi\alpha'}\nabla_i X,
\eea
and this function gives a location of the brane. Let us discuss the 
symmetries of the problem. Since the brane is curved in one of the 
transverse directions, the rotations around the brane are broken to 
$SO(8-p)$ and for a generic profile of $X$ this is the only non--abelian 
symmetry of the configuration\footnote{Since the system is static, it is 
also invariant under time translations.}. However for special functions $X$ 
there 
might be additional symmetry coming from the worldvolume of the brane. 
For example, figure \ref{FigSpike}a suggests an $SO(p)$ rotational symmetry 
around the string, so it is natural to consider a single spike which is 
invariant under such rotations. Thus we see that the maximally symmetric 
spike 
has an $SO(8-p)\times SO(p)\times U(1)$ symmetry, in particular both D3 and 
D5 branes are invariant under 
$SO(5)\times SO(3)\times U(1)$. This symmetry will be further explored in 
section \ref{SectSpike}. 

Let us now make a comment about emergence of fundamental strings. 
Due to nontrivial value of the electric field, the action (\ref{CMact}) 
sources a bulk Kalb--Ramond field even 
in the linear order \cite{bachas}. The simplest way to find the relevant 
coupling is to make a 
substitution 
$2\pi\alpha' F\rightarrow 2\pi\alpha' F-P[B]$ in (\ref{CMact}) and compute 
the first 
correction:
\bea
\delta S_p&=&-\int d^{p+1}\xi \frac{\delta S_p}{\delta (2\pi\alpha' 
F_{\mu\nu})}
P[B]_{\mu\nu}=T_p\int d^{p+1}\xi P[B]_{ti}\nabla_i X\nonumber\\
&=&T_p\int d^{p+1}\xi (B_{ti}+(\nabla_i X)B_{t(p+1)})\nabla_i X.
\eea
Here $P[B]$ is a pullback of the B field to the worldvolume of the brane. 

It is interesting to look at a single spherically symmetric spike which has 
$X=Qr^{-(p-2)}$, for this configuration 
the coupling to the Kalb--Ramond field becomes:
\bea
\delta S_p&=&T_p\Omega_{p-1}\int dX \left[r^{p-1}B_{tr}-Q(p-2)B_{t(p+1)}
\right].
\eea
As expected, in the region where the spike becomes thin (i.e. close to the 
origin in $r$ coordinate) this term sources 
strings stretching in $(t,X^{p+1})$ directions with a density which is 
uniform in $X^{p+1}=X$. 

\subsection{Bions in brane backgrounds}
\label{SectSbCrvBion}

In the previous subsection we recalled the description of spikes on Dp 
branes assuming that these branes are placed in the flat space. In 
particular we observed that a single spike attached to a 
D3 brane has the same 
$SO(5)\times SO(3)\times U(1)$ symmetry as a spike attached to D5. This 
leads to a natural proposal to consider these two types of branes together.
In the probe approximation superposition of branes leads to addition of 
their DBI actions, so the analysis of the previous subsection goes through. 
However branes with different orientations preserve different 
supersymmetries, so in general a combination of branes would break SUSY 
completely. Of course, in the exceptional cases some SUSY is still 
preserved and as we reviewed in section \ref{SectSbScan}, the combination of 
D3, D5 branes and 
fundamental strings preserves eight supercharges. Moreover, by building 
configuration (\ref{BraneScan}) from 
one stack of D3 branes and one stack of D5s, we can also preserve 
$SO(5)\times SO(3)\times U(1)$ bosonic symmetry.
In a more general case when we have several D3 and D5 branes at various 
positions, the 
$SO(5)\times SO(3)$ symmetry would be broken, but eight supersymmetries will 
still be preserved as long as the orientations of the branes are the same as 
in (\ref{BraneScan}).

To probe this picture one can consider the following setup. Suppose one 
starts with large number of D3 branes without strings attached to them. 
These branes would lead to the modification on the geometry, and the 
resulting metric is well--known \cite{HorowStrom}. Then to describe an 
additional D3 brane with string ending on it, one needs to solve the 
equations of motion coming from the DBI action on curved background. It 
would be interesting to find a profile of the spike in this case. One can 
also look at the D5 brane on D3 background and solve equations in this case 
as well. The D branes in the geometry produced by multiple D5's can be 
analyzed in the same way. While these exercises are very straightforward, 
it seems useful to outline them here since we will need to compare the 
results with the outcome of computations in supergravity. 

{\bf D3 spike in the geometry of D3.} This case has been previously analyzed 
in \cite{GKMTZ}, so we will be very brief. The background geometry is given 
by the metric of $N$ coincident D3 branes\footnote{In this section we use 
the string conventions which has a different normalization of $F_5$ compared 
to standard supergravity notation. We discuss this difference in more detail 
in Appendix \ref{AppSugra} (see also \cite{granPolDil}).} \cite{HorowStrom}:
\bea
ds^2&=&H^{-1/2}ds_{3,1}^2+H^{1/2}(dz^2+d{\bf y}_5^2),\nonumber\\
F_5&=&d\left[H^{-1}d^4 x\right]-^*_6dH,\quad 
H=1+\frac{Q}{(z^2+{\bf y}^2)^2},\quad Q=4\pi gN\alpha'.
\eea
One can study dynamics of a probe D3 brane assuming that its worldvolume is 
described by a profile $r=0$, $z=X(x_1,x_2,x_3)$. 
In the static gauge ($\xi_0=x_0$, \dots, $\xi_3=x_3$) the induced metric and 
the pullback of the RR potentials are
\bea
G_{ab}=H^{-1/2}\eta_{ab}+H^{1/2}\d_a X\d_b X,\quad
C_{abcd}=H^{-1}\eps_{abcd}
\eea
and the action governing the dynamics of the probe brane 
becomes\footnote{Here we again defined $E_i=2\pi\alpha' F_{i0}=\nabla_i A$.}:
\bea\label{D3inD3Act}
S_{D3}&=&-T_{3}\int d^{4}\xi\sqrt{-\mbox{det}(G+2\pi\alpha' F)}+
\frac{T_3}{4!}\int C_{abcd} d\xi^{abcd}
\nonumber\\
&=&
-T_{3}\int d^{4}\xi H^{-1}
\sqrt{(1-HE^2)(1+H(\nabla X)^2)+H^2({\bf E}\nabla X)^2}\nonumber\\
&&+T_3\int d^4\xi H^{-1}.
\eea
One can see that equations of motion for $A$ and $X$ are satisfied if the relations (\ref{ProbeHarmonic}) are imposed\footnote{Notice that the Chern--Simons term is the action is crucial for enforcing a condition 
$\frac{\d{\cal L}}{\d X}=0$.}.
We conclude that even in the background produced by other D3s, the spike of D3 brane should still follow the harmonic profile in $\xi_1$, \dots, $\xi_3$.

{\bf D5 spike in the geometry of D3.} 
Next we put a D5 brane in the background written above\footnote{Such brane is relevant for the description of baryons in AdS/CFT \cite{WittBar} and its DBI dynamics has been discussed in \cite{CalGui}. Unfortunately, the 
coordinate system used in \cite{CalGui} is not very convenient for comparison with gravity solutions, so we need an alternative derivation presented below.}.
Using the static gauge $\xi_0=x_0$, \dots, $\xi_5=y_5$ and writing 
$z=Y(y_1,\dots,y_5)$, we find the induced metric $G_{AB}$ and the DBI action:
\bea
&&G_{00}=-H^{-1/2},\quad 
G_{ab}=H^{1/2}\left[\delta_{ab}+\d_a X\d_b X\right],\quad E_a=\nabla_a A,\nonumber\\
\label{D5DBI}
&&S^{DBI}_{D5}=
-T_{5}\int d^{6}\xi H
\sqrt{(1-E^2)(1+(\nabla X)^2)+({\bf E}\nabla X)^2}.
\eea
While in the absence of the electric field there is no direct coupling between D5 brane and four--form potential, 
in the present case we do have a nontrivial contribution to the Chern--Simons term:
\bea\label{D5CSin}
S^{CS}_{D5}&=&T_5\int_{D5}  2\pi\alpha'F\wedge P[{\tilde C}_4]=
-T_5\int d\xi^0\int_{\mbox{vol}} dA\wedge P[{\tilde C}_4].
\eea
Here ${\tilde C}_4$ is defined by the relation
\bea
d{\tilde C}_4=^*_6dH=-\frac{4Q}{(z^2+{\bf y}^2)^3}
~^*_6\left[zdz+{\bf y}d{\bf y}\right].\nonumber
\eea
In particular if we choose a convenient gauge where
\bea
{\tilde C}_4=4Qy_a
\frac{\eps_{abcde}}{4!}dy_{bcde}\int_0^{z}
\frac{dw}{(w^2+{\bf y}^2)^3},\nonumber
\eea
then the pullback is especially simple. Plugging this expression in (\ref{D5CSin}), we can simplify the Chern--Simons coupling:
\bea
S^{CS}_{D5}=
-T_5\int d\xi^0\int_{\mbox{vol}} d\left[A {\tilde C}_4\right]+
4Q T_5\int d^6\xi A\d_a\left[\int ^X\frac{y_adw}{(w^2+{\bf y}^2)^3}\right].\nonumber
\eea
Using the relation
\bea
4\d_a\left[\int^X \frac{y_adw}{(w^2+{\bf y}^2)^3}\right]=
-\d_a\d_a\int ^X \frac{dw}{(w^2+{\bf y}^2)^2}+
\d_a\frac{\d_a X}{(z^2+{\bf y}^2)^2}
\nonumber
\eea
and dropping total derivatives from the Chern--Simons action\footnote{We are looking for configurations where gauge potential $A$ decays sufficiently fast as we go to infinity on the D5 brane, so the boundary terms do not contribute.}, we arrive at the final expression:
\bea\label{D5CS}
S^{CS}_{D5}=-T_5\int d^6\xi \nabla^2 A\int_0^X dw(H-1)|_{z=w}
-T_5\int d^6\xi (H-1)|_{z=X}\d_a A\d_a X.
\eea

To summarize, the action for the D5 brane is given by a sum of 
(\ref{D5DBI}) and (\ref{D5CS}). 
Writing equations of motion for $X$ and $A$ and setting $A=X$ in the result, we find
\bea
&&-\d_X H+\nabla(H\nabla X)-(H-1)\nabla^2 X-\d_X H(\nabla X)^2+
\nabla((H-1)\nabla X)=0,\nonumber\\
&&-\nabla(H\nabla X)-\nabla^2\int_0^X dw(H-1)_w+
\nabla((H-1)\nabla X)=0.\nonumber
\eea
To simplify the second equation, we rewrite the term containing the 
integral in a more transparent form\footnote{Here we defined 
${\tilde\nabla}_i H$ as a derivative taken at fixed value of $z$. Its 
relation to a total derivative is given by 
${\tilde\nabla}_i H=\nabla_i H-\d_X H\nabla_i X$. We also used the fact that $H$ is harmonic.
}
\bea
\nabla^2\int_0^X dw(H-1)_w&=&\nabla((H-1)\nabla X)+
\nabla X {\tilde\nabla}(H-1)+\int_0^X dw {\tilde\nabla}^2(H-1)|_{z=w}
\nonumber\\
&=&(H-1)\nabla^2 X+2\nabla(H-1)\nabla X-
((\nabla X)^2+1) \d_X H.\nonumber
\eea
Using this expression, one concludes that equations for $X$ and $A$ collapse to the same relation:
\bea\label{EOMD5inD3}
-(1+(\nabla X)^2)\d_X H+H\nabla^2 X+2
\nabla H\nabla X=0.
\eea
Even though this relation looks more complicated than the Laplace equation (\ref{ProbeHarmonic}), in section 
\ref{SectSbBcnd} we will see that (\ref{EOMD5inD3}) has a very simple interpretation once it is rewritten in different coordinates.

{\bf D3 spike in the geometry of D5.} Let us now consider probes in the geometry produced by $N$ coincident D5 branes\footnote{In this paper most of the metrics are written in the Einstein frame. However since the DBI action 
is usually written in terms of the string metric, we use this frame in 
(\ref{MetrD5}).}:
\bea\label{MetrD5}
&&ds_S^2=H^{-1/2}(-dt^2+d{\bf y}_5^2)+H^{1/2}(dz^2+d{\bf x}_3^2),\\
&&e^{2\Phi}=H^{-1},\quad 
F_3=^*_{10}(dH^{-1}\wedge dt\wedge d^5 y)=^*_{3}dH,\quad
H=1+\frac{Q}{(z^2+{\bf x}^2)}.\nonumber
\eea
We begin with putting a D3 brane on this background. As one goes to infinity, the effect of D5 branes become 
negligible, and in this region we expect the D3 brane to stretch along $t$ and $x_1,x_2,x_3$. This suggests a natural static gauge which can be imposed everywhere: 
$\xi^0=t$, $\xi_i=x_i$. The action describing D3 brane contains the DBI piece, and, in the presence of the electric field, there is also a Chern--Simons coupling with two--form potential. We analyze these two terms separately starting with DBI contribution:
\bea
S^{DBI}_{D3}&=&-T_{3}\int d^{4}\xi e^{-\Phi}
\sqrt{-\mbox{det}(G+2\pi\alpha' F)}\nonumber\\
&=&-T_{3}\int d^{4}\xi H
\sqrt{(1-E^2)(1+(\nabla X)^2)+({\bf E}\nabla X)^2}.
\eea
The evaluation of the Chern--Simons term follows the same steps as 
the derivation of (\ref{D5CS}), so we will be brief here. If one chooses a convenient gauge for $C_2$:
\bea
C_2=2Qx_a\frac{\eps_{abc}}{2}dx_{bc}\int_0^z\frac{dw}{(w^2+{\bf x}^2)^2},
\eea
then up to total derivatives, the Chern-Simons action becomes:
\bea
S^{CS}_{D3}&=&T_3\int_D3 2\pi\alpha'\wedge P[C_2]=
2QT_3\int d^4\xi A\d_a\left[\int_0^X\frac{x_a dw}{(w^2+{\bf x}^2)^2}\right]
\nonumber\\
&=&-T_3\int d^4\xi \left\{\nabla^2 A\int_0^X dw(H-1)+
(H-1)\d_a A\d_a X\right\}.
\eea
We observe that the action for D3 brane superficially looks the same as the sum of (\ref{D5DBI}) and (\ref{D5CS}), although the harmonic functions 
and the number of independent variables appearing in these two cases are different. In spite of this differences, 
one can see that the same manipulations that led to the (\ref{EOMD5inD3}) can be repeated here, and we 
conclude that for configurations with $A=X$ there is only one independent equation of motion:
\bea\label{EOMD3inD5}
-(1+(\nabla X)^2)\d_X H+H\nabla^2 X+2
\nabla H\nabla X=0.
\eea

{\bf D5 branes in D5 background.} Finally we analyze the D5 brane in the 
geometry (\ref{MetrD5}). To do this it is convenient to describe the Ramond--Ramond field in terms of the dual six--form potential:
\bea
C_6=H^{-1}dt\wedge d^5 y.
\eea
Then the action for D5 brane becomes:
\bea
S_{D5}&=&
-T_{5}\int d^{5}\xi H^{-1}
\sqrt{(1-HE^2)(1+H(\nabla X)^2)+H^2({\bf E}\nabla X)^2}\nonumber\\
&&+T_5\int d^4\xi H^{-1}.
\eea
Equations of motion are satisfied by ${\bf E}=\nabla X$ as long as $X$ is a 
harmonic function.


\subsection{Spikes in M theory.}
\label{SbctMtchPrb}


In the last two subsections we discussed various configurations of branes with fluxes in type IIB string theory. A similar analysis can be performed in M theory as well and we will outline it here.

M theory has two fundamental objects: M2 and M5 branes. In string theory we looked at fundamental strings ending on D brane, and the closest analog of this configuration in eleven dimensions is a set 
of membranes ending on M5 brane. To preserve supersymmetry, the branes should intersect on a line,
and the third object can be added without breaking additional supersymmetry:
\bea\label{MemBraneScan}
\begin{array}{c|cccccccccc}
&1&2&3&4&5&6&7&8&9&10\\
M5&\bullet&\bullet&\bullet&\bullet&\bullet&&&&&\\
M2&&&&&\bullet&\bullet&&&&\\
M5'&&&&&\bullet&&\bullet&\bullet&\bullet&\bullet\\
\end{array}
\eea
Notice that one can arrive at configuration (\ref{MemBraneScan}) by 
starting from (\ref{BraneScan}), T dualizing along $x_5$ 
and lifting to eleven dimensions. 

To analyze the dynamics of various branes in (\ref{MemBraneScan}), 
it is convenient to start with a metric produced by a stack of $N$ 
five--branes which have the same orientation as M5 in 
(\ref{MemBraneScan}). Then 
we can probe this geometry by either M5 or M5' with M2 branes attached 
to them. The M5--M2 configuration in flat space will be recovered if we set $N=0$. 

{\bf M5 spike in the M5 geometry.} We begin with quoting geometry produced by a stack of M5 branes
\cite{gueven}:
\bea\label{M5geometry}
ds_{11}&=&H^{-1/3}(-dt^2+d{\bf x}_4^2+dw^2)+H^{2/3}(dz^2+d{\bf y}_4^2),\nonumber\\
F_4&=&*d\left[H^{-1}dt\wedge d^4 x\wedge dw\right],\quad H=1+\frac{Q}{(z^2+{\bf y}^2)^{3/2}}.
\eea
To study dynamics of an additional M5 brane with flux we need an analog of the DBI action, where instead of the one--form gauge field one has a two--form potential 
on the worldvolume. Since the three--form field strength has to be self--dual, finding of such action is a 
nontrivial task and there have been various proposals in the literature \cite{MDBI,PST}.  We will use a 
formalism based on introduction of one auxiliary field $a$ \cite{PST}:
\bea\label{PSTAct}
S_{PST}&=&
-\int d^6\xi \left[\sqrt{-\mbox{det}(g_{mn}+i{\tilde F}_{mn})}+
\frac{\sqrt{-g}}{4(\nabla a)^2}\d_m a F^{*mnl}F_{nlp}\d^p a\right].
\eea
The dynamical variable is a two--form $B_{mn}$ and following \cite{PST}
we introduced
\bea\label{PSTDual}
F=2dB,\quad F^{*mnl}=\frac{1}{6\sqrt{-g}}\eps^{mnlabc}F_{abc},\quad
{\tilde F}_{mn}=\frac{1}{\sqrt{(\nabla a)^2}}F^*_{mnl}\d^l a.
\eea
As usual, we can fix the invariance under diffeomorphisms by choosing the 
static gauge $\xi^0=t$, $\xi_1=x_1$, \dots $\xi_4=x_4$, $\xi_5=w$, but 
(\ref{PSTAct}) has an additional gauge invariance and to fix it one can make $a$ to be any convenient function of the worldvolume coordinates (see \cite{PST} for further discussion). In the present case the natural choice is
$a=w$. From (\ref{MemBraneScan}) it is clear that in the absence of M5' branes we expect to have a translational invariance in 
$x_5\equiv w$ and in time, moreover since the worldvolume of M2 always contains these two directions, it is natural to parameterize $F$ in terms of a one--form $\omega$ as $F^*=\omega\wedge dt\wedge dw$. 
For this set of fields the relations (\ref{PSTDual}) become
\bea
\omega\wedge dt\wedge dw=2*dB,\quad 
{\tilde F}=H^{1/6}\omega\wedge dt.
\eea
Here the Hodge dual is computed using the six dimensional metric induced on the M5 brane:
\bea\label{M5DefGtil}
g_{mn}=H^{-1/3}\eta_{mn}+H^{2/3}\d_m X\d_n X\equiv H^{-1/3}{\tilde g}_{mn}.
\eea
In our gauge the last term in (\ref{PSTAct})
drops out and the action can be rewritten in terms of the vector $\omega$:
\bea\label{PST1}
S_{PST}&=&-\int d^6\xi H^{-1}\sqrt{(1+H\omega^2)(1+H(\nabla X)^2)-
H^2(\omega_m \nabla_m X)^2}.
\eea
The remaining part of the action comes from the direct coupling of $C_6$ with M5 brane and it has a very simple form:
\bea
S_{CS}=\int d^6\xi H^{-1}.
\eea
It is now convenient to perform a dualization similar to the one discussed in 
\cite{tseytlin}. To do so we introduce two Lagrange multipliers: one to enforce the relation between $\omega$ and $B$ and another one to make the action quadratic in $\omega$:
\bea\label{PST2}
S&=&S_{CS}-\frac{1}{2}\int d^6\xi H^{-1}(V+PV^{-1})+\int d^6\xi \Lambda^m(\omega_m-
2(^*_6 dB)_{mtw}),\\
P&\equiv&-\mbox{det}({\tilde g}_{ab}+2iH^{1/2}\omega_{[a}\delta^t_{b]})=
-(1+H{\tilde g}^{mn}\omega_m\omega_n)~\mbox{det}{\tilde g}_{ab}.
\eea 
Notice that $P$ is the expression which appears under the square root in 
(\ref{PST1}), but to simplify the discussion below we wrote it in terms of the metric 
${\tilde g}_{ab}$ defined by (\ref{M5DefGtil}).

Taking variation with respect to $\omega_m$ and $B$, we find two equations:
\bea\label{LambdaM}
&&\Lambda^m=-V^{-1}\mbox{det}{\tilde g}~{\tilde g}^{mn}\omega_n,\\
&&d\left(\frac{\Lambda}{\sqrt{-\mbox{det}{\tilde g}}}\right)=0:\qquad 
\Lambda=\sqrt{-\mbox{det}{\tilde g}}~dA.
\eea
These relations allow one to eliminate $\omega_m$ from the action (\ref{PST2}): 
\bea
S=S_{CS}-\frac{1}{2}\int d^6\xi \left[V(H^{-1}+\frac{1}{\mbox{det}{\tilde g}}\Lambda^2)-
\mbox{det}{\tilde g}~ (HV)^{-1}\right].
\eea
Integrating out the auxiliary field $V$ and substituting the expressions for 
${\tilde g}_{mn}$ and $\Lambda^m$, we arrive at the final action:
\bea\label{LastM5M5}
S&=&S_{CS}-\int d^6\xi \sqrt{-\mbox{det}{\tilde g}~H^{-1}(H^{-1}-
{\tilde g}^{mn}\d_m A\d_n A)}\\
&=&\int d^6\xi H^{-1}-\int d^6\xi H^{-1}\sqrt{(1+H(\nabla X)^2)(1-H(\nabla A)^2)+
H^2(\nabla A\nabla X)^2}.\nonumber
\eea
This action has been encountered before (see equation (\ref{D3inD3Act})) and we showed that any 
harmonic function 
$X$ leads to a solution if one also sets $A=X$. Previously this action 
arose from the analysis of D3 or D5 branes, and in the present context  
(\ref{LastM5M5}) can be viewed as a DBI action for D4 branes: we started with a set of M5 and effectively did a dimensional reduction along $w$.

{\bf M5' spike in M5 geometry.} We again use the metric (\ref{M5geometry}), however in this case it is 
more convenient to use a magnetic three--form potential instead of an electric six--form:
\bea
d{\tilde C}_3=-\frac{3Q}{(z^2+{\bf y}^2)^{5/2}}~^*_5[zdz+{\bf y}d{\bf y}]:\quad
{\tilde C}_3=3Qy_a\frac{\eps_{abcd}}{3!}dy_{bcd}\int_0^z
\frac{d\zeta }{(\zeta^2+{\bf y}^2)^{5/2}}.
\eea
According to \cite{PST} the magnetic potential couples to M5' brane both through PST action and through Chern--Simons term. The former coupling is accomplished by a replacement $F_{abc}\rightarrow F_{abc}-C^{(3)}_{abc}$ in 
(\ref{PSTAct}), (\ref{PSTDual}), and the latter is given by
\bea\label{ZeroChern}
S_{CS}=\frac{1}{2}\int F\wedge C^{(3)}.
\eea
In the present context we can impose a static gauge with worldvolume coordinates
$(t,w,{\bf y})$, then the profile of M5' would be described by\footnote{We did not look at a more general profile 
$z=X({\bf y},w)$ since configuration (\ref{MemBraneScan}) has a translational invariance in $w$.} $z=X({\bf y})$ 
and the induced metric becomes
\bea
g_{00}=-H^{-1/3},\quad g_{ww}=H^{-1/3},\quad 
g_{mn}=H^{2/3}\left[\eta_{mn}+\d_m X\d_n X\right]\equiv H^{2/3}{\tilde g}_{mn}.
\eea
Taking into account the orientation of M2 brane given in (\ref{MemBraneScan}), it is reasonable to choose a 
gauge $a=w$ and to assume that the only non--vanishing component 
of $F^*$ is $F^*_{twm}\equiv \omega_m$. In particular we observe that for for this class of configurations, the second term in the PST action (\ref{PSTAct}) does not contribute.
The differential equation for 
$\omega_m$ is given by
\bea
\omega_m=\left[^*_6(2dB-C^{(3)})\right]_{tzm}
\eea
and as before we will enforce this relation via Lagrange multiplier. Introducing another multiplier to eliminate the square root as in (\ref{PST2}), we find the PST action:
\bea\label{PST3}
S_{PST}&=&-\frac{1}{2}\int d^6\xi H(V+PV^{-1})+\int d^6\xi \Lambda^m(\omega_m-
[^*_6 (2dB-C^{(3)})]_{mtw}),\\
P&\equiv&-\mbox{det}({\tilde g}_{ab}+2iH^{-1/6}(H^{1/6}\omega)_{[a}\delta^t_{b]})=
-(1+{\tilde g}^{mn}\omega_m\omega_n)~\mbox{det}{\tilde g}_{ab}.
\nonumber
\eea 
With our choice of gauge, $F\wedge C^{(3)}=0$, so the Chern--Simons term (\ref{ZeroChern}) does not contribute. We can integrate out $B$ and 
$\omega$ using their equations of motion\footnote{To arrive at the equation for 
$\Lambda$ one should notice that 
${\eps_{tw}}^{mabc}\Lambda_m (dB)_{abc}=
H^{-5/3}{\tilde\eps}^{mabc}\Lambda_m (dB)_{abc}$.}:
\bea
&&d\left(\frac{H^{-5/3}\Lambda}{\sqrt{-{\tilde g}}}\right)=0:\quad 
\Lambda_m=H^{5/3}\sqrt{-{\tilde g}}~\d_m A,\nonumber\\
&&\omega_m=\frac{H^{-1}V}{(-{\tilde g})}{\tilde g}_{mn}\Lambda^n
=\frac{1}{\sqrt{-{\tilde g}}}\d_m A,\nonumber
\eea
and rewrite (\ref{PST3}) as an action for $A$:
\bea\label{PST4}
S_{PST}&=&-\frac{1}{2}\int d^6\xi H\left[
V\left(1-\frac{(\d A)^2}{(-{\tilde g})}\right)-
{\tilde g}V^{-1}\right]+\int d^6\xi \Lambda^m (^*_6 C^{(3)})_{mtz}\nonumber\\
&=&-\int d^6\xi H\sqrt{-{\tilde g}(1-{\tilde g}^{mn}\d_m A\d_n A)}-
\int d^6\xi A\d_a\int_0^Xd\zeta \d_a H({\bf y},\zeta).
\eea
To arrive at the last term we used the following transformations:
\bea
\Lambda^m(^*_6 C^{(3)})_{mtz}=(H\sqrt{-{\tilde g}}{\tilde g}^{mn}\d_n A)
\frac{H^{-1}{{\tilde\eps}_{mtz}}^{\ \ \ abc}}{3!}C^{(3)}_{abc}=
3Qy^m\d_m A \int_0^X\frac{d\zeta }{(\zeta^2+{\bf y}^2)^{5/2}}.\nonumber
\eea
We observe that the action (\ref{PST4}) looks similar to the sum of (\ref{D5DBI}) and (\ref{D5CS}), the difference is hidden in the harmonic function $H$. To derive equations of motion coming from 
(\ref{D5DBI}), 
(\ref{D5CS}) we only used a harmonicity of $H$, so repeating the similar steps here and setting $A=X$ we arrive at the equation
\bea\label{EOM5M5pr}
-(1+(\nabla X)^2)\d_X H+H\nabla^2 X+2
\nabla H\nabla X=0.
\eea

\subsection{Summary}

\begin{figure}[tb]
\begin{center}
\epsfysize=2.2in \epsffile{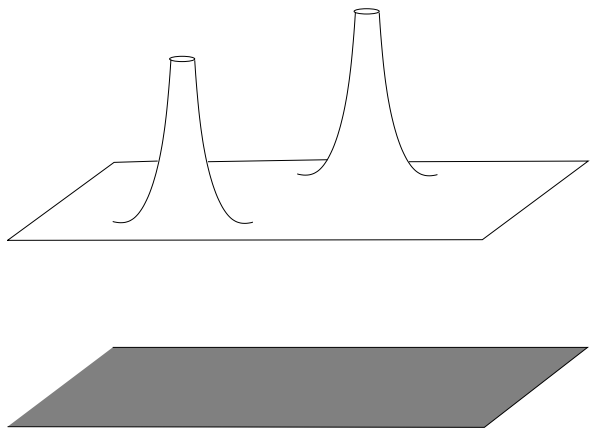}
\end{center}
\caption{A probe Dp brane with electric flux in the presence of $N$ parallel Dp branes.
} \label{FigStack}
\end{figure}

Let us summarize the results of this section. We looked at geometries produced by stacks of various branes and studied dynamics of various probe objects on such backgrounds. If the probe branes have the same type as the objects which created geometry, then their profiles in transverse coordinates 
are governed by a harmonic equation:
\bea
\nabla^2 X=0.
\eea
Notice that the probe branes become parallel to the original stack only at infinity: in the interior of the space the probes are curved (see figure \ref{FigStack}) and they have a nonvanishing electric field. 
This field is responsible for breaking eight out of $16$ supersymmetries which would be preserved by the parallel branes. We considered three examples of such setup: D3--D3, D5--D5 systems in type IIB theory and M5--M5 configuration on M theory. In the first two cases the worldvolume flux sources fundamental strings, while in eleven dimensions it mimics M2 branes.  


We also looked at other configurations preserving eight supercharges in ten dimensions: they were constructed by putting D3 branes on a D5 geometry or by putting D5 branes on a D3 geometry. 
In both cases the brane profiles 
$X({\bf x})$ were governed by the same nonlinear equation
\bea\label{NonlLapl}
-(1+(\nabla X)^2)\d_X H+H\nabla^2 X+2
\nabla H\nabla X=0,
\eea
where $H({\bf x},X)$ was a harmonic function describing the background. The same equation was found to describe a profile of M5' brane on the M5 geometry (see (\ref{MemBraneScan})) and for future reference we summarize the harmonic functions for the three situations:
\bea
H_{D3}=1+\frac{Q_3}{(X^2+{\bf x}_5^2)^2},\quad H_{D5}=1+\frac{Q_5}{X^2+{\bf x}_3^2},\quad 
H_{M5}=1+\frac{Q_{M5}}{(X^2+{\bf x}_4^2)^{3/2}}.
\eea
Here a subscript of ${\bf x}$ denotes the number of components of this vector. It might be somewhat 
counterintuitive that positions of supersymmetric probes are described by a nonlinear equation like (\ref{NonlLapl}): 
looking at configuration (\ref{BraneScan}), one would expect that the branes can be freely superposed. In section 
\ref{SbctMtchPrb} we will show that this expectation is correct and equation (\ref{NonlLapl}) can be linearized 
by a change of variables.

Notice that not only equation (\ref{NonlLapl}) is nonlinear, it also has a term which does not have derivatives of 
$X$, so constant $X\ne 0$ is not a solution. Thus if one starts from a geometry produced by D3 branes and adds a probe D5 passing through some point $(X,{\bf x})$, then to be supersymmetric, this probe must have nontrivial electric field $E_i=\nabla_i X$ on the worldvolume and fundamental strings must be sourced. Of course, as one goes to infinity in ${\bf x}$ directions, (\ref{NonlLapl}) reduces to a usual Laplace equation and flat D5 branes are allowed for any value of $X$. However, as we just argued, unless such brane is placed at $X=0$, it will become curved in the interior and it will have fundamental string attached to it. This situation should be contrasted to the 
case of supersymmetric D3 probes which can be placed anywhere and still remain flat.

While the discussion of the last paragraph pertains to a geometry created by a stack of D3 branes, the same argument can be made for metrics produced by D5 and M5 branes since the probe objects are still described by 
the equation (\ref{NonlLapl}). 

Once we established that the branes with fluxes are supersymmetric and they can be superposed, it is natural to look at configurations which contain many such branes on top of each other. As usual, when the brane charge becomes large, such stacks are expected to modify the geometry and the remaining part of this paper is devoted to finding the appropriate gravity solutions. 


\section{Single spike in IIB supergravity}
\label{SectSpike}

\renewcommand{\theequation}{3.\arabic{equation}}
\setcounter{equation}{0}


In the previous section we reviewed the construction of branes with fluxes in the probe approximation and our next task is to find the geometries which are generated by such branes. We will start with analysis on the type IIB side and first we assume a large bosonic symmetry which is present in the case of a single spike. Then we will be able to {\it derive} the form of the supergravity solution and express it in terms of two functions which obey three differential equations. In the next section we will generalize the solution to the case of multiple spikes and discuss the boundary conditions.

\subsection{Summary of the solution.}

Let us consider a stack of D3 branes and a stack of fundamental strings with orientations described in 
(\ref{BraneScan}). This diagram suggest that the configuration has a rotational symmetry between 
$(x_1,x_2,x_3)$ and between $(x_5,\dots,x_9)$. From the point of view of brane probes described in subsection 
\ref{SectSbBion}, the 
$SO(5)$ symmetry is automatic, while the $SO(3)$ symmetry implies that in 
(\ref{ProbeHarmonic}) we choose a function $X$ which depends only on the radial coordinate along D3 brane. Once the number of branes becomes large, the geometry is modified, but one expects the $SO(5)\times SO(3)$ symmetry 
to remain unbroken. Moreover, the solution corresponding to BPS branes is expected to be static, so we arrive at 
the following ansatz for the metric:
\bea\label{IIBMetrAns}
ds^2=-e^{2A}dt^2+e^{2B}d\Omega_2^2+e^{2C}d\Omega_4^2+h_{ij}dx^idx^j.
\eea
Here and below the indices $i,j$ are running over the three remaining coordinates and all scalars are taken 
to be functions of $x^i$. To describe a configuration of fundamental strings ending on D3 brane, we need to have a nontrivial $F_5$ and an electric component of the NS--NS flux:
\bea
H_3&=&2\omega_2\wedge dt,\quad
F_5=df_3\wedge d\Omega_4+dual,\quad e^{\phi}.
\eea
Here $\omega_2$ is a closed two--form in three--dimensional space spanned by $x_i$. The equation of motion\footnote{Our conventions for the supergravity fields are summarized in the Appendix \ref{AppSugra}.} for $F_3$:
\bea
d*(e^\phi F_3)=4F_5\wedge H_3
\eea
implies that we should excite at least one component of this three form: 
$F_3=df_2\wedge d\Omega_2$. One can see that the dilaton will be generated 
as well. While there are other fields consistent with 
$SO(5)\times SO(3)\times U(1)$ symmetry, the set which we just described gives 
a consistent truncation of type IIB supergravity: to see this one should look at a $Z_2$ symmetry which acts by reversing the signs of all RR fields and changes the 
orientations of $S^2$ and $S^4$ as well\footnote{This symmetry was also used to restrict the ansatz in \cite{myWils}}. It is clear that the only fields that are invariant under this $Z_2$ and 
$SO(5)\times SO(3)\times U(1)$ symmetries are
\bea\label{IIBFluxAns}
H_3&=&2\omega_2\wedge dt,\quad F_3=df_2\wedge d\Omega_2,\quad
F_5=df_3\wedge d\Omega_4+dual,\quad e^{\phi}.
\eea
One can write the equations for the Killing spinors for the geometry 
(\ref{IIBMetrAns}), (\ref{IIBFluxAns}), these equations are solved in the appendix \ref{AppSpike} and here we 
just quote the result:
\bea
ds^2=e^H\left[-e^{3\phi/2}dt^2+e^{-\phi/2}(dv^2+v^2 d\Omega_2^2)\right]+
e^{-H-\phi/2}(du^2+u^2 d\Omega^2_4)+e^{3H+3\phi/2}(dw+{\cal A})^2\nonumber
\eea
\bea\label{SymSltnSum}
&&{\cal A}=A_u du=\frac{\d_u F}{\d_w F},\quad e^{2H}=\d_w F,\quad
F_5=-\frac{1}{4}d(u^4A_u)\wedge d\Omega_4+dual\nonumber\\
&&H_3=d\left[e^{2H+2\phi}(dw+{\cal A})\right]dt,\quad F_3=d(v^2\d_v F)\wedge d\Omega_2.
\eea
The solution is parameterized by two functions $F$, $e^\phi$ and they obey differential equations:
\bea\label{IIBeqn1}
&&\d_w e^{-2\phi}+v^{-2}\d_v(v^2\d_v F)=0,\\
\label{IIBeqn2}
&&u^4 e^{-2H}\d_w e^{-2H-2\phi}-(\d_u-A_u\d_w)(u^4 A_u)=0.
\eea
The last relation can also be rewritten in terms of the coordinates $(u,v,F)$:
\bea\label{IIBeqn2a}
u^4\d_F e^{-2H-2\phi}+\d_u(u^4 \d_u w)|_{v,F}=0.
\eea

It turns out that the equations for the Killing spinors are not sufficient to determine $F$ and dilaton completely. The simplest way to see this is to observe that the system
(\ref{SymSltnSum}) should be applicable for the description of fundamental strings 
in the absence of D3 branes. Requiring $F_5$ and $F_3$ to vanish, we find that 
$F$ can depend only on $w$, then equations (\ref{IIBeqn1}) and (\ref{IIBeqn2}) reduce to two simple statements: $H$ has to be a constant and the 
dilaton is an arbitrary function of $(u,v)$. Of course we do not expect the dilaton to be arbitrary for the fundamental string, this shows that (\ref{IIBeqn1}) and (\ref{IIBeqn2}) do not give a complete set of equations. In the case of fundamental string, the missing relation comes from the equation of motion for $H_3$, so one may suspect that this equation should be added for a general solution as well. 

The only nontrivial component of the equation for 
the Kalb--Ramond two--form is evaluated in the appendix \ref{AppSbEOM}:
\bea\label{IIBeqn3}
v^2\d_u\left[u^4\d_u e^{-2\phi}\right]+\d_v\left[v^2 u^4\d_ve^{-2\phi-2H}\right]+
u^4 v^4\Delta_u(e^{2H}\d_v w\d_v w)|_{v,F}=0
\eea
and it turns out that (\ref{IIBeqn1}), (\ref{IIBeqn2}), (\ref{IIBeqn3}) form a complete 
set of equations. We postpone the proof of this fact until 
subsection \ref{SectSbPert}, here we just notice that for a fundamental string relation
(\ref{IIBeqn3}) leads to a standard harmonic equation for the dilaton:
\bea
u^{-4}\d_u\left[u^4\d_u e^{-2\phi}\right]+e^{-2H}v^{-2}\d_v\left[
v^2\d_ve^{-2\phi}\right]=0.
\eea

To summarize, we have shown that for the ansatz (\ref{IIBMetrAns}), 
(\ref{IIBFluxAns}), the equations for the Killing spinors can be solved to yield 
the result (\ref{SymSltnSum}) which is parameterized  by two functions 
$F,e^\phi$ satisfying (\ref{IIBeqn1}) and (\ref{IIBeqn2}). We also argued that in general 
these two equations should be supplemented by (\ref{IIBeqn3}) to give a complete {\it local} description of the geometry. To specify the unique solution, one 
should also impose boundary conditions at infinity and at the points where one of the spheres contacts to zero size. To avoid repetition, we will not discuss these boundary conditions here, but perform the analysis for more general solutions in subsection \ref{SectSbBcnd}.


\subsection{Comparison with geometries dual to Wilson lines.}
\label{SecSbWils}


In the previous subsection we analyzed the geometries with $SO(5)\times SO(3)\times U(1)$ symmetry. 
The motivation came from studying D3 branes and fundamental strings in flat space, so the most interesting solutions are asymptotically flat. In section \ref{SectSbScan} we saw that a combination 
of D3 and F1 in flat space can preserve at most eight real supercharges, and it was precisely 
such 1/4-BPS configuration that was analyzed in section \ref{SectSbBion} and in the previous subsection. 
It turns out that the situation is different if space asymptotes to $AdS_5\times S^5$. In this case one can find D3 branes with fluxes which preserve 16 supercharges \cite{rey,Drukker} and the set of fluxes in the corresponding supergravity solutions is similar to (\ref{IIBFluxAns}) \cite{yama,myWils}. The solutions described in \cite{myWils} preserve twice as many supersymmetries as 
(\ref{SymSltnSum}), and they also have a bigger set of bosonic symmetries: 
$SO(5)\times SO(3)\times U(1)$ is enhanced to $SO(5)\times SO(3)\times SO(2,1)$. In this subsection we discuss the relation between these two classes of geometries. We will only present the results and the details of computations can be found in the Appendix \ref{AppCmpr}.

To embed the solutions of \cite{myWils} with $SO(5)\times SO(3)\times SO(2,1)$ symmetry 
into a more general class of geometries described by (\ref{SymSltnSum}), we need to recall the metric found in 
\cite{myWils}:
\bea\label{BWilsMetric}
ds^2=ye^{S-\phi/2}dH_2^2+ye^{G-\phi/2}d\Omega_2^2+ye^{-G-\phi/2}d\Omega_4^2+
\frac{e^{-\phi}}{2y\cosh G}(dx^2+dy^2).
\eea
The warp factors entering this expression are specified in terms of one harmonic function, but since these 
relations are fairly complicated we refer to \cite{myWils} for details. 

Starting from the geometry (\ref{BWilsMetric}) one can look for a change of coordinates which puts the metric in the form (\ref{SymSltnSum}). It is natural to identify the spheres in these two descriptions, so one only needs to find the map for the remaining four coordinates. To extract time, we write the metric on $AdS_2$ as\footnote{One should use Poincare patch rather than global coordinates, since in derivation of (\ref{SymSltnSum}) the spinor was assumed to be $t$--independent.}
\bea
dH_2^2=-z^2 dt^2+\frac{dz^2}{z^2}.
\eea
Then matching the  coefficients in front of $dt^2$, $d\Omega_2^2$, $d\Omega_4^2$ in 
(\ref{BWilsMetric}) and (\ref{SymSltnSum}), we arrive at the relations
\bea\label{BWlsToNew}
e^{H}=yz^2 e^{S-2\phi},\quad
u^2=ye^{H-G}=y^2 z^2 e^{S-G-2\phi}\quad
v^2=ye^{G-H}=z^{-2} e^{-S+G+2\phi}.
\eea
This leaves only one undetermined coordinate $w$ and in Appendix \ref{AppCmpr} we derive the 
expression for its differential  (\ref{WlsToNew1}):
\bea\label{BWlsToNew1}
dw+A_udu&=&e^{-2H}\left[y^{1/2}e^{(S-4\phi)/2+\phi/4}{\cal F}dz+ze^{-2\phi}dx\right],\\
{\cal F}&=&\sqrt{ye^{-\phi/2}(e^S-e^G-e^{-G})}.\nonumber
\eea  
Equations (\ref{BWlsToNew}), (\ref{BWlsToNew1}) allow one to recover a unique set of coordinates $(u,v,w)$ starting from any solution with $SO(5)\times SO(3)\times SO(2,1)$ symmetry. 

To give an example of a more explicit map from $(x,y,z)$ to $(u,v,w)$ coordinates, we look at
$AdS_5\times S^5$.
In this case it is convenient to parameterize $x$, $y$ in terms of the radial coordinate $\rho$ on AdS and an angle 
$\theta$ on the sphere (see \cite{myWils} for details):
\bea
x=\cosh\rho\cos\theta,\quad y=\sinh\rho\sin\theta.
\eea
In terms of these variables one finds
\bea
e^S=y^{-1}\cosh^2\rho,\quad e^G=y^{-1}\sinh^2\rho,\quad {\cal F}=\cos\theta.
\eea
Substituting this into (\ref{BWlsToNew}), we find the expressions for $u$, $v$ and $e^H$:
\bea
u=z\sin\theta\cosh\rho,\quad v=z^{-1}\tanh\rho,\quad e^H=z^2\cosh^2\rho.
\eea
Since are looking at a solution with $e^\phi=1$, the relation (\ref{BWlsToNew1}) can be simplified:
\bea
dF-\d_v F dv=d(zx).
\eea  
This equation can be easily solved ($F=zx+{\tilde F}(v)$), then recalling the definition 
\bea
e^{-2H}=\d_F w,\nonumber
\eea
we find the expression for the differential of $w$:
\bea
dw=e^{-2H}dF|_{u,v}=e^{-2H}d(zx)=-d\left(
\frac{2\theta-\sin 2\theta}{4(z\cosh\rho\sin\theta)^3}\right).
\eea
Similarly, starting from any other solution of \cite{myWils}, one can use (\ref{BWlsToNew}) and 
(\ref{BWlsToNew1}) to find $w$ as a function of $(x,y,z)$.

To summarize, we showed that the solutions (\ref{BWilsMetric}) can be embedded into the coordinate system defined by (\ref{SymSltnSum}). Of course, the geometries (\ref{BWilsMetric}) represent only a small subclass of the metric discussed in this section, in particular they 
preserve 16 supercharges, rather than eight which were used to construct (\ref{SymSltnSum}).


\subsection{Relation to non--commutative theories.}
\label{SecSbNonCm}


Solution (\ref{SymSltnSum}) describes a geometry produced by D3 (or D5) branes with worldvolume fluxes 
and similar 
systems have been studied in connection with non--commutative field theories. To introduce non--commutativity 
on the field theory side, one turns on a constant 
Kalb--Ramond field on the brane \cite{SeibWitNC}, and on the bulk side the relevant geometries
have been constructed in \cite{HasItz,MalRus}. In this subsection we will recover these solutions 
by taking a certain limit of (\ref{SymSltnSum}). 

We begin with recalling that for a flat D3 brane without fluxes one has a simple expression for 
$e^{-2H}$:
\bea
e^{-2H}=1+\frac{Q}{(u^2+F^2)^2}.
\eea
The worldvolume of the brane is parameterized by $(t,v,S^2)$ and to zoom in on some point 
on D3 one should introduce a rescaling
\bea\label{RecZoom}
t=\eps{\tilde t},\quad v=v_0+\eps{\tilde v},\quad d\Omega^2_2=\eps^2d{\bf x}_2^2
\eea
and send $\eps$ to zero. In a more complicated case of branes with fluxes, the worldvolume 
is still parameterized by $(t,v,S^2)$, but now the position of the brane in $F$--direction can depend on $v$. 
However even in that situation the rescaling (\ref{RecZoom}) can be used to zoom in on a particular point on D3, 
and in the limit $\eps\rightarrow 0$ the brane becomes flat. To recover regular solution from (\ref{SymSltnSum}), redefinition (\ref{RecZoom}) should be supplemented by additional 
rescalings:
\bea
e^H=\eps^{-2}e^{\tilde H},\quad u=\eps^{-1}{\tilde u},\quad F=\eps^{-1}{\tilde F},\quad
w=\eps^3{\tilde w}.
\eea
Notice that the dilaton has a trivial $\eps$ dependence, in particular this implies that 
$\d_{\tilde v}e^\phi=0$. Since we started with regular $g_{tt}$, the zooming procedure eliminates ${\tilde v}$--dependence from this component of the metric, then we conclude that 
$\d_{\tilde v}e^{\tilde H}|_{{\tilde u},{\tilde F}}=0$ in the limit $\eps\rightarrow 0$. Similarly, a regularity of 
$F_3$ implies that 
\bea\label{MRdFdV}
0=\d_{\tilde v}F_3|_{{\tilde u},{\tilde F}}=\d_{\tilde v}\left[
d(\d_{\tilde v}{\tilde F}|_{{\tilde u},{\tilde w}})\wedge d^2{\bf x}\right]_{{\tilde u},{\tilde F}}:\qquad
\d_{\tilde v}{\tilde F}|_{{\tilde u},{\tilde w}}=h({\tilde u},{\tilde F}).
\eea
Let us rewrite the equations (\ref{IIBeqn1}) and (\ref{IIBeqn2a}) in the $\eps\rightarrow 0$ limit:
\bea\label{MReqn1}
&&\d_{\tilde w} e^{-2\phi}+\d_{\tilde v}^2 {\tilde F}|_{{\tilde u},{\tilde w}}=0,\\
\label{MReqn2}
&&{\tilde u}^4\d_{\tilde F} e^{-2{\tilde H}-2\phi}+\d_{\tilde u}
({\tilde u}^4 \d_{\tilde u} {\tilde w})|_{{\tilde v},{\tilde F}}=0.
\eea
Using the relation
\bea
\d_{\tilde v}|_{{\tilde u},{\tilde w}}=\d_{\tilde v}|_{{\tilde u},{\tilde F}}+
\d_{\tilde v}{\tilde F}|_{{\tilde u},{\tilde w}}~\d_{\tilde F}|_{{\tilde u},{\tilde v}}=
\d_{\tilde v}|_{{\tilde u},{\tilde F}}+h\d_{\tilde F}|_{{\tilde u},{\tilde v}}
\eea
and definition of $e^{\tilde H}$, one can simplify equation (\ref{MReqn1}):
\bea\label{MReqn3}
e^{2\tilde H}\d_{\tilde F} e^{-2\phi}+h\d_{\tilde F} h=0.
\eea
To relate $e^{2\tilde H}$ and ${\tilde F}$, we differentiate (\ref{MRdFdV}) with respect to 
${\tilde w}$ and compare the result with an expression for 
$\d_{\tilde v}e^{2\tilde H}|_{{\tilde u},{\tilde w}}$: 
\bea
\d_{\tilde w}(\d_{\tilde v}{\tilde F})=e^{2\tilde H}\d_{\tilde F}h,\quad
\d_{\tilde v}(\d_{\tilde w}{\tilde F})=h\d_{\tilde F}e^{2\tilde H}.\nonumber
\eea
Integrability condition for these two equations requires a particular combination of $h$ and 
$e^{\tilde H}$ to 
be ${\tilde F}$--independent:
\bea
he^{-2\tilde H}=h_1(\tilde u).
\eea
Finally we can solve (\ref{MReqn3}) and substitute the result into an ${\tilde F}$--derivative of (\ref{MReqn2}):
\bea\label{MReqn4}
&&e^{-2\phi}=-h_1^2(\tilde u)e^{2{\tilde H}}+h_2({\tilde u}),\quad h=e^{2{\tilde H}}h_1(\tilde u),\\
\label{MReqn5}
&&\d^2_{\tilde F} (h_2e^{-2\tilde H})+\Delta_{\tilde u}e^{-2\tilde H}=0.
\eea
The functions $h_1$ and $h_2$ are not arbitrary and to find the restrictions imposed on them, we begin with 
rewriting (\ref{MRdFdV}) in terms of $h_1$:
\bea
\d_{\tilde v}{\tilde F}-h_1({\tilde u})\d_{\tilde w}{\tilde F}=0:\quad 
{\tilde F}={\tilde F}({\tilde w}+h_1{\tilde v},{\tilde u}),\quad
{\tilde w}=-h_1{\tilde v}+{\hat w}({\tilde F},{\tilde u}).
\eea
Combining the last relation with (\ref{MReqn2}), we conclude that $\Delta_u h_1=0$. Then equation for the flux 
(\ref{IIBeqn3}) reduces to a simple relation $\Delta_u h_2=0$, so we arrive at the complete solutions for 
$h_1$, $h_2$ in terms of four constants $g_1,g'_1,g_2,g'_2$:
\bea
h_1=g_1+\frac{g'_1}{{\tilde u}^3},\quad h_2=g_2+\frac{g_2'}{{\tilde u}^3}.
\eea
To avoid singularity at ${\tilde u}=0$, one must set $g'_1=g'_2=0$. 

We can now rewrite the complete solution (\ref{SymSltnSum}) in terms of two constants $g_1$ and $g_2$ and 
a function $e^{-2{\tilde H}}$ which satisfied a Laplace equation (\ref{MReqn5}):
\bea\label{MRD3soln}
ds^2&=&e^{-\phi/2}\left[e^{\tilde H+2\phi}\left\{-d{\tilde t}^2+e^{-2\phi}v_0^2 d{\bf x}_2^2+
g_2(d{\tilde v}-\frac{g_1}{g_2}d{\tilde F})^2\right\}\right.\nonumber\\
&&\qquad\left.+
e^{-{\tilde H}}
\left(d{\tilde u}^2+{\tilde u}^2d\Omega_4^2+\frac{d{\tilde F}^2}{g_2}\right)\right]
\nonumber\\
e^{-2\phi}&=&-g_1^2 e^{2{\tilde H}}+g_2,\quad
H_3=d\left[e^{2\phi}(d{\tilde F}-\frac{g_2}{g_1}d{\tilde v})\right]\wedge dt,\quad
F_3=v_0^2g_1~de^{2{\tilde H}}\wedge d^2{\bf x},\nonumber\\
&&F_5=\frac{1}{4}\left[u^4\int \d_u e^{-2{\tilde H}}dF\right]+dual,\qquad 
\d^2_{\tilde F} (g_2e^{-2\tilde H})+\Delta_{\tilde u}e^{-2\tilde H}=0.
\eea
This is precisely the geometry produced by flat D3 branes with fluxes, which was constructed 
in \cite{MalRus}. The standard D3 brane corresponds to $g_1=0$.

To obtain the solution (\ref{MRD3soln}), we looked at a vicinity of some point on D3 brane. Similar analysis can be performed for D5 brane as well, in this case one should introduce a rescaling
\bea\label{RecD5Zoom}
t=\eps{\tilde t},\ u=u_0+\eps{\tilde u},\ d\Omega_4^2=\eps^2d{\bf y}_4^2,\
e^H=\eps^{4}e^{\tilde H},\ e^\phi=\eps^{-4}e^{\tilde\phi},~
\left(\begin{array}{c}v\\ w\end{array}\right)=
\eps^{-3}\left(\begin{array}{c}{\tilde v}\\ {\tilde w}\end{array}\right)
\eea
and send $\eps$ to zero. Assuming that we started with a regular metric, we conclude that after taking the limit, the dilaton, $e^{\tilde H}$ and $C_4=-\frac{1}{4}u_0^4 A_u~d^4{\bf y}$ should not depend on ${\tilde u}$, this leads to the relation
\bea
{\tilde F}=c{\tilde u}+{\hat F}({\tilde v},{\tilde w}):\qquad A_u=ce^{-2{\tilde H}}.
\eea
Rewriting the differential equations (\ref{IIBeqn1}), (\ref{IIBeqn2}) in terms of rescaled variables, we find two relations
\bea
&&\d_{\tilde w}e^{-2\tilde\phi}+\Delta_{\tilde v} {\hat F}=0,\\
&&e^{-2{\tilde H}}\d_{\tilde w} e^{-2{\tilde H}-2\tilde\phi}+
c^2e^{-2{\tilde H}}\d_{\tilde w}e^{-2{\tilde H}}=0.
\eea
The second equation can be solved in terms of a function $h(\tilde v)$, then the first relation 
leads to the Laplace equation for $e^{2{\tilde H}}$:
\bea
&&e^{-2{\tilde \phi}}=-c^2+h(\tilde v) e^{2\tilde H},\nonumber\\
&&h\d_{\tilde w}^2 e^{2{\tilde H}}+\Delta_{\tilde v}e^{2{\tilde H}}=0.
\eea
As before, we find that (\ref{IIBeqn3}) reduces to a harmonic equation for $h(\tilde v)$ and requiring regularity, we conclude that $h$ must be a constant. This leads to the final solution describing D5 branes in the presence of the Kalb--Ramond field:
\bea
ds^2&=&e^{-{\tilde\phi}/2}\left[e^{\tilde H+2\tilde\phi}\left\{-d{\tilde t}^2+
h(d{\tilde u}+\frac{c}{h}d{\tilde w})^2\right\}+e^{-\tilde H}u_0^2 d{\bf y}^2_4+
e^{\tilde H}(d\tilde v^2+\tilde v^2 d\Omega_2^2+\frac{d\tilde w^2}{h})
\right]\nonumber\\
&&e^{-2{\tilde \phi}}=-c^2+he^{2\tilde H},\quad
F_5=-\frac{cu_0^4}{4}de^{-2\tilde H}\wedge d^4{\bf y}+dual,\\
&&H_3=c~de^{2\tilde\phi}\wedge (d{\tilde u}+\frac{c}{h}d{\tilde w})\wedge dt,
\quad F_3=v^2d(-\d_w e^{-2{\tilde \phi}} dv+\d_v e^{2\tilde H} dw)\wedge d\Omega_2,
\nonumber\\
&&h\d_{\tilde w}^2 e^{2{\tilde H}}+\Delta_{\tilde v}e^{2{\tilde H}}=0.
\eea
This solutions has been constructed in \cite{MalRus} using T duality and shift.


\section{General solution in ten dimensions}
\label{SecGnrSln}

\renewcommand{\theequation}{4.\arabic{equation}}
\setcounter{equation}{0}


In the previous section we derived a geometry produced by a single spike which is attached to
either D3 or D5 brane. From the brane probe analysis of section \ref{SectProbe} we know that such spikes can 
be linearly superposed and in this section we will present supergravity solutions which describes such superpositions. Previously we had a large symmetry group 
($SO(3)\times SO(5)\times U(1)$) which allowed us to {\it derive} the solution. Unfortunately for a 
general superpositions of D3, D5 branes and fundamental strings we do not expect to have any nonabelain symmetry, so it seems that one would need to find the most general static 1/4--BPS solution of type IIB supergravity.
Rather than facing this complicated problem, we will {\it guess} the solution using geometries constructed in the previous section as a guide. 
In this section we will {\it propose} a very natural generalization of the solution (\ref{SymSltnSum}) which has all the required properties and 
then we will {\it check} that the geometry indeed preserves $8$ supercharges. Then we will analyze various properties of the new solution, in particular we will show that the new geometries have the right number of degrees 
of freedom to account for all D3--D5--F1 intersections. We will also see that the regularity of the supergravity 
solution requires that one can place the brane sources only on specific curves. 
It turns out that this restriction coming from {\it closed strings} gives exactly the same profiles 
of the branes as we derived in section \ref{SectProbe} using the {\it open string} language.


\subsection{Summary of the solution}
\label{SectSbGenSum}


We begin with writing a guess for the solution which generalizes (\ref{SymSltnSum}) and does 
not rely on having nonabelian symmetries:
\bea
ds^2&=&e^H\left[-e^{3\phi/2}dt^2+e^{-\phi/2}d{\bf x}^2_3\right]+
e^{-H-\phi/2}d{\bf y}^2_5+e^{-H+3\phi/2}(\d_w Fdw+\d_{\bf y}F d{\bf y})^2\nonumber\\
\label{GenSltnSum}
&&e^{2H}=\d_w F,\quad
F_5=-\frac{1}{4\cdot 4!}d\left[e^{-2H}\eps_{ijklm}\d^{y_m}Fdy^{ijkl} \right]+dual,\\
&&H_3=d\left[e^{2\phi}(\d_w Fdw+\d_{\bf y}F d{\bf y})\right]dt,\quad 
F_3=\frac{1}{2}d(\eps_{ijk}\d^k Fdx^{ij}).\nonumber
\eea
Starting with this ansatz, one can solve equations for Killing spinors and show that if $F$ and $e^\phi$ obey certain differential equations, then the geometry preserves eight 
supercharges\footnote{We perform this check in the Appendix \ref{AppGenSln} while still assuming $SO(5)$ symmetry, and an extension to the most general case is trivial. We also notice that 
to arrive at (\ref{GenSltnSum}) it is sufficient to postulate the form of the metric and $F_5$, 
while only requiring 
$H_3$ to be electric and $F_3$ to be magnetic.},
and the Killing spinor $\eps$ can be expressed in terms of a constant eight--component object 
$\eps_0$:
\bea
\eps=\exp\left[\frac{1}{4}(H+\frac{3\phi}{2})\right]\eps_0:\quad 
\Gamma_w\Gamma_{45678}\eps_0=-i\eps_0,\quad
\Gamma_w\Gamma_{123}\eps_0^*=i\eps_0,\quad \Gamma_{11}\eps_0=-\eps_0.
\eea
As before, the solutions are parameterized by two functions $F$, $e^\phi$ which  
obey differential equations\footnote{Here we introduced the Laplace operators in flat spaces: $\Delta_{\bf x}=\sum_1^3 \d^2_{x_i}$, $\Delta_{\bf y}=\sum_1^5 \d^2_{y_i}$}:
\bea\label{GenDifUr1}
&&\d_w e^{-2\phi}+\Delta_{\bf x} F|_{y,w}=0,\\
\label{GenDifUr2}
&&\d_F e^{-2H-2\phi}+(\Delta_{\bf y} w)|_{x,F}=0.
\eea
While it seems unusual to write two equations using different variables ($(x,y,w)$ in the first equation and $(x,y,F)$ in the second one), this "mixed notation" makes the relations compact and 
more importantly, it is more natural for finding the positions of the branes. Of course one can always go to a more consistent notation which uses $(x,y,w)$ everywhere, this can be accomplished via translation rules:
\bea
\d_{\bf x}|_{w,y}=\d_{\bf x}|_{F,y}-\frac{\d_{\bf x}w}{\d_F w}\d_F|_{x,y},\quad
\d_{\bf y}|_{w,x}=\d_{\bf y}|_{F,x}-\frac{\d_{\bf y}w}{\d_F w}\d_F|_{x,y},\quad
\d_F=e^{-2H}\d_w.
\eea
As before, one also needs an equation of motion for the Kalb--Ramond field:
\bea\label{GenDifUr3}
\Delta_{\bf y} e^{-2\phi}|_F+\Delta_{\bf x} e^{-2\phi-2H}|_F+
\Delta_{\bf y}(e^{2H}\d_{x_i} w\d_{x_i} w)|_{x,F}=0.
\eea

\subsection{Boundary conditions}
\label{SectSbBcnd}

So far we presented the results of local analysis which led to the conclusion that the geometry was 
parameterized in terms of two functions $e^\phi$, $F$ satisfying equations 
(\ref{GenDifUr1}), (\ref{GenDifUr2}), (\ref{GenDifUr3}). These equations were derived assuming absence 
of sources and if this assumption
holds everywhere, then $R^{1,9}$ is the only asymptotically--flat solution\footnote{This statement is familiar in a 
case of D3 branes where the system (\ref{GenDifUr1})--(\ref{GenDifUr3}) reduces to a Laplace equation 
$(\d_F^2+\Delta_{\bf y})e^{-2H}=0$ and the sourceless solution is unique due to the maximum principle. In general one has a more complicated elliptic system, but the sourceless solution is 
still expected to be unique.}. To describe nontrivial geometries we will need to introduce the branes into the system.
 It turns out that a consistency of supergravity equations leads to strong restrictions on the positions of the branes 
 and we will derive these restrictions below. 

The conventional way of accounting for branes in supergravity is an introduction of sources into the equations of motion for various NS--NS or RR fields. Unfortunately this approach is not convenient in the present context since we were solving not the equations of motion, but rather 
the conditions for supersymmetry. Luckily there is an alternative way of looking at D branes: 
if the metric is known, then the location of the branes can be found by looking at the points where 
metric becomes degenerate. For example, a metric produced by a stack of flat Dp branes:
\bea
ds_S^2=H^{-1/2}ds_{1,p}+H^{1/2}d{\bf x}_{10-p}^2
\eea 
becomes singular at the locations of the branes (i.e. at the poles of the harmonic function $H$), 
in particular the warp factor multiplying the worldvolume goes to zero. If positions of the branes are characterized 
in this fashion, then one can still use the sourceless equations, but impose certain boundary conditions on the warp factors. 
In the Einstein frame the situation becomes slightly more complicated and depending on the value of $p$, the warp factor could either go to zero or to infinity. Due to this non--universality and since we are only interested in the case of D3 and D5 branes, it is convenient to consider these two types of sources separately.

{\bf Geometric description of D3 branes.} Near D3 brane sources it is convenient to use coordinates 
$(x,y,F)$, then the metric becomes
\bea
ds^2=e^H\left[-e^{3\phi/2}dt^2+e^{-\phi/2}d{\bf x}^2_3\right]+
e^{-H-\phi/2}d{\bf y}^2_5+e^{-H+3\phi/2}(dF-\d_{\bf x}F d{\bf x})^2.\nonumber
\eea
The worldvolume of the brane can be parameterized by $t$ and $x_i$, then the brane position can
be specified as $y_i=y_i^{(0)}({\bf x})$, $F=f({\bf x})$. In the case of a single spike (or multiple spikes preserving $SO(5)$ symmetry) the equation $y_i=y_i^{(0)}({\bf x})$ should be replaced by
$u=u_0({\bf x})$ and to describe a three--brane rather than a $7+1$ dimensional object we must set
$u_0({\bf x})=0$. A natural generalization of this statement to the spikes without the symmetry is to 
require the D3 branes to be located at constant values of $y_i^{(0)}$, so the profile should be 
given by
\bea\label{D3ProfSugra}
y_i=y_i^{(0)}, \qquad F=f({\bf x}).
\eea
Since one expects the gradient of $e^H$ to point in the directions orthogonal to D3 brane, we 
conclude that in the leading order both $e^H$ and $w$ are constant along the brane worldvolume, implying that at this order
\bea\label{D3Profile}
w={\tilde w}({\bf y},{\tilde F}),\qquad {\tilde F}\equiv F-f({\bf x}).
\eea

We expect that near D3 brane the dilaton remains finite and $e^H$ goes to zero. In particular this 
implies that $e^{2H}\d_F e^{-2\phi}$ goes to zero as we approach the brane, then rewriting the equation 
(\ref{GenDifUr1}) in terms of $(x,y,F)$ and taking the near--brane limit, we find
\bea
\left(\d_{\bf x}-\frac{\d_{\bf x}w}{\d_F w}\d_F\right)\frac{\d_{\bf x}w}{\d_F w}=0.
\eea
Substituting the leading term in the expression for $w$ (\ref{D3Profile}), we find a simple harmonic equation for $f({\bf x})$:
\bea\label{funHarm}
\d_{\bf x}\d_{\bf x}f=0.
\eea
Then the leading terms in (\ref{GenDifUr2}) give a harmonic equation for ${\tilde w}$ (notice that the dilaton is constant in this approximation):
\bea
e^{-2\phi_0}\d_{\tilde F}^2 {\tilde w}+\Delta_{\bf y}{\tilde w}=0.
\eea
Of course this equation is only satisfied away from the brane and the correct relation has sources at 
${\tilde F}=0$, $y_i=y_i^{(0)}$. Since the source is located at a point 
in six dimensional space\footnote{Notice that it is the source for $e^{-2H}=\d_F w$ that 
should be localized in 
both ${\bf y}$ and ${\tilde F}$, this is the origin of the theta function in (\ref{HarmNearD3}).}, 
it is completely characterized by one number $Q$:
\bea\label{HarmNearD3}
e^{-2\phi_0}\d_{\tilde F}^2 {\tilde w}+\Delta_{\bf y}{\tilde w}=-Q\theta({\tilde F})
\delta({\bf y}-{\bf y}^{(0)}).
\eea
This coefficient should be interpreted as a number of branes in the stack. 

To summarize, we started with very natural assumption about positions of D3 branes (namely we assumed that the branes are located at fixed values of $y_i$) and showed that consistency of supergravity equations requires the profiles to be
\bea\label{D3ProfA}
{\bf y}={\bf y}^{(0)},\quad F=f({\bf x}),\quad \Delta_{\bf x}f=0.
\eea
As already mentioned, for the solutions with $SO(5)$ symmetry
no assumption is needed and we suspect that the condition $y_i=y_i^{(0)}$ can be extracted from the equations of motion even in the general case, but we will not discuss this further. Once the profile $f({\bf x})$ is specified, the harmonic equation (\ref{HarmNearD3}) allows one to recover function $w$ in the vicinity of D3 brane. Thus it appears that if we only have D3 sources, then the 
boundary conditions are completely specified by the harmonic functions $f_a({\bf x})$ and coordinates ${\bf y}_a^{(0)}$ giving the positions of the branes, and the set of charges $Q_a$ 
characterizing the stacks.  

{\bf Geometric description of D5 branes.} If one looks for the geometries without singularities, 
D3 branes are the only allowed sources. Indeed, the necessary condition for avoiding the singularities is the requirement for the dilaton to remain finite. This condition can only be satisfied by D3 branes: for the other two objects (D5 and fundamental strings) $e^\phi$ goes to zero as we 
approach the branes, so the metric must be singular. However these singularities of supergravity are resolved by string theory since D5 branes and fundamental strings are allowed sources. 

As one approaches D5 brane the dilaton $e^\phi$ goes to zero, however combination 
$g_{tt}e^{-\phi/2}$ remains finite in the limit. For the solution (\ref{GenSltnSum}) it means that 
$e^{\Psi}=e^{H+\phi}$ should remain finite as one approaches the brane. Let us rewrite the metric 
in terms of this function and coordinates $(x,y,w)$:
\bea
ds^2=e^{-H/2}(-e^{3\Psi/2}dt^2+e^{-\Psi/2}d{\bf y}^2_5)+e^{3H/2-\Psi/2}d{\bf x}^2_3+
e^{-5H/2+3\Psi/2}(\d_w Fdw+\d_{\bf y}F d{\bf y})^2.\nonumber
\eea
Notice that in the vicinity of the brane $e^{-H}$ goes to zero. 

As in the case of D3 branes, we assume that the worldvolume of D5 is parameterized by 
$(t,{\bf y})$ and the profile is given by
\bea\label{D5ProfA}
{\bf x}={\bf x}^{(0)}, \quad w=h({\bf y}),\quad \Delta_{\bf y}h=0.
\eea
Then near the brane the function $F$ depends upon $w$ and ${\bf y}$ only through their combination ${\tilde w}$:
\bea
F={\tilde F}({\bf x},{\tilde w}),\quad {\tilde w}\equiv w-h({\bf y}).
\eea
Then eliminating $e^\phi$ from (\ref{GenDifUr1}), (\ref{GenDifUr2}) and neglecting the term 
$e^{-2H}\d_w e^{-2\Psi}$ in the leading order of (\ref{GenDifUr2}), we arrive at the relations which hold in the vicinity of D5 branes:
\bea
&&\Delta_{\bf x} {\tilde F}+e^{-\Psi_0}\d^2_w {\tilde F}=0,\nonumber\\
&&
(\d_{\bf y}-\frac{(\d_{\bf y}F)}{\d_w F}\d_w)(\frac{\d_{\bf y}F}{\d_w F})=0.
\nonumber
\eea
The second equation is equivalent to harmonicity of $h({\bf y})$, and the first equation allows one to recover $F$ once the D5 charges are known. We see a direct analogy with description of D3 branes which was discussed above: to be consistent with SUGRA equations, the sources should be specified in terms of the harmonic profiles $h_a({\bf y})$, positions in the transverse directions 
${\bf x}_a^{(0)}$ and charges $Q_a$. 

{\bf Geometric description of strings.} Although our goal is to describe strings dissolved in D3 or D5 branes, 
for completeness we also mention 
a possibility of having a "freestanding" string in the geometry. As one approaches such object, 
$e^\phi$ goes to zero, but $e^H$ remains finite. This implies that $F(w,{\bf x},{\bf y})$ is finite as well, then 
equations (\ref{GenDifUr1}), (\ref{GenDifUr2}) are equivalent to the statement that the divergent part of 
$e^{-2\phi}$ is $w$--independent. The leading contribution to (\ref{GenDifUr3}) implies 
harmonicity of the dilaton in the transverse directions:
\bea\label{BCString}
e^{-2H}\Delta_{\bf x}e^{-2\phi}+\Delta_{\bf y}e^{-2\phi}=0,
\eea
which is not very surprising. As usual, to describe the strings we have to add some sources to the last equation. 
We see that for fundamental strings there is no issue of finding the "profile": since there are eight transverse coordinates, the string can only do along $w$ direction. 

\begin{figure}[tb]
\begin{center}
\epsfxsize=5.2in \epsffile{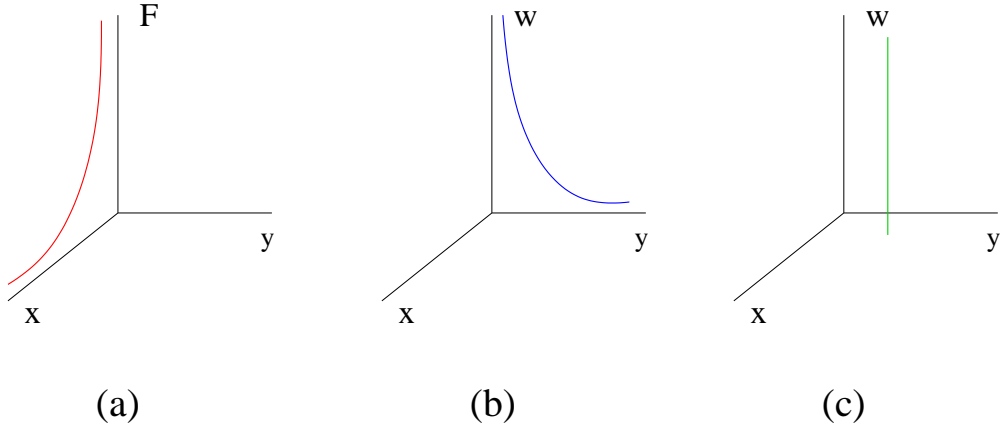}
\end{center}
\caption{Boundary conditions are imposed along the harmonic profiles corresponding to D3 branes (a)
or D5 branes (b). One can also have freestanding strings which do not end on branes (c).
} \label{FigBndCn}
\end{figure}

{\bf Summary of the boundary conditions.} By adding sources to the gravity equations and analyzing 
consistency conditions, we found that the branes cannot be introduced arbitrarily, but rather they should follow specific profiles. In particular, D3 branes can only be stretched along the surfaces (\ref{D3ProfA}) with harmonic 
function $f({\bf x})$, while D5 branes must follow (\ref{D5ProfA})\footnote{Of course the branes also extend along time direction.}. We also found that near free--standing fundamental string, the equation for divergent part of the dilaton becomes linear (\ref{BCString}) and the sources can easily be added to it:
\bea\label{BCStringSr}
e^{-2H}\Delta_{\bf x}e^{-2\phi}+\Delta_{\bf y}e^{-2\phi}=-\sum Q_1^a\delta({\bf x}-{\bf x}_a)\delta({\bf y}-{\bf y}_a).
\eea
The pictorial representation of boundary conditions is given in figure \ref{FigBndCn}.

In section \ref{SectSbPert} we will show that starting from any set of allowed boundary conditions, one can 
construct a  unique geometry produced by corresponding brane configuration. But first it might be useful to 
compare the results of this subsection with probe analysis presented in section \ref{SectProbe}.


\subsection{Relation to the brane probes}

In the previous subsection we derived two sets of boundary conditions which are consistent with 
supergravity: the geometry can end either on D3 or on D5 branes. Moreover, the profiles of such branes cannot be arbitrary, but rather they are parameterized in terms of harmonic functions. 
Let us now compare these boundary conditions which brane profiles which were 
derived in section \ref{SectProbe}.

{\bf Probes in flat space} As a warm-up we will recover the profiles discussed in subsection \ref{SectSbBion}.
The flat space can be easily embedded in the solution (\ref{GenSltnSum}) by setting
\bea
F=w,\quad e^{2\phi}=1.
\eea
Clearly this solves all equations. Few D3 branes added to flat space can be described in two alternative ways: one can either use an open string picture (as we did in the subsection \ref{SectSbBion}), or one can look at the changes in the geometry produced by branes. If the number of branes is small, we expect the metric to be flat everywhere except the small vicinity of the branes, and the consistency of SUGRA in this vicinity leads to the restriction on the brane profile (\ref{funHarm}):
\bea
F=f({\bf x}),\quad \Delta_{\bf x}f=0.
\eea
This consequence of {\it closed string} analysis is in complete agreement with {\it open string} result 
(\ref{ProbeHarmonic}). The agreement for D5 branes works in the same way.

{\bf D3 brane in D3 geometry}. Next we start with a stack of $N$ D3 branes without worldvolume 
fluxes and introduce $k$ additional branes. We will assume that $N$ is large and replace the stack 
of branes by the geometry that they produce, while for the $k$ branes we compare the DBI and SUGRA descriptions. The DBI analysis of section \ref{SectSbCrvBion} led to conclusion that in the geometry
\bea
ds^2=H_3^{-1/2}ds_{3,1}^2+H_3^{1/2}(dz^2+d{\bf y}_5^2)
\eea
the profile of the probe brane is $z=X({\bf x})$ with a harmonic function $X$. 
To recover the metric of flat D3 branes from (\ref{GenSltnSum}) one has to take a dilaton to be a constant and assume that $\d_{\bf x}F=0$:
\bea
ds^2=e^H ds_{3,1}^2+e^{-H}(dF^2+d{\bf y}_5^2).
\eea
Then the SUGRA profile (\ref{D3ProfA}) with harmonic function $f({\bf x})$ gives a perfect 
agreement with DBI analysis. 

{\bf D5 brane in D3 geometry}. In this case the DBI equation (\ref{EOMD5inD3}) looked somewhat complicated:
\bea\label{D5biCond}
-(1+(\nabla X)^2)\d_z H_3+H_3\nabla^2 X+2
\nabla H_3\nabla X=0
\eea
and now we understand the reason: while the coordinates $(x,y,F)$ are natural for describing D3 branes, the boundary conditions for D5 branes look simpler in $(x,y,w)$ variables, so we need to perform a translation. 

From the analysis of the previous subsection we know that supergravity requires the profile of 
D5 brane to be $w=h({\bf y})$ with harmonic function $h$. To compare with (\ref{D5biCond}) we recall the relation:
\bea
\d_F w=e^{-2H}:\qquad w=\int^F e^{-2H}dF=\int^F H_3(z,{\bf y}) dz.
\eea
Then writing the profile of D5 in $F$ coordinate as $F=X({\bf y})$, we arrive at the relation:
\bea\label{D5altBC}
h({\bf y})=\int^{X({\bf y})} H_3(z,{\bf y}) dz.
\eea
It turns out that this relation (which is a consequence of SUGRA analysis) is equivalent to the equation (\ref{D5biCond}). To see this we apply the Laplace operator $\Delta_{\bf y}$ to both sides of the last equation:
\bea
0&=&H_3\Delta_{\bf y}X+\d_{\bf y}X \d_{\bf y}H_3(X,{\bf y})+
\d_{\bf y}X \d_{\bf y}H_3(z,{\bf y})|_{z=X}+\int^{X({\bf y})} \Delta_{\bf y}H_3 dz\nonumber\\
&=&H_3\Delta_{\bf y}X+2\d_{\bf y}X \d_{\bf y}H_3(X,{\bf y})-
(\d_{\bf y}X)^2 \d_zH_3(z,{\bf y})|_{z=X}-\d_z H_3|_{z=X}.
\eea
Here we used the harmonicity of $H_3$ ($(\d^2_z+\Delta_{\bf y})H_3(z,{\bf y})=0$) and we assumed that the low limit of integration in (\ref{D5altBC}) is chosen to be along the hypersurface 
where $\d_z H_3=0$\footnote{Notice that the same choice was made in section \ref{SectSbCrvBion} to derive 
(\ref{D5biCond}).}. The last equation is exactly the same as (\ref{D5biCond}), so we demonstrated a perfect agreement between the results of open string analysis and SUGRA computations in the 
geometry produced by D3 branes.

{\bf Branes in D5 geometry} can be analyzed in the same way and one would find that both the DBI analysis and SUGRA computations require that the brane profiles are described by harmonic functions. However these functions should be written in appropriate variables and in particular, to recover a harmonic function governing the profile of D3 brane one needs to rewrite 
(\ref{EOMD3inD5}) in terms of coordinates $(x,y,F)$. This involves essentially the same computations that were used to show that the profile (\ref{D5biCond}) is equivalent to 
$w=h({\bf y})$ with harmonic $h$.

To summarize, we compared two descriptions of D branes with fluxes: one is given by open strings and the other one involves closed strings. At low energies the physics of open strings
is well described by the DBI action and we analyzed the 1/4 BPS solutions of such theory on the backgrounds produced by D3 or D5 branes. In the closed string picture, the consistency of supergravity led to restrictions on the brane profiles, and by looking at this restrictions on D3 or D5 background, we found a perfect agreement with DBI analysis. This provides a nontrivial check of the DBI/SUGRA duality in the 1/4 BPS sector. If one further takes a decoupling limit, this duality reduces to a more conventional gauge/gravity correspondence. Let us discuss the decoupling limits which are relevant in the present case.


\subsection{Near--horizon limits}
\label{SectSbNrhrz}


In this paper we have been studying the brane configurations preserving eight supersymmetries. Out main goal was to describe branes embedded in flat space, so at infinity the geometry approaches 
$R^{9,1}$ and the number of supersymmetries is enhanced to $32$. It might also be interesting to look for geometries which asymptote to different solutions with enhanced symmetry (such as $AdS_5\times S^5$). In particular, it is natural to ask whether solutions with $AdS_5\times S^5$ asymptotics can be recovered from asymptotically--flat geometries, just like the $AdS_5\times S^5$ itself is recovered from the metric produced by D3 branes.

To address this question we introduce a generalization of the near horizon limit which would 
work for any asymptotically--flat solution (\ref{GenSltnSum}). The decoupling limit of the geometry produced by D3 brane is obtained by zooming in on the vicinity of the brane 
\cite{malda}:
\bea
e^{-2H}=1+\frac{Q}{({\bf y}^2+F^2)^2}\rightarrow  \frac{Q}{({{\bf y}}^2+{F}^2)^2},\nonumber
\eea
in particular one goes to small values of $|{\bf y}|$. Notice that in this limit both the equation for the
 harmonic function and the expression for the metric in terms of this function remain the same. Let us start from a general solution (\ref{GenSltnSum}) and make a rescaling 
 ${\bf y}\rightarrow \eps{\tilde {\bf y}}$, then to keep 
the form of the solution (\ref{GenSltnSum}) unchanged, additional redefinitions should be implemented:
\bea\label{BSymmNearHor}
{\bf x}=\eps^{-1}{\tilde {\bf x}},\quad {\bf y}=\eps{\tilde {\bf y}},\quad  e^H=\eps^2 e^{\tilde H}, 
\quad  w=\eps^{-3}{\tilde w},\quad t=\eps^{-1}{\tilde t} ,\quad
F=\eps{\tilde F}.
\eea
With these changes the metric written in terms of variables with tildes looks exactly the same as 
the original one. Moreover one can see that equations (\ref{GenDifUr1})--(\ref{GenDifUr3}) are invariant under 
such rescaling. 

Starting from the metric of D3 branes and introducing a change of variables 
(\ref{BSymmNearHor}), one extracts the decoupling limit as $\eps$ goes to zero.
This can be seen by looking at the harmonic function for that case:
\bea
e^{-H}=1+\frac{Q}{({\bf y}^2+F^2)^2}:\qquad
e^{-{\tilde H}}=\eps^2+\frac{Q}{({\tilde {\bf y}}^2+{\tilde F}^2)^2}\rightarrow 
\frac{Q}{({\tilde {\bf y}}^2+{\tilde F}^2)^2}.
\eea
In this limit the $U(1)$ symmetry is enhanced to $SO(2,1)$: 
\bea
ds_5^2=r^2(-dt^2+dv^2+v^2 d\Omega_2^2)+\frac{dr^2}{r^2}=
\cosh^2\rho\left[-z^2 dt^2+\frac{dz^2}{z^2}\right]+d\rho^2+\sinh^2\rho d\Omega_2^2\nonumber
\eea
and the map between the coordinates is given by
\bea
r=z\cosh\rho,\quad v=z^{-1}\tanh\rho. \nonumber
\eea

For a general asymptotically flat solution (\ref{GenSltnSum}), the rescaling 
(\ref{BSymmNearHor}) accompanied by the limit $\eps\rightarrow 0$ gives a new geometry 
with different asymptotics, but it seems impossible to 
have an interesting solution with enhanced symmetry in this case (we discuss this in more 
detail in the Appendix \ref{AppCmpr}). Thus the solutions produced by the near--horizon limit 
(\ref{BSymmNearHor}) asymptote to $AdS_5\times S^5$, but they preserve only $8$ supercharges. An analogous situation has been encountered for the metrics describing a Coulomb branch \cite{KrLarTriv}: they preserved $16$ supercharges in asymptotically-flat space
and symmetry was not enhanced in the near--horizon region. 

An alternative near--horizon limit can be defined by zooming in on a vicinity of D5 branes. 
By starting with asymptotically flat solution and introducing a rescaling
\bea
t=\frac{{\tilde t}}{\eps},\quad
{\bf x}=\eps^3{\tilde {\bf x}},\quad {\bf y}=\eps^{-1}{\tilde {\bf y}},\quad  e^H=\eps^{-4} e^{\tilde H}, 
\quad e^\phi=\eps^4 e^{\tilde\phi},
\quad  w=\eps^{3}{\tilde w},\quad F=\eps^{-5}{\tilde F},\nonumber
\eea
one ends up with a new solution of (\ref{GenSltnSum})--(\ref{GenDifUr2}), and for  
$\eps= 0$ the resulting geometry has a linear dilaton in the asymptotic region.


\subsection{Existence of the solution: perturbative proof.}
\label{SectSbPert}


Let us summarize what we learned so far. Imposing the ansatz (\ref{GenSltnSum}) and looking at supersymmetry variations we showed that {\it locally} the geometry preserves eight supercharges 
if functions $F$, $e^\phi$ satisfy (\ref{GenDifUr1}), (\ref{GenDifUr2}), (\ref{GenDifUr3}). We also know that to construct nontrivial asymptotically flat solutions, one needs to add certain sources to these three equations, and in subsection \ref{SectSbBcnd} we showed that a consistency of supergravity requires the brane sources to follow harmonic curves. Suppose one chooses such curves and assigns certain D3/D5 charges to them. Would this lead to a unique asymptotically flat solution? 
For the flat D3 branes it is easy to show that the answer is yes: since one deals with Laplace equation, the sources
fix the solution uniquely. Moreover such solution can be easily constructed. In a more general case we cannot solve the nonlinear equations, but one can show that any allowed distribution of sources leads to a unique solution. We will outline the argument in this subsection.

Our starting point is flat space which has constant dilaton (to simplify the formulas below we will set $e^{\phi_0}=1$, although this relation can be easily relaxed) and $w=F$. To formulate a 
perturbation theory around flat space, we introduce a small parameter $\eps$ and write 
\bea\label{PertSeries}
w=F+\sum_{k=1} \eps^k w_k,\quad e^{-2\phi}=1+\sum_{k=1}\eps^k \Phi_k.
\eea
Next we substitute these expansions into (\ref{GenDifUr1}), (\ref{GenDifUr2}), (\ref{GenDifUr3}) and look at those equations order by order in $\eps$. For the first terms we find:
\bea
&&\d_F \Phi_1-\Delta_{\bf x} w_1=0,\nonumber\\
&&\d_F^2 w_1+\d_F \Phi_1+\Delta_{\bf y} w_1=0,\nonumber\\
&&(\Delta_{\bf y}+\Delta_{\bf x})\Phi_1+\Delta_{\bf x}\d_F w_1=0.
\eea
One can combine the first two equations to write an equation for $w_1$:
\bea\label{PertEqn1}
\d_F^2 w_1+\Delta_{\bf x} w_1+\Delta_{\bf y} w_1=0
\eea
and solve it, then $\Phi_1$ can be determined by looking at the system:
\bea\label{PertEqn2}
\d_F \Phi_1=\Delta_{\bf x} w_1,\qquad 
(\Delta_{\bf y}+\Delta_{\bf x})\Phi_1+\Delta_{\bf x}\d_F w_1=0.
\eea
Notice that integrability condition is satisfied due to (\ref{PertEqn1}). 

The requirement of 
asymptotic flatness translates into the boundary conditions for $w_1$ and $\Phi_1$: they should vanish as one goes to infinity. Thus in the absence of sources, the maximum principle can be used to argue 
that $w_1=\Phi_1=0$, this demonstrates that unless the branes are put in, the 
flat space is the only solution of our equations. 
To add D3 and D5 branes we introduce of sources to (\ref{PertEqn1}):
\bea\label{PertD3D5e1}
&&\d_F^2 w_1+\Delta_{\bf x} w_1+\Delta_{\bf y} w_1=\\
&&\qquad\qquad=-\sum_a Q^a_3\delta({\bf y}-{\bf y}_a)\theta(F-p_a({\bf x}))+
\sum_a Q^a_5\delta({\bf x}-{\bf x}_a)\theta(F-{\tilde p}_a({\bf y})).\nonumber
\eea 
Notice that these sources are non--local in $F$ direction (they appear with $\theta$ instead of $\delta$--function), however the branes do lead to pointlike sources for $e^{-2H}=\d_F w$. This justifies interpretation of $Q^a_3$ 
and $Q^a_5$ as brane charges.

At the linear order there are no restrictions on the profiles $p_a$, ${\tilde p}_a$, but keeping in mind the consistency 
of the nonlinear equations, we choose these functions to be harmonic. This will allow us to assume that the sources are introduced only at the linearized level, and the higher orders of perturbation theory are included just to correct this seed solution (see below). 

Now we look at the equations (\ref{PertEqn2}). The first of these equations is a first order ODE for 
$\Phi_1$, so it is very unnatural to introduce sources there. Introduction of sources in the second equation is possible, but they must be $F$--independent for consistency:
\bea
(\Delta_{\bf y}+\Delta_{\bf x})\Phi_1+\Delta_{\bf x}\d_F w_1=-\sum Q_1^a\delta({\bf x}-{\bf x}_{Fa})
\delta({\bf y}-{\bf y}_{Fa}).
\eea
Such sources correspond to "freestanding" fundamental strings located at 
$({\bf x}_{Fa},{\bf y}_{Fa})$, and, as already mentioned, such objects are covered by our ansatz. 
Using the properties of the Laplace equation, we conclude that for any distribution of D3, D5 and F1 sources, one finds a unique solution $(w_1,\Phi_1)$ in the first order of perturbation theory. 

Suppose $k-1$ orders in perturbation theory have been constructed. Let us look at the terms 
in (\ref{GenDifUr1}), (\ref{GenDifUr2}), (\ref{GenDifUr3}) which multiply $\eps^k$:
\bea
&&\d_F \Phi_k-\Delta_{\bf x} w_k=\Psi_k^{(1)},\nonumber\\
&&\d_F^2 w_k+\d_F \Phi_k+\Delta_{\bf y} w_k=\Psi_k^{(2)},\nonumber\\
&&(\Delta_{\bf y}+\Delta_{\bf x})\Phi_k+\Delta_{\bf x}\d_F w_k=\Psi_k^{(3)}.
\eea
The expressions in the right hand sides contain backreaction of the previous orders, but we 
{\it do not} add extra sources for $k\ge 2$. Then we arrive at a Poisson equation for $w_k$:
\bea
\d_F^2 w_k+\Delta_{\bf x} w_k+\Delta_{\bf y} w_k=\Psi_k^{(2)}-\Psi_k^{(1)}
\eea
and it has a unique solution once we require $w_k$ to vanish at infinity (this is necessary for the asymptotic flatness). The remaining two equations become
\bea
\d_F \Phi_k=\Delta_{\bf x} w_k+\Psi_k^{(1)},\qquad 
(\Delta_{\bf y}+\Delta_{\bf x})\Phi_k=-\Delta_{\bf x}\d_F w_1+\Psi_k^{(3)}.
\eea
The integrability condition is satisfied since the three original equations (\ref{GenDifUr1}), 
(\ref{GenDifUr2}), (\ref{GenDifUr3}) were compatible, so one finds a unique solution $\Phi_k$.

We see that starting from some set of D3, D5 and F1 sources and requiring the solution to be asymptotically flat, one can construct a unique perturbative expansions (\ref{PertSeries}) for the  dilaton and $w$. Since the first term in the series $(w_1,\Phi_1)$ is regular everywhere away from the 
sources, we expect all $\Psi^{(a)}_k$ to be regular away form sources as well, and the same would be true for $(w_k,\Phi_k)$. Thus at any point away from the brane the perturbative expansions 
(\ref{PertSeries}) are well--defined. We also know that these series converge in the asymptotic region, and it is natural to assume the convergence everywhere away from the sources. We do not give a rigorous proof of this fact, but rather appeal to the analogy with multipole expansion. Thus one ends up 
with a geometry which solves "vacuum" equations of type  IIB supergravity everywhere away from the location of the sources. Fortunately the vicinity of the branes was already analyzed before, so we know that staring from harmonic $p_a({\bf x})$ and ${\tilde p}_a({\bf y})$, one constructs a solution which is sourced by allowed D3 and D5 branes. 

One can ask what would happen if the functions $p_a$ and ${\tilde p}_a$ were not chosen to be harmonic. The perturbation theory can be constructed in this case as well, and the sources would still be at $F=p({\bf x})$ or 
$F={\tilde p}({\bf y})$ and SUGRA solution would be valid away from the branes. However such "branes" are not 
a part of string theory: as we showed in subsection 
\ref{SectSbBcnd} supergravity leads to standard D3 and D5 only for harmonic profiles. We conclude that for any 
other profile SUGRA is sourced by some other "strange matter" and since we do not want to couple string theory to external degrees of freedom, such solutions should be declared unphysical. 

To summarize, in this subsection we showed that starting from an allowed configuration of sources, one can recover the complete solution (\ref{GenSltnSum}) using perturbation theory, and while this may not be useful in practice, the procedure demonstrates an existence and uniqueness of a solution for any allowed distribution of branes.  Of course, we have developed a perturbation theory around flat space and to demonstrate an existence of the solution with different asymptotics one should repeat the analysis for that case. For the geometries which asymptote to $AdS_5\times S^5$ or a linear dilaton, one might also use the limits discussed in the previous subsection. 


\subsection{Example: smeared intersection}
\label{SectSbExplc}


While the general solution (\ref{GenSltnSum}) has a relatively simple form, the two functions $(F,e^{-2\phi})$ parameterizing it satisfy a system of nonlinear equations (\ref{GenDifUr1})--(\ref{GenDifUr3}), so the metric 
(\ref{GenSltnSum}) is not very explicit. It turns out that equations (\ref{GenDifUr1})--(\ref{GenDifUr3}) can be 
solved if one assumes that the brane sources are uniformly smeared along $w$ 
(or $F$) direction. In this subsection we will present such solutions.


\begin{figure}[tb]
\begin{center}
\epsfxsize=4.6in \epsffile{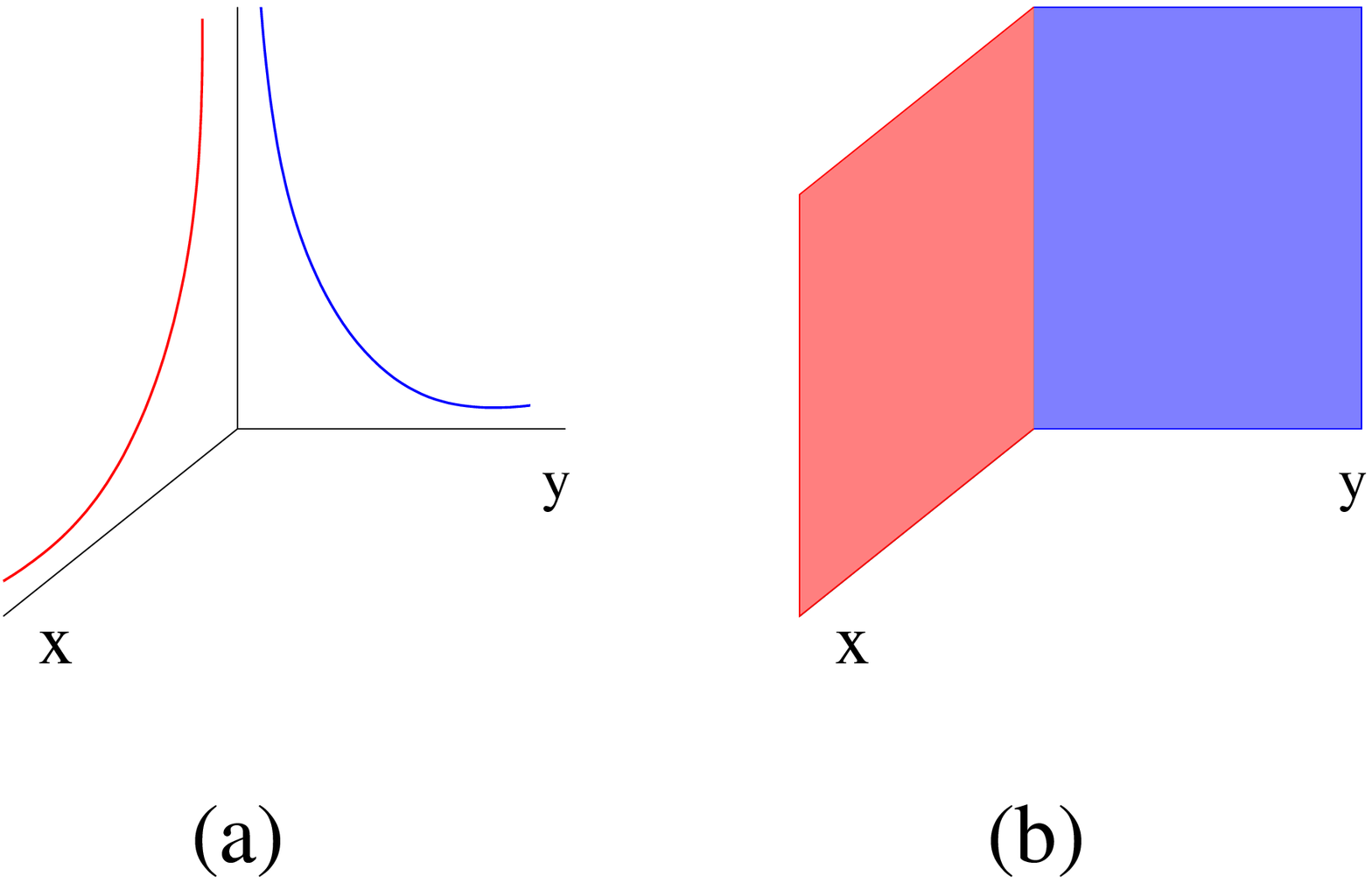}
\end{center}
\caption{Smearing D3--D5 intersection:\newline
(a) profiles for localized D3 (red) and D5 (blue) branes;\newline
(b) hypersurfaces corresponding to boundary conditions for smeared intersections.
} \label{FigSmear}
\end{figure}


We begin by looking at a perturbative solution discussed in the previous subsection. The right--hand side of 
equation (\ref{PertD3D5e1}) contains D--brane sources and their location is shown in figure 
\ref{FigSmear}a. 
Let us now smear the branes along coordinate $F$ (see figure \ref{FigSmear}b): this can be accomplished by integrating over $F$ in (\ref{PertD3D5e1})\footnote{Such procedure leads 
to $F$--independence of $e^{-2H}$.}:
\bea
\d_F^2 w_1+\Delta_{\bf x}w_1+\Delta_{\bf y}w_1=-\sum_a Q^a_3\delta({\bf y}-{\bf y}_a)|F-p_a({\bf x})|+
\sum_a Q^a_5\delta({\bf x}-{\bf x}_a)|F-{\tilde p}_a({\bf y})|.\nonumber
\eea
The last equation can be easily solved, in particular in the region where $F>f_a({\bf x}),h_a({\bf y})$ one finds:
\bea
w_1&=&(-\sum_a \phi_a({\bf x})+\sum_a {\tilde \phi}_a({\bf y}))F-\sum_a p_a
({\bf x}) {\tilde \phi}_a({\bf y})+
\sum_a {\tilde p}_a({\bf y}) {\phi}_a({\bf x})+h({\bf x},{\bf y}),\nonumber\\
&&\Delta_{\bf x}\phi_a({\bf x})=-Q^a_5\delta({\bf x}-{\bf x}_a),\quad
\Delta_{\bf y}{\tilde\phi}_a({\bf y})=-Q^a_3\delta({\bf y}-{\bf y}_a),\nonumber\\ 
&&(\Delta_{\bf x}+\Delta_{\bf y})h({\bf x},{\bf y})=0.
\eea
By construction, function $w_1$ should vanish at infinity of $({\bf x},{\bf y})$ space, this implies that 
$h({\bf x},{\bf y})=0$. Combining this result with zeroes order solution ($w=F$), we find a relation
\bea
\left[1+\eps\sum_a \phi_a({\bf x})\right](w-\sum {\tilde p}_a({\bf y}))=
\left[1+\eps\sum_a {\tilde \phi}_a({\bf y})\right](F-\sum p_a({\bf x}))+O(\eps^2).
\eea
It turns out that this linearized expression can be easily promoted into an exact solution of the system 
(\ref{GenDifUr1})--(\ref{GenDifUr3}): we 
begin with assuming the following relation:
\bea
w=\frac{{\tilde q}({\bf y})}{q({\bf x})}(F-p({\bf x}))+{\tilde p}({\bf y}),
\eea
where $p,{\tilde p},q,{\tilde q}$ are harmonic functions in appropriate variables which are also allowed to have pointlike sources. With this ansatz one can simplify equations (\ref{GenDifUr1}), (\ref{GenDifUr2}) away from
the sources:
\bea
\d_w e^{-2\phi}=0,\quad \d_Fe^{-2\phi}=0,
\eea
so the dilaton is a function of ${\bf x}$ and ${\bf y}$. 
Finally, equation (\ref{GenDifUr3}) becomes\footnote{We also added string sources to the right--hand side of 
that equation.}:
\bea\label{SmearFundSrc}
q\Delta_{\bf y} (q^{-1}e^{-2\phi})+{\tilde q}\Delta_{\bf x}(q^{-1}e^{-2\phi})=-\sum Q_1^a\delta({\bf x}-{\bf x}_{F}^{(a)})
\delta({\bf y}-{\bf y}_{F}^{(a)}).
\eea
We conclude that the geometry is specified by four harmonic functions $(p,{\tilde p},q,{\tilde q})$ and a 
dilaton satisfying (\ref{SmearFundSrc}):
\bea
ds^2&=&e^H\left[-e^{3\phi/2}dt^2+e^{-\phi/2}d{\bf x}^2_3\right]+
e^{-H-\phi/2}d{\bf y}^2_5+e^{-H+3\phi/2}(dF+e^{2H}\d_{\bf x}w d{\bf x})^2\nonumber\\
\label{SmearedGeom}
F_5&=&\frac{1}{4}d\left[^*_5d_y w \right]+dual,\quad F_3=-d(e^{2H}~^*_3d_x w),\quad
H_3=d\left[e^{2\phi}(dF+e^{2H}\d_{\bf x}w d{\bf x})\right]dt,\nonumber\\
&&e^{2H}=\frac{q({\bf x})}{{\tilde q}({\bf y})},\quad w=\frac{{\tilde q}({\bf y})}{q({\bf x})}(F-p({\bf x}))+{\tilde p}({\bf y}).
\eea
Notice that function ${\tilde p}$ has no effect on the geometry and function $p$ can be eliminated by shifting $F$. Thus without loss of generality we can set $p={\tilde p}=0$, then the solution is parameterized by three harmonic functions $q,{\tilde q}$ and $q^{-1}e^{-2\phi}$ which are sourced by D5, D3 branes and fundamental strings. D 
branes can be superposed freely, then equation (\ref{SmearFundSrc}) allows one to find the dilaton for any distribution of fundamental strings. 


\subsection{Other intersecting branes in IIB supergravity}
\label{SectSbOthIIB}


The solution (\ref{GenSltnSum}) can be easily modified to describe other 1/4--BPS brane intersections in IIB supergravity. While generically the metric in (\ref{GenSltnSum}) has no isometries (apart from time translation which is a consequence of supersymmetry), one can also look at particular solutions which are invariant under translations in some $x_i$ or $y_i$.
Starting from such solutions, one can apply various dualities to find geometries produced by some other configuration of intersecting branes. The branes in the resulting solutions are partially smeared, but from the structure of the geometries it will be clear how to generalize them to the completely localized intersections. In this subsection we will write the geometries produced by such brane configurations.

The brane intersections preserving $8$ supercharges have been classified in section \ref{SectSbScan} and here we will give a geometric description of the configurations appearing in the first two lines of equation (\ref{IntersBraneIIB}). 
We already did it for the $(D3_{123},D5_{56789},F1_{4})$ intersections and it turns out that all other cases can be found by using various dualities:
\bea
\begin{array}{cccl}
\left(\begin{array}{c}D3_{123}\\ {D5}_{56789}\\ F1_4\end{array}\right)&
\stackrel{S}{\longrightarrow}&
\left(\begin{array}{c}D3_{123}\\ \mbox{NS5}_{56789}\\ D1_4\end{array}\right)&
\stackrel{ T_{56}}{\longrightarrow}
\left(\begin{array}{c}D5_{12356}\\ \mbox{NS5}_{56789}\\ D3_{456}\end{array}\right)
\stackrel{ T_{78}}{\longrightarrow}
\left(\begin{array}{c}D7_{1235678}\\ {\rm NS5}_{56789}\\ D5_{45678}\end{array}\right)\\
\begin{array}{c}\ \\ \stackrel{T_{23}}{\ }\downarrow\\ \ \end{array}&&
\begin{array}{c}\ \\ \stackrel{T_{25}}{\ }\downarrow\\ \ \end{array}&\\
\left(\begin{array}{c}D1_1\\ D7_{2356789}\\ F1_4\end{array}\right)&&
\left(\begin{array}{c}D3_{135}\\ KK_{1234}\\ D3_{245}\end{array}\right)&\\
\end{array}
\eea
Let us summarize the resulting geometries\footnote{To perform T duality, we are using conventions summarized in \cite{johnson}. However, one should notice that in this paper we 
use normalization of fluxes which is conventional in supergravity \cite{schwarz}, while the T duality rules are more natural in the string frame. Apart from the usual rescaling of the metric 
($ds_S^2=e^{\phi/2}ds_E^2$), one should also recall that $F_5^{(string)}=4F_5^{(SUGRA)}$, 
$G_3^{(string)}=G_3^{(SUGRA)}$ (see \cite{granPolDil} and Appendix \ref{AppSugra} for details).}:

\noindent
{\bf D1--D7--F1 solution:}
\bea
ds_S^2&=&e^H\left[-e^{2\phi}dt^2+d{x}^2\right]+
e^{-H}d{\bf y}^2_7+e^{-H+2\phi}(\d_w Fdw+\d_{\bf y}F d{\bf y})^2\nonumber\\
&&e^{\Phi}=e^{\phi-H},\quad H_3=d\left[e^{2\phi}(\d_w Fdw+\d_{\bf y}F d{\bf y})\right]dt,\quad 
F_1=\frac{1}{2}d(\d_x F),\nonumber\\
&&F_7=d\left[e^{-2H}~^*_7d_y F\right],\quad F_3\equiv dC_2+C_0H_3=-*F_7.
\eea
{\bf D1--D3--NS5 solution:}
\bea
ds_S^2&=&e^{H-\phi}\left[-e^{2\phi}dt^2+d{\bf x}^2_3\right]+
e^{-H-\phi}d{\bf y}^2_5+e^{-H+\phi}(\d_w Fdw+\d_{\bf y}F d{\bf y})^2\nonumber\\
&&e^{\Phi}=e^{-\phi},\quad
F_5=-\frac{1}{4}d\left[e^{-2H}~^*_5d_y F\right]+dual\\
&&F_3=-d\left[e^{2\phi}(\d_w Fdw+\d_{\bf y}F d{\bf y})\right]dt,\quad 
H_3=d(~^*_3d_x F).\nonumber
\eea
{\bf D3--D3--KK solution:}
\bea
ds_S^2&=&e^{H-\phi}\left[e^{2\phi}(-dt^2+dz^2)+d{\bf x}^2_2\right]+
e^{-H-\phi}d{\bf y}^2_4+e^{-H+\phi}(\d_w Fdw+\d_{\bf y}F d{\bf y})^2\nonumber\\
&&+e^{\phi-H}(du+\eps_{ij}\d_{x_i}F dx_j)^2,\qquad
e^\Phi=1\\
F_5&=&-\frac{1}{4}\left\{
d\left[e^{-2H}~^*_4d_y F\right]+d\left[e^{2\phi}(\d_w Fdw+\d_{\bf y}F d{\bf y})\right]dtdz\right\}
(du+\eps_{ij}\d_{x_i}F dx_j)+dual.
\nonumber
\eea
{\bf D3--D5--NS5 solution:}
\bea
ds_S^2&=&e^{H-\phi}\left[e^{2\phi}dz_{1,2}^2+d{\bf x}^2_3\right]+
e^{-H-\phi}d{\bf y}^2_3+e^{-H+\phi}(\d_w Fdw+\d_{\bf y}F d{\bf y})^2\nonumber\\
&&e^{\Phi}=e^{H},\quad
F_3=-d\left[e^{-2H}~^*_3d_y F\right], \quad H_3=d(~^*_3d_x F)\\
&&F_5=-\frac{1}{4}d\left[e^{2\phi}(\d_w Fdw+\d_{\bf y}F d{\bf y})\right]d^3 z+dual.
\nonumber
\eea
{\bf D5--D7--NS5 solution:}
\bea
ds_S^2&=&e^{H-\phi}\left[e^{2\phi}dz_{1,4}^2+d{\bf x}^2_3\right]+
e^{-H-\phi}dy^2+e^{-H+\phi}(\d_w Fdw+\d_{y}F d{y})^2\nonumber\\
&&e^{\Phi}=e^{2H+\phi},\quad
F_1=d\left[e^{-2H}\d_y F\right], \quad H_3=d(~^*_3d_x F)\\
&&F_7=-d\left[e^{2\phi}(\d_w Fdw+\d_{y}F d{y})\right]d^5 z,\quad 
F_3\equiv dC_2+C_0H_3=-*F_7.
\nonumber
\eea
In all solutions written above, $F$ and $e^\phi$ depend on appropriate numbers of $x_i$ and $y_j$ and these functions satisfy the generalizations of equations (\ref{GenDifUr1}), 
(\ref{GenDifUr2}), (\ref{GenDifUr3}):
\bea\label{GenIntrsDifEqn}
&&\d_w e^{-2\phi}+\Delta_{\bf x} F|_{y,w}=0,\qquad e^{2H}=\d_w F|_{\bf x,\bf y},\nonumber\\
&&\d_F e^{-2H-2\phi}+(\Delta_{\bf y} w)|_{x,F}=0,\\
&&\Delta_{\bf y} e^{-2\phi}+\Delta_x e^{-2\phi-2H}+
\Delta_{\bf y}(e^{2H}\d_{x_i} w\d_{x_i} w)|_{x,F}=0.\nonumber
\eea
The classification of boundary conditions follows the logic that was used in section 
\ref{SectSbBcnd}, and we will not repeat that analysis here. The arguments of section \ref{SectSbPert} show that 
once the brane sources are accounted for by the proper boundary conditions, the solution exists and it is unique. 
The geometries involving D7 branes have the standard problem associated with low co--dimension: 
for example in the case of D5--D7--NS5 intersection, the linearized equation 
for $w$ becomes
\bea
\d_F^2 w_1+\Delta_{\bf x} w_1+\d_y^2 w_1=
-\sum_a Q^a_7\delta({y}-{y}_a)\theta(F-p_a({\bf x}))+
\sum_a Q^a_5\delta({\bf x}-{\bf x}_a)\theta(F-{\tilde p}_a({\bf y}))\nonumber
\eea 
and non--zero values of $Q^a_7$ lead to $e^{-2H}$ which logarithmically diverges at infinity.
While the argument about existence of solution still goes through, it is clear that D7 branes modify flat asymptotics, but we will not discuss this further.

To summarize, we constructed the gravity solutions for all intersecting branes appearing in the first two lines of (\ref{IntersBraneIIB}). All such solutions are characterized by two functions satisfying coupled differential equations (\ref{GenIntrsDifEqn}). The situation with intersections in the last line of (\ref{IntersBraneIIB}) (which can be interpreted as branes inside branes) is slightly different. While it is very easy to find solutions describing {\it smeared} intersections, it appears that the {\it localized} intersections do not exist 
\cite{MarlfPeet}\footnote{The only Dp--D(p+4) configuration for which localization is possible is D2--D6 in type IIA and the relevant gravity solution has been constructed in \cite{CherkHash}.}. The smeared D1--D5 intersection has a very peculiar property: in addition to the standard flat D1, one can find more general solutions which preserve the same amount of SUSY, but describe arbitrary profiles of "D1--D5 string" \cite{lmmm}. It would be interesting to see whether there is a similar generalization of the solutions presented here. In the case of D1--D5 system one can also add a momentum charge to produce geometries preserving $4$ supercharges\footnote{The equations describing such system were written in 
\cite{GMR,BW} and some particular solutions were constructed in \cite{AddMom,BW}.} and it would be nice to find a counterpart of such  
$1/8$--BPS configurations for the setup discussed here.


\section{Solutions in M theory}
\label{SectMsoln}

\renewcommand{\theequation}{5.\arabic{equation}}
\setcounter{equation}{0}


So far we looked at the geometries produced by D3--D5--F1 system in type IIB SUGRA. 
However as we discussed in section \ref{SbctMtchPrb} this setup has a natural counterpart in M theory 
which contains M5 and M2 branes. One can start from scratch and look for geometries describing such M2--M5 configurations, but since we already know the type IIB solutions, one can get eleven dimensional 
geometries by following the duality chains. It turns out, we can proceed in three directions: 
one gives M2--M5--M5' which was described before, and the other two lead to 
M2--M2'--KK and to M5--KK--P systems. In this section we will discuss all three cases.

Let us look at eight brane intersections which appear in (\ref{MthryInter}). The geometries corresponding to five 
of them can be fund by superposing harmonic functions, and the remaining three configurations are related to 
D3--D5--F1 system by the following dualities:
\bea\label{11DTdual}
\begin{array}{cccl}
\left(\begin{array}{c}M2_{12}\\ {KK}_{124(10)}\\ M2_{4(10)}\end{array}\right)&
\stackrel{T_3L}{\longleftarrow}&
\left(\begin{array}{c}D3_{123}\\ {D5}_{56789}\\ F1_4\end{array}\right)&
\stackrel{ T_{5}L}{\longrightarrow}
\left(\begin{array}{c}M5_{1235(10)}\\ {M5}_{6789(10)}\\ M2_{4(10)}\end{array}\right)
\\
&&
\begin{array}{c}\ \\ \stackrel{T_{4}L}{\ }\downarrow\\ \ \end{array}&\\
&&
\left(\begin{array}{c}M5_{1234(10)}\\ KK_{123(10)}\\ P_{4}\end{array}\right)&\\
\end{array}
\eea
Here $L$ labels a lift from ten to eleven dimensions. To be more precise, after T duality and a 
lift one finds a smeared intersection in M theory, however as we will see below, in some cases the unsmeared form 
of the solutions can be easily guessed and checked. In this section we will discuss three eleven dimensional solutions appearing in (\ref{11DTdual}) and discuss some of their properties.

\subsection{M2--M5--M5' geometry}

Let us go back to the general solution (\ref{GenSltnSum}) and assume a translational invariance in 
$z=y_5$. Then one can perform a T duality along this direction to get the following metric in the string frame\footnote{So far in this paper we have been using Einstein frame and normalization of fluxes which comes from type IIB supergravity \cite{schwarz}. To perform T duality one has to rewrite (\ref{GenSltnSum}) in string frame ($ds_S^2=e^{\phi/2}ds_E^2$) and use "stringy" normalization of fluxes: $F_5^{(string)}=4F_5^{(SUGRA)}$ (see \cite{granPolDil} for details).}:
\bea
ds_{IIA}^2&=&e^H\left[-e^{2\phi}dt^2+d{\bf x}_3^2+dz^2\right]+e^{-H}\left[d{\bf y}_4^2+
e^{2\phi}(dF-e^{2H}\d_{\bf x}w d{\bf x})^2\right]\nonumber\\
&&e^{\Phi}=e^{\phi+H/2},\quad H_3=d\left[e^{2\phi}(\d_w Fdw+\d_{\bf y}F d{\bf y})\right]dt,
\nonumber\\
&&F_4=\frac{1}{2}~d(\eps^{ijk}\d_{x_k}Fdx_{ij})\wedge dz+
\frac{1}{3!}d\left[e^{-2H}\eps_{ijkm}\d^{y_m}Fdy^{ijk} \right].
\eea
Performing a lift to M theory\footnote{We recall the general type IIA --M theory relation
$ds_{11}^2=e^{4\Phi/3}(dx_{11}-C)^2+e^{-2\Phi/3}ds_{IIA}^2$.}, we find a geometry
\bea
ds_{11}^2&=&e^{-2\phi/3}\left[
e^{2H/3}\left[e^{2\phi}ds_{1,1}^2+d{\bf x}_3^2+dz^2\right]+e^{-4H/3}\left[d{\bf y}_4^2+
e^{2\phi}(dF-e^{2H}\d_{\bf x}w d{\bf x})^2\right]\right]\nonumber\\
F_4&=&\frac{1}{2}~d(\eps^{ijk}\d_{x_k}Fdx_{ij})\wedge dz+
\frac{1}{3!}d\left[e^{-2H}\eps_{ijkm}\d^{y_m}Fdy^{ijk} \right]\nonumber\\
&&+
d\left[e^{2\phi}(\d_w Fdw+\d_{\bf y}F d{\bf y})\right]\wedge d^2 s_{1,1}.
\eea
This solution is derived assuming translational invariance along $z$ direction, but one can see that 
this restriction can be relaxed, so we find a more general eleven dimensional geometry:
\bea\label{M5M5M2}
ds_{11}^2&=&e^{-2\phi/3}\left\{
e^{2H/3}\left[e^{2\phi}ds_{1,1}^2+d{\bf x}_4^2\right]+e^{-4H/3}\left[d{\bf y}_4^2+
e^{2\phi}(\d_w Fdw+\d_{\bf y}F d{\bf y})^2\right]\right\}\nonumber\\
F_4&=&d(^*_xdF)+d\left[e^{-2H}~^*_ydF \right]+
d\left[e^{2\phi}(\d_w Fdw+\d_{\bf y}F d{\bf y})\right]\wedge d^2 s_{1,1}.
\eea
The equations (\ref{GenDifUr1})--(\ref{GenDifUr3}) still hold, but now both ${\bf x}$ and ${\bf y}$ are four--component vectors.  To detect the sources we look at the points where the warp factor in front of $R^{1,1}$ goes to zero. Since there are only two types of branes in M theory, one should look for the objects whose worldvolume is either 3-- or 6--dimensional. This leads to the following 
three possibilities:

{\bf I. Boundary conditions for M5 branes.} We assume that $R^{1,1}$ combines with ${\bf x}$ to give a worldvolume of M5 brane. This implies that $e^\phi$ remains finite and in this case the boundary conditions for 
(\ref{GenDifUr1})--(\ref{GenDifUr3}) were analyzed in section \ref{SectSbBcnd}: we concluded that to have 
regular branes one needs
\bea\label{BCMth1}
y_i=y_i^{(0)},\quad F=f({\bf x}),\quad \Delta_{\bf x}f({\bf x})=0.
\eea
The only difference in the present case is the dimensionality of vectors ${\bf x}$ and ${\bf y}$. 

{\bf II. Boundary conditions for M5' branes.} Assuming that the worldvolume is spanned by 
$R^{1,1}$ and ${\bf y}$, one concludes that $e^\Psi=e^{H+\phi}$ must remain finite and the boundary conditions in this case are
\bea\label{BCMth2}
x_i=x_i^{(0)},\quad w=h({\bf y}),\quad \Delta_{\bf y}h({\bf y})=0.
\eea

Just as in the type IIB setup, one can show that the conditions (\ref{BCMth1}), (\ref{BCMth2}) are in a perfect agreement with probe analysis presented in subsection \ref{SbctMtchPrb}. 

{\bf III. Boundary conditions for M2 branes.} In this case the worldvolume is $R^{1,1}$, 
then $e^H$ remains finite and one needs to specify the sources in (\ref{BCStringSr}). Again the only difference in the present case is that both ${\bf x}$ and ${\bf y}$ in (\ref{BCStringSr}) should be understood as four--vectors. This set of boundary conditions gives freestanding M2 branes.

Once the appropriate boundary conditions are specified, one can use the perturbative construction discussed in subsection \ref{SectSbPert} to argue the existence and uniqueness of M theory solution. 


\subsubsection{Near--horizon limits and $1/2$--BPS states in $AdS_p\times S^q$.}
\label{SecSbSbs}


Following the logic of section \ref{SectSbNrhrz}, one can define the near--horizon limits by zooming in on the 
vicinity of 
membranes or M5 branes. To arrive at solutions asymptoting to $AdS_4\times S^7$, we perform a rescaling 
\bea
M2:\quad ds_{1,1}^2=\eps^{-4}d{\tilde s}_{1,1}^2,~
\left(\begin{array}{c}{\bf x}\\ {\bf y}\end{array}\right)=\eps\left(\begin{array}{c}{\tilde{\bf x}}\\ {\tilde{\bf y}} \end{array}\right),~\left(\begin{array}{c}w\\ F\end{array}\right)=
\frac{1}{\eps^{4}}\left(\begin{array}{c}{\tilde w}\\ {\tilde F} \end{array}\right),~
e^\phi=\eps^3e^{\tilde\phi}\nonumber
\eea
and send $\eps$ to zero. Notice that this parameter drops out from the equations 
(\ref{GenDifUr1})--(\ref{GenDifUr3}), so the only difference between the old and new solutions is in the boundary conditions at infinity: both $e^\phi$ and $e^H$ go to one for asymptotically flat space, while 
\bea
e^H\rightarrow 1,\quad e^{-2\phi}\rightarrow\frac{Q}{({\bf x}^2+{\bf y}^2)^{3}}
\eea
for the solutions with $AdS_4\times S^7$ asymptotics. 

To zoom in on the vicinity of M5 branes we perform the rescaling
\bea
M5:\quad ds_{1,1}^2=\eps^{-2}d{\tilde s}_{1,1}^2,\quad {\bf x}=\frac{\tilde{\bf x}}{\eps},\quad 
{\bf y}=\eps^2{\tilde{\bf y}},\quad
e^H=\eps^3e^{\tilde H},\quad F=\eps^2{\tilde F},\quad w=\frac{\tilde w}{\eps^4}\nonumber
\eea
and send $\eps$ to zero. The $AdS_7\times S^4$ asymptotics correspond to the following boundary 
conditions:
\bea
e^{\phi}\rightarrow 1,\quad e^{-2H}\rightarrow \frac{Q}{({\bf y}^2+F^2)^{3/2}}.
\eea
An alternative way of getting solutions with $AdS_7\times S^4$ asymptotics involves the near horizon limit of M5' branes:
\bea
M5':\quad ds_{1,1}^2=\frac{d{\tilde s}_{1,1}^2}{\eps^2},~ 
\left(\begin{array}{c}{\bf x}\\ w\end{array}\right)=\eps^2\left(\begin{array}{c}{\tilde{\bf x}}\\ w \end{array}\right),~ 
{\bf y}=\frac{\tilde{\bf y}}{\eps},~
\left(\begin{array}{c}e^{\phi}\\ e^{-H}\end{array}\right)=
\eps^3\left(\begin{array}{c}e^{\tilde \phi}\\ e^{-\tilde H}\end{array}\right),~ F=\frac{\tilde F}{\eps^4}.
\nonumber
\eea
In fact, one can see that geometry (\ref{M5M5M2}) as well as equations (\ref{GenDifUr1})--(\ref{GenDifUr3}) 
are invariant under $Z_2$ symmetry which exchanges two types of M5 
branes\footnote{Notice that this is the symmetry of equations, and 
while it is broken by individual solutions, it can be used to relate different geometries.} and allows one to get one near horizon limit from the other:
\bea
{\bf x}\leftrightarrow {\bf y},\quad w\leftrightarrow F,\quad e^H\leftrightarrow e^{-H},\quad
e^{-2\phi}\leftrightarrow e^{-2\phi-2H}.
\eea

Generic solutions discussed in this section preserve eight supercharges, but in special cases the 
supersymmetry can be enhanced. For example, the geometries produced by parallel membranes 
(or by M5 branes alone) preserve $16$ supercharges. The probe analysis presented in section 
\ref{SectSbScan}
suggests that among asymptotically--flat geometries, these are the only solutions with enhanced supersymmetry. However other configurations preserving 
$16$ supersymmetries are possible, but they must have different asymptotics. In the most interesting cases the amount of SUSY is further enhanced at infinity, so we will look at solutions which asymptote to 
$AdS_4\times S^7$ or  
$AdS_7\times S^4$. The $1/2$--BPS solutions for these spaces were constructed in \cite{LLM,myDefect} and it is easy to embed the geometries of \cite{myDefect} into the more general 
setup (\ref{M5M5M2}). 

First we observe that $1/2$--BPS configurations in $AdS_p\times S^q$ preserve 
$SO(2,2)\times SO(4)^2$ 
symmetry\footnote{There are also solutions with $SO(6)\times SO(3)$ symmetry \cite{LLM}, but they do not correspond to intersecting branes and thus do not fit into the ansatz (\ref{M5M5M2}).}, so to match them we 
will assume the rotational invariance in ${\bf x}$ and ${\bf y}$ subspaces. We also assume that $R^{1,1}$ is promoted into $AdS_3$:
\bea
dH_3^2=-z^2ds_{1,1}^2+\frac{dz^2}{z^2}.
\eea
Then one can easily embed the solutions of \cite{myDefect} into the general form (\ref{M5M5M2}) (similar task 
for type IIB geometries was accomplished by equations (\ref{BWlsToNew}), (\ref{BWlsToNew1})), and it appears that 
among these solutions only $AdS_4\times S^7$ and $AdS_7\times S^4$ can be obtained as some near--horizon 
limits of asymptotically--flat geometries.

\subsection{Smeared M2--M2'--KK intersection}

For an alternative dualization we assume that nothing depends on $z=x_3$, then T duality along that direction gives a type IIA solution and a further lift produces a geometry in M theory.
To find the type IIA description in terms of $F_4$ (rather than dual six--form), it is convenient 
to find a more explicit four--form RR potential corresponding to (\ref{GenSltnSum}). Such potential obeys 
an equation:
\bea\label{DualC4}
dC_4&=&-\frac{1}{4}(d+^*_{10}d)\left[e^{-2H}~^*_5 d_y F\right]+\frac{1}{4}C_2\wedge H_3=
-\frac{1}{4}d\left[e^{-2H}~^*_5 d_y F\right]\nonumber\\
&&-\frac{1}{4}\left(e^{H+\phi}~^{\tilde *}_8d\left[e^{-2H}~^*_5 d_y F\right]+
\eps_{ij}\d_j Fdx^i\wedge 
d\left[e^{2\phi}(\d_w Fdw+\d_{\bf y}F d{\bf y})\right]\right)dt~dz\nonumber\\
&\equiv&-\frac{1}{4}d\left[e^{-2H}~^*_5 d_y F\right]-\frac{1}{4}G_3\wedge dt~dz.
\eea
This relation defines a useful three--form $G_3$ and 
eight dimensional Hodge dual appearing in it is taken with respect to the string metric:
\bea
ds_8=e^H d{\bf x}_2^2+e^{-H}\left[d{\bf y}_5^2+
e^{2\phi}(dF+e^{2H}\d_{\bf x}w d{\bf x})^2\right].\nonumber
\eea
Now it is easy to dualize (\ref{GenSltnSum}) along $z$ direction:
\bea
ds_{IIA}^2&=&e^H\left[-e^{2\phi}dt^2+d{\bf x}_2^2\right]+e^{-H}\left[d{\bf y}_5^2+dz^2+
e^{2\phi}(dF+e^{2H}\d_{\bf x}w d{\bf x})^2\right]\nonumber\\
e^{\Phi}&=&e^{\phi-H/2},\quad F_2=d\left[\eps_{ik}\d_k Fdx^{i}\right],\quad
H_3=d\left[e^{2\phi}(\d_w Fdw+\d_{\bf y}F d{\bf y})\right]dt,\nonumber\\
F_4&=&-G_3\wedge dt,\quad F_6=-d\left[e^{-2H}~^*_5d_y F\wedge dz\right],\nonumber
\eea
and a lift to M theory produces a geometry describing smeared intersecting branes:
\bea
ds_M^2&=&e^{-2\phi/3}\left\{
e^{4H/3}\left[-e^{2\phi}dt^2+d{\bf x}_2^2\right]+e^{-2H/3}\left[d{\bf y}_5^2+dz^2\right]\right\}\nonumber\\
&&+e^{-2H/3+4\phi/3}\left[(dF+e^{2H}\d_{\bf x}w d{\bf x})^2+(dx_{11}+\eps_{ik}\d_k Fdx^{i})^2\right],\nonumber\\
F_4&=&d\left[e^{2\phi}(\d_w Fdw+\d_{\bf y}Fd{\bf y})dt\wedge dx_{11}\right]-G_3\wedge dt.
\nonumber
\eea
While it is easy to guess the solution which is not smeared along $z$ direction:
\bea\label{M2M2KKsln}
ds_M^2&=&e^{-2\phi/3}\left\{
e^{4H/3}\left[-e^{2\phi}dt^2+d{\bf x}_2^2\right]+e^{-2H/3}d{\bf y}_6^2\right\}\nonumber\\
&&+e^{-2H/3+4\phi/3}\left[(dF+e^{2H}\d_{\bf x}w d{\bf x})^2+(dx_{11}+\eps_{ik}\d_k Fdx^{i})^2\right],\nonumber\\
F_4&=&-\left\{d\left[e^{2\phi}(\d_w Fdw+\d_{\bf y}Fd{\bf y})\right]\wedge 
(dx_{11}+\eps_{ik}\d_k Fdx^{i})+{\tilde G}_3\right\}\wedge dt,\\
{\tilde G}_3&\equiv&e^{H/2+\phi}~^{\tilde *}_9d\left[e^{-2H}~^*_6 d_y F\right],\nonumber
\eea
the translational invariance in $x_{11}$ seems to be a crucial property of the geometry, and 
we will not try to relax it. Notice that the nine dimensional duality in (\ref{M2M2KKsln}) is performed using the metric
\bea
ds_9&=&e^H d{\bf x}_2^2+e^{-H}\left[d{\bf y}_6^2+
e^{2\phi}(dF+e^{2H}\d_{\bf x}w d{\bf x})^2\right]
\eea
and functions $\phi,w,F$ satisfy the system (\ref{GenDifUr1})--(\ref{GenDifUr3}) with two--component ${\bf x}$ 
and six--component ${\bf y}$. Generically the metric (\ref{M2M2KKsln}) is expected to have only 
$U(1)\times U(1)$ isometry (which corresponds to the translations in time and in the direction of smearing $x_{11}$), but more symmetric solutions can also be found. For example, requiring that all functions depend only on the radial directions in ${\bf x}$ and ${\bf y}$, one finds an enhanced $SO(6)\times U(1)^3$ symmetry. However this isometry should be distinguished from the symmetry of a pointlike intersection of two membranes: in the first case two of the $U(1)$s 
correspond to translations and one to rotation, and in the second case the roles are reversed:
\bea\label{M2IntrsNlcl}
\begin{array}{c|ccccc}
&1&2&3&4&5-10\\
\hline
M2_{smeared}&\bullet&\bullet&\sim&&\\
M2&&&\bullet&\bullet&\\
\end{array}\qquad \qquad
\begin{array}{c|ccccc}
&1&2&3&4&5-10\\
\hline
M2&\bullet&\bullet&&&\\
M2&&&\bullet&\bullet&\\
\end{array}
\eea
In the near horizon limit the geometry produced by two intersecting branes has an enhanced 
$SO(6)\times SO(2,1)$ symmetry and the corresponding metrics can be specified in terms of solutions of Toda equation \cite{myDefect}. It would be very interesting to find an asymptotically flat solution describing the localized intersection (i.e. the second configuration in (\ref{M2IntrsNlcl})) and compare with \cite{myDefect}. We will not attempt to do this here.

\subsection{M5 brane, KK monopole and a plane wave}

Let us now discuss the last possible duality mentioned in (\ref{11DTdual}). To proceed we need to assume an extra isometry in the direction orthogonal to ${\bf x}$ and ${\bf y}$ and  it appears that there are two natural possibilities: we can require a translational invariance in either $F$ or $w$. Let us consider these cases separately.

{\bf T duality along $w$.} Assuming a translational isometry in $w$, we find restrictions on $F$:
\bea
F=wq({\bf x})+{\tilde F}({\bf x},{\bf y}),\quad \Delta_{\bf x}q({\bf x})=0:\quad
e^{2H}=q.
\eea 
This leads to significant simplifications in the equations (\ref{GenDifUr1}), (\ref{GenDifUr2}):
\bea\label{M5KK1}
\Delta_{\bf x}e^{2H}=0,\quad \d_{\bf y}e^{2H}=0,\quad \Delta_{\bf x}{\tilde F}=0,\quad
\nabla_{\bf y}(e^{-2H}\nabla_{\bf y}{\tilde F})=0,
\eea
Of course, one should add sources to some of these equations and the relevant analysis for smeared branes 
was performed in section \ref{SectSbExplc}. Applying it to the present case, we arrive at a special case of 
(\ref{SmearedGeom}), (\ref{SmearFundSrc}) corresponding to ${\tilde q}({\bf y})=1$:
\bea
ds^2&=&e^H\left[-e^{3\phi/2}dt^2+e^{-\phi/2}d{\bf x}^2_3\right]+
e^{-H-\phi/2}d{\bf y}^2_5+e^{3H+3\phi/2}dw^2,\nonumber\\
F_5&=&0,\quad F_3=^*_3d_x q\wedge dw,\quad
H_3=de^{2\phi+2H}\wedge dw\wedge dt,\nonumber\\
e^{2H}&=&q({\bf x}),\quad F=q({\bf x})w,\quad 
q\Delta_{\bf y} (q^{-1}e^{-2\phi})+\Delta_{\bf x}(q^{-1}e^{-2\phi})=0.
\eea
Unfortunately, this system does not contain D3 branes. However it is still interesting to perform a T duality 
along $w$ and an M theory lift to produce a pure metric in eleven dimensions:
\bea\label{M5KK00}
ds_{M}^2&=&e^{2H}\left[-e^{2\phi}dt^2+d{\bf x}^2_3\right]+e^{-2H}(dx_{11}-\omega_1)^2+
e^{-2H-2\phi}(dw-e^{2H+2\phi} dt)^2+d{\bf y}^2_5\nonumber\\
d\omega_1&\equiv&^*_3d_x e^{2H}\quad \Delta_{\bf x}e^{2H}=0,\quad 
e^{2H}\Delta_{\bf y} (e^{-2H-2\phi})+\Delta_{\bf x}(e^{-2\phi-2H})=0.
\eea
This solution corresponds to a smeared configuration of a plane wave and KK monopole:
\bea\label{KKPscan}
\begin{array}{c|cccccccccc}
&1&2&3&4&5&6&7&8&9&11\\
\hline
KK&&&&\bullet&\bullet&\bullet&\bullet&\bullet&\bullet& \sim\\
P&&&&\bullet&&&&&&\sim \\
\end{array}
\eea
and it is easy to guess a non--smeared solution corresponding to an arbitrary hyper--Kahler base in four 
dimensions\footnote{This solution can be generalized even further by replacing $R^5_y\rightarrow HK'\times R^1$, but such $(KK_{1234},KK_{5678},P_9)$ system preserves only four supercharges, and in this paper we are interested in $1/4$--BPS intersections.}:
\bea\label{M5KK01}
ds_{M}^2&=&-2dw dt+e^{-2H-2\phi}dw^2+\left[ds_{HK}^2+d{\bf y}_5^2\right],\nonumber\\
&&\Delta_{\bf y} (e^{-2H-2\phi})+\Delta_{HK}(e^{-2\phi-2H})=0.
\eea
Let us now discuss another possible dualization.

{\bf T duality along $F$.} Assuming that (\ref{GenSltnSum}) is invariant under translations in $F$, we
find that $w$ is linear in $F$ and $e^{-2H}$ is a harmonic function which depends only on ${\bf y}$. Then the analysis of section \ref{SectSbExplc} implies that the D3--D5--F1 geometry is a particular case of 
(\ref{SmearedGeom}):
\bea
ds^2&=&e^H\left[-e^{3\phi/2}dt^2+e^{-\phi/2}d{\bf x}^2_3\right]+
e^{-H-\phi/2}d{\bf y}^2_5+e^{-H+3\phi/2}dF^2,\nonumber\\
F_5&=&\frac{1}{4}dF\wedge ^*_5d_y e^{-2H} +dual,\quad F_3=0,\quad
H_3=de^{2\phi}\wedge dF\wedge dt,\nonumber\\
&&e^{-2H}={\tilde q}({\bf y}),\quad w={\tilde q}({\bf y})F.
\eea
Dualizing along $F$ direction and lifting to M theory, we arrive at geometry describing M5 brane with longitudinal momentum:
\bea\label{M5KKp0}
ds^2_{M}&=&e^{2H/3}\left[-2dt~dF+e^{-2\phi}dF^2+d{\bf x}^2_3+dx_{11}^2\right]+
e^{-4H/3}d{\bf y}^2_5,\nonumber\\
F_4&=&^*_5 d_y e^{-2H},\quad \Delta_{\bf y}e^{-2H}=0,\quad
\Delta_{\bf y} e^{-2\phi}+e^{-2H}\Delta_{\bf x} e^{-2\phi}=0.
\eea
While this solution was obtained assuming translational invariance in $x_{11}$, this requirement can be relaxed since $x_{11}$ appears on the same footing as three other $x_i$. 

To summarize,  we found the geometries produced by either $(KK,P)$ or $(M5,P)$ systems, i.e. we were able to 
put together two elements of the triple:
\bea
\begin{array}{c|cccccccccc}
&1&2&3&4&5&6&7&8&9&10\\
\hline
M5&\bullet&\bullet&\bullet&\bullet&&&&&&\bullet\\
KK&&&&\bullet&\bullet&\bullet&\bullet&\bullet&\bullet& \\
P&&&&\bullet&&&&&&\\
\end{array}
\eea
It appears that a more general geometry containing both KK monopoles and M5 branes cannot be found by applying dualities to D5--D3--F1 system and one needs to solve the equations of motion in eleven dimensions. We leave this problem for future publication.

\subsection{Summary}

Let us summarize the results of this section. By applying various dualities to D5--D3--F1 solution, we have constructed geometries produced by various brane intersections which preserve eight supercharges in eleven dimensions. One of such intersections (M5--M5--M2) was completely localized and in this case we found a perfect agreement between gravity picture and probe analysis presented in section \ref{SbctMtchPrb}. The other two 
intersections were partially delocalized: M2--M2--KK was smeared in one of the directions along M2 brane and M5--KK--P system was smeared in the direction orthogonal to the monopole and to the momentum.  It would be very interesting to find the localized version of the last two solutions. 
In the case of M5--M5--M2 intersection one can go to the near--horizon limit of one of the M5 branes, then an enhancement of the (super)symmetry is possible. The relevant geometries preserve $16$ supercharges along with 
$SO(2,2)\times SO(4)^2$ bosonic symmetries and corresponding metrics were constructed in 
\cite{myDefect}. In this section we saw how such symmetric solution can be embedded in a general solution (\ref{M5M5M2}). Notice that a generically the supersymmetric solution 
(\ref{M5M5M2}) is only expected to have $ISO(2,1)$ isometry and the geometries constructed in 
\cite{myDefect} present a very special class of solutions.


\section{Discussion}
\renewcommand{\theequation}{6.\arabic{equation}}
\setcounter{equation}{0}


D branes are essential part of string theory so it is very important to understand their dynamics.
There are two ways of looking at branes: one is based on open string physics and another uses a picture in terms of closed strings. While both methods have been equally successful in describing branes preserving $16$ supercharges, for less symmetric branes the situation is more complicated. At low energies the open string physics is well--described by the DBI action and in this approximation various brane intersections have been extensively studied in the past. However from the point of view of closed strings, the low--energy dynamics is governed by supergravity and in the past very few 
$1/4$--BPS configurations have been described using this language. In the known solutions 
the positions 
of the branes were specified from the beginning and the geometries were constructed using
so--called "harmonic rule": different branes obeyed independent linear equations.  In this paper we constructed a large class of supersymmetric solutions which are governed by two functions satisfying a system of nonlinear PDEs and the positions of the branes are determined dynamically. Of course, 
for BPS objects one expects to have a superposition principle, 
so it is possible that the nonlinear equations (\ref{GenDifUr1})--(\ref{GenDifUr3}) are integrable. If this is indeed the 
case, it would be very nice to find a map to the appropriate variables in which this system becomes linear. Despite the lack of such map at the moment, the superposition principle did manifest itself in the boundary conditions: in section \ref{SectSbBcnd} we showed that for consistency the branes should follow harmonic profiles (in a perfect agreement with probe analysis) and the construction of section 
\ref{SectSbPert} demonstrates that any 
combination of such branes leads to a unique solution.

It is very natural to consider strings ending on branes which were discussed in this paper: 
looking at $1/4$--BPS 
configurations (\ref{IntersBraneIIB}) in IIB string theory one observes that they fall into two categories: the geometries corresponding to the last two lines can be constructed using "harmonic rule" and they have been studied in the past, while all intersections appearing in the first two lines are captured by the ansatz presented in this paper. To be more precise, we explicitly derived the D3--D5--F1 solution in section \ref{SectSbGenSum} and other geometries were obtained from it in section \ref{SectSbOthIIB}. 

While we were not able to solve equations for the most general distribution of branes,
some special solutions can be constructed. In particular, in section \ref{SecSbNonCm} we showed that the 
geometric duals of non--commutative field theories \cite{HasItz,MalRus} can be recovered from our ansatz. 
In section 
\ref{SectSbExplc} we also found an explicit solution for a smeared D3--D5 intersection. Although we were mostly 
interested in asymptotically flat geometries, the system (\ref{GenDifUr1})--(\ref{GenDifUr3}) is also applicable 
to solutions 
embedded in different spaces, in particular in sections \ref{SecSbWils} and \ref{SecSbSbs} we showed that 
$1/2$--BPS geometries $AdS_p\times S^q$ asymptotics \cite{myWils,myDefect} are included as very special 
cases into the ansatz discussed here. 

The results of our investigation are very encouraging. We were able to find solutions preserving only eight supercharges and no bosonic isometries (apart from the time translation which is a consequence of supersymmetry).
One may hope that similar techniques can be applied to situations with lower supersymmetry and all brane intersections preserving four supercharges can also be classified. In fact, the equations governing some of such
configurations are known \cite{GMR,BW}, and it would be nice to describe other intersections as well.

\section*{Acknowledgments}

It is a pleasure to thank Jeff Harvey and David Kutasov for useful discussions. 
This work is supported by DOE grant DE-FG02-90ER40560.

\appendix


\section{Conventions}
\label{AppSugra}

\renewcommand{\theequation}{A.\arabic{equation}}
\setcounter{equation}{0}


The main goal of this paper is to find a geometric description of intersecting branes, and one needs to solve equations coming from supergravity to accomplish this task. In this appendix we collect some basic facts about 
Type IIB supergravity following the the standard notation of \cite{schwarz}.

Since we are looking for bosonic solutions preserving supersymmetry, so we begin by summarizing the 
SUSY variations for such geometries
\bea\label{GenSUSYVar}
\delta\la&=&i{\not P}\eps^*-\frac{i}{24}\gamma^{mnp}G_{mnp}\eps,\nonumber\\
\delta \psi_m&=&(\nabla_m-\frac{i}{2}Q_M)\eps+\frac{i}{480}{\not F}_5\gamma_m\eps+
\frac{1}{96}(-\gamma_m{\not G}-2{\not G}\gamma_m)\eps^*.
\eea
Supersymmetry parameter $\eps$ is a complex Weyl spinor ($\Gamma_{11}\eps=-\eps$), 
and the general expressions for two vectors $Q_m, P_m$ and a scalar $B$ can be found in 
\cite{schwarz} (see also \cite{granaPolch}). As explained in section \ref{SectSpike}, we are interested in  
solutions with vanishing axion $C^{(0)}$, this implies that $\tau=ie^{-\phi}$, $Q_\mu=0$, and
\bea
&&P_m=\frac{1}{2}\d_m\phi,\quad
B=\frac{1-e^{-\phi}}{1+e^{-\phi}},\quad f^{-2}=
\frac{4e^{-\phi}}{(1+e^{-\phi})^2},\nonumber\\
&&G_3=f(H_3+iF_3-BH_3+iBF_3)=e^{-\phi/2}H_3+ie^{\phi/2}F_3.
\eea
Substituting these expressions into (\ref{GenSUSYVar}) and requiring the variations to vanish, we arrive at 
the equations which will be analyzed in the next two appendices:
\bea\label{TheSUSYVar}
&&\delta\la=\frac{i}{2}{\not\d}\phi\eps^*-\frac{i}{24}\gamma^{mnp}G_{mnp}\eps=0,\\
&&\delta\psi_M=\nabla_M\eps+\frac{i}{480}{\not F}_5\gamma_M\eps+\frac{1}{96}
(-\gamma_M
{\not G}-2{\not G}\gamma_M)\eps^*=0.\nonumber
\eea
While some of the equations of motion of type IIB supergravity follow from the last two relations, a generic background satisfying (\ref{TheSUSYVar}) might not be a solution of the theory. In particular, one always has to supplement SUSY variations with Bianchi identities for the field strengths, but sometimes even this system is not complete and some equations of motion should be solved explicitly. Let us summarize these equations for 
$C^{(0)}=0$ (see \cite{schwarz} for the discussion of the general case):
\bea
&&\nabla^2\phi=-\frac{e^{-\phi}}{12}H_{mnp}H^{mnp}+\frac{e^{\phi}}{12}F_{mnp}F^{mnp},\quad 
dF_5=-\frac{1}{4}F_3\wedge H_3,\nonumber\\
&&d*(e^{-\phi}H_3)=-4F_5\wedge F_3,\quad d*(e^{\phi}F_3)=4F_5\wedge H_3,\\
&&R_{mn}=\frac{1}{2}\d_m\phi\d_n\phi+\frac{1}{24}F_{mabcd}{F_n}^{abcd}+
(\delta_m^p\delta_n^q-\frac{g_{mn}g^{pq}}{12})(\frac{e^{-\phi}}{4}H_{pab}{H_q}^{ab}+
\frac{e^{\phi}}{4}F_{mpq}{F_n}^{pq}).\nonumber
\eea
The field strengths appearing in these equations are related to gauge potentials in the following way:
\bea
H_3=dB_2,\quad F_3=dC_2,\quad F_5=dC_4-\frac{1}{4}C_2\wedge H_3.
\eea
Moreover, the five--form fiend strength must be self--dual: $F_5=*F_5$.

Throughout this paper we use normalization which is common in supergravity literature, but it is slightly different 
from conventions which are natural from the point of view of string theory. It is well--known that a metric in the Einstein frame (which is used in supergravity) is different from a metric seen by a fundamental string propagating on the geometry:
\bea
ds_S^2=\frac{e^{\phi/2}}{\sqrt{g}}ds^2_E.
\eea
It turns out that there are also differences in normalization of RR fluxes and the detailed discussion can be found in \cite{granPolDil}. To simplify the calculations, we will set both string coupling constant and 
Newton's constant 
$\kappa$ to be equal to one (although they can be easily restored), then the map between string and gravitational quantities found in \cite{granPolDil} simplifies:
\bea
H^{(s)}_3=H_3,\quad F^{(s)}_3=F_3,\quad F^{(s)}_5=4F_5.\quad
\eea


\section{Solutions with $SO(5)\times SO(3)$ symmetry}
\label{AppSpike}

\renewcommand{\theequation}{B.\arabic{equation}}
\setcounter{equation}{0}


In this appendix we study supersymmetry variations for configurations in type IIB supergravity 
which have $SO(3)\times SO(5)$ symmetry. This symmetry was motivated in section \ref{SectSpike} 
by looking at 
a single spherically symmetric spike, and as we will see, once the symmetric solution is obtained, it is very easy to generalize it to the case of multiple spikes.

\subsection{Formulation of the problem}

We begin with metric and fluxes given by (\ref{IIBMetrAns}), (\ref{IIBFluxAns}):
\bea\label{SngSpkStrt}
ds^2&=&-e^{2A}dt^2+e^{2B}d\Omega_2^2+e^{2C}d\Omega_4^2+h_{ij}dx^idx^j,\\
H_3&=&2\omega_2\wedge dt,\quad F_3=df_2\wedge d\Omega_2,\quad
F_5=df_3\wedge d\Omega_4+dual,\quad e^{\phi}.\nonumber
\eea
Here $\omega_2$ is a closed two--form in three--dimensional space spanned by $x_i$ and 
all scalars are assumed to be functions of these three coordinates.

Equations (\ref{SngSpkStrt}) guarantee that all bosonic fields have the required symmetry, but we also need 
to impose the symmetry on the spinor.
To do this we need to review a 
construction of invariant spinors on even--dimensional spheres. First we recall that a covariant 
derivative $\nabla_m$ along one of the directions on $S^2$ can be rewritten in terms of a 
derivative ${\tilde\nabla}_m$ on a unit sphere:
\bea
\nabla_m^S={\tilde\nabla}_n^S-\frac{1}{2}{\gamma^\mu}_m\d_\mu B.
\eea
There are various ways of expressing ${\tilde\nabla}_m$ in terms of gamma matrices, here we will follow 
the approach of \cite{myWils} where it was shown that 
\bea
{\tilde\nabla}_m\eps=-\frac{i}{2}e^{-B}\gamma_m P_S\eps,
\eea
where $P_S$ is a hermitean matrix which anticommutes with chirality operator $\Gamma_S$ on $S^2$ and 
with gamma matrices along the directions orthogonal to this sphere. The derivatives along 
$S^4$ directions can be computed in an analogous way. Notice that equations (\ref{TheSUSYVar}) 
are valid only in the basis where all 
gamma matrices are real, this imposes certain reality conditions on four hermitean matrices: 
$\Gamma_\Omega$ is real while 
$\Gamma_S$, $P_S$, $P_\Omega$ are pure imaginary. 

It is convenient to split ${\not G}$ into the real and imaginary pieces:
\bea
\frac{1}{24}{\not G}=G_++G_-,\qquad (G_\pm)^*= \pm G_\pm,
\eea
and an explicit computation gives
\bea
G_+=-\frac{1}{4}e^{-\phi/2-A}{\not\omega}_2\Gamma_t,&& G_-=-\frac{1}{4}
e^{\phi/2-2B}{\not\d}f_2\Gamma_S.
\eea
The last remaining ingredient that enters the equations is
\bea
\frac{1}{480}{\not F}_5\eps=\frac{e^{-4C}}{2}{\not \d}f_3\Gamma_\Omega\eps.
\eea
Combining all this information, we arrive at the system:
\bea\label{App1eqnDltn}
&&\frac{1}{2}{\not\d}\phi\eps^*-(G_++G_-)\eps=0,\\
\label{App1eqnA}
&&{\not\d}A\eps-ie^{-4C}{\not\d}f_3\Gamma_{\Omega}\eps+\frac{1}{2}
(-3G_++G_-)\eps^*=0,\\
\label{App1eqnB}
&&(-ie^{-B}P_S+{\not\d}B)\eps-ie^{-4C}{\not\d}f_3\Gamma_{\Omega}\eps+\frac{1}{2}
(G_+-3G_-)\eps^*=0,\\
\label{App1eqnC}
&&(-ie^{-C}P_\Omega+{\not\d}C)\eps+ie^{-4C}{\not\d}f_3\Gamma_{\Omega}\eps+
\frac{1}{2}(G_++G_-)\eps^*=0,\\
\label{App1eqnDiff}
&&\nabla_\mu\eps+i\frac{e^{-4C}}{2}{\not\d}f_3\gamma_\mu\Gamma_\Omega\eps+
\frac{1}{96}(\gamma_\mu
{\not G}-2\{{\not G},\gamma_\mu\})\eps^*=0.
\eea
For future reference we write the complex conjugate of the dilatino equation:
\bea\label{App1eqnE}
\frac{1}{2}{\not\d}\phi\eps-(G_+-G_-)\eps^*=0.
\eea
To evaluate the spinor bilinears we will also need the hermitean conjugates of the relations 
(\ref{App1eqnDltn})--(\ref{App1eqnE}):
\bea\label{App1HermCnj}
&&\frac{1}{2}\eps^T{\not\d}\phi-\eps^\dagger(G_++G_-)=0,\qquad 
\frac{1}{2}\eps^\dagger{\not\d}\phi-
\eps^T(G_+-G_-)=0,\nonumber\\
&&\eps^\dagger {\not\d}A+ie^{-4C}\eps^\dagger{\not\d}f_3\Gamma_{\Omega}
+
\frac{1}{2}\eps^T(-3G_++G_-)=0,\nonumber\\
&&\eps^\dagger(ie^{-B}P_S+{\not\d}B)+ie^{-4C}\eps^\dagger{\not\d}f_3
\Gamma_{\Omega}+
\frac{1}{2}\eps^T(G_+-3G_-)=0,\\
&&\eps^\dagger(ie^{-C}P_\Omega+{\not\d}C)-ie^{-4C}\eps^\dagger{\not\d}f_3
\Gamma_{\Omega}+
\frac{1}{2}\eps^T(G_++G_-)=0,\nonumber\\
&&\nabla_\mu\eps^\dagger-i\frac{e^{-4C}}{2}\eps^\dagger\gamma_\mu{\not\d}f_3
\Gamma_\Omega+
\frac{1}{96}\eps^T({\not G}^\dagger\gamma_\mu-2\{{\not G}^\dagger,\gamma_\mu\})=0.
\nonumber
\eea
The remaining part of this appendix will be devoted to solving the system 
(\ref{App1eqnDltn})--(\ref{App1eqnE}).

\subsection{Looking at the projectors and choosing the coordinates}
\label{AppSubsProj}

We begin by combining the projectors appearing in (\ref{App1eqnA})--(\ref{App1eqnE}) to construct 
the equations 
that do not contain fluxes:
\bea\label{GeomProj1}
&&\left[-ie^{-C}P_\Omega+{\not\d}(A+C-\frac{\phi}{2})\right]\eps=0,\\
\label{GeomProj2}
&&\left[-ie^{-B}P_S+{\not\d}(B-A+\phi)\right]\eps=0.
\eea
Depending on a choice of coordinates, each of these two projectors can contain up to four gamma matrices, 
however by choosing some special set of coordinates one can simplify both projectors. Namely we introduce
two functions
\bea\label{DefUV}
u=e^{A+C-\frac{\phi}{2}},\quad v=e^{B-A+\phi}
\eea
and use them as two of the coordinates. In principle one can worry that $u$ and $v$ are not independent functions (then they cannot be used as 
two coordinates), and to show the independence we evaluate a commutator in two different ways:
\bea
&&\eps^\dagger\{{\not\d}(A+C-\frac{\phi}{2}),{\not\d}(B-A+\phi)\}\eps=
-e^{-B-C}\eps^\dagger\{P_\Omega,P_S\}\eps=0,\nonumber\\
&&\eps^\dagger\{{\not\d}(A+C-\frac{\phi}{2}),{\not\d}(B-A+\phi)\}\eps=2\eps^\dagger\eps 
g^{\mu\nu}\d_\mu \log u~\d_\nu\log v.
\eea
We see that not only $u$ and $v$ are independent, but also $g^{uv}=0$, so we can choose 
frames\footnote{Here and below we use ordinary indices $u,v,\dots$ in curves spacetime and 
we use bold letters ${\bf u},{\bf v},\dots$ to denote frame indices. The exception is made for the gamma matrices, where we have $\gamma_\mu$ with spacetime index and $\Gamma_a$ with frame index.} 
\bea\label{SemiFrame1}
&&e^{\bf u}=e_udu,\quad e^{\bf v}=e_vdv,\quad e^{\bf w}=e_w(dw+A_u du+A_v dv),\nonumber\\
&&e_{\bf u}=e_u^{-1}(\d_u-A_u\d_w),\quad e_{\bf v}=e_v^{-1}(\d_v-A_v \d_w),\quad 
e_{\bf w}=e^{-1}_w \d_w.
\eea
Here $w$ is introduced as the third coordinate and we still have some freedom in choosing it. In particular, it is convenient to impose a gauge $A_v=0$. Such choice still leaves reparameterizations 
$w\rightarrow w'(w,u)$ and we will fix 
this freedom later. 

With this choice of frames the geometric projectors (\ref{GeomProj1}), (\ref{GeomProj2}) become
\bea
&&\left[-ive^{-B}P_S+e_v^{-1}\Gamma_v\right]\eps=0,\quad
\left[-iue^{-C}P_\Omega+e_u^{-1}\Gamma_u\right]\eps=0.
\eea
We conclude that the spinor satisfies two projections:
\bea
(1-i\Gamma_v P_S)\eps=(1-i\Gamma_v P_S)\eps^*=0,\quad
(1-i\Gamma_u P_\Omega)\eps=(1-i\Gamma_u P_\Omega)\eps^*=0,
\eea
and we also extract the expressions for $e_u$, $e_v$:
\bea\label{SemiFrame2}
e_v=e^{A-\phi},\quad e_u=e^{-A+\phi/2}.
\eea
Let us construct a projector which does not contain $G_+$:
\bea
&&{\not\d}(A-\frac{3}{4}\phi)\eps-ie^{-4C}{\not\d}f_3\Gamma_{\Omega}\eps-
G_-\eps^*=0,\nonumber\\
&&{\not\d}(A-\frac{3}{4}\phi)\eps-ie^{-4C}{\not\d}f_3\Gamma_{\Omega}\eps+
\frac{1}{4}e^{\phi/2-2B}{\not\d}f_2\Gamma_S\eps^*=0,
\eea
and apply various projectors to this relation: 
\bea\label{June17Syst}
(1+i\Gamma_v P_S)(1-i\Gamma_u P_\Omega):&&
\Gamma^u{\d}_{\bf u}(A-\frac{3}{4}\phi)\eps-ie^{-4C}\Gamma^w{\d}_{\bf w}f_3\Gamma_{\Omega}\eps=0,\nonumber\\
(1-i\Gamma_v P_S)(1+i\Gamma_u P_\Omega):&&
\Gamma^v{\d}_{\bf v}(A-\frac{3}{4}\phi)\eps+
\frac{1}{4}e^{\phi/2-2B}\Gamma^w{\d}_{\bf w}f_2\Gamma_S\eps^*=0,\\
(1+i\Gamma_v P_S)(1+i\Gamma_u P_\Omega):&&
\Gamma^w{\d}_{\bf w}(A-\frac{3}{4}\phi)\eps-ie^{-4C}\Gamma^u{\d}_{\bf u}f_3\Gamma_{\Omega}\eps+
\frac{1}{4}e^{\phi/2-2B}\Gamma^v{\d}_{\bf v}f_2\Gamma_S\eps^*=0\nonumber\\
(1-i\Gamma_v P_S)(1-i\Gamma_u P_\Omega):&&
-ie^{-4C}\Gamma^v{\d}_{\bf v}f_3\Gamma_{\Omega}\eps+
\frac{1}{4}e^{\phi/2-2B}\Gamma^u{\d}_{\bf u}f_2\Gamma_S\eps^*=0.\nonumber
\eea
Let us assume that the derivatives appearing in the first equation do not vanish (this assumption is true 
even for the flat D3 branes without fluxes), then we find a projector
\bea
\Gamma_u\Gamma_w\Gamma_\Omega\eps=ia\eps.
\eea
The third equation can be rewritten as 
\bea
{\d}_{\bf w}(A-\frac{3}{4}\phi)\eps-ae^{-4C}{\d}_{\bf u}f_3\eps+
\frac{1}{4}e^{\phi/2-2B}\Gamma_w\Gamma^v{\d}_{\bf v}f_2\Gamma_S\eps^*=0,
\eea
and assuming a nontrivial $v$--dependence in $f_2$, we arrive at a projection:
\bea
\Gamma_w\Gamma_v\Gamma_S\eps^*=b\eps,\quad \Gamma_w\Gamma_v\Gamma_S\eps=-b\eps^*,\quad b^2=1.
\eea
Imposing the projections listed above, we reduce the system (\ref{June17Syst}) to a set of scalar equations:
\bea\label{AppEqnAu}
&&{\d}_{\bf u}(A-\frac{3}{4}\phi)+ae^{-4C}{\d}_{\bf w}f_3=0,\\
\label{AppEqnAw}
&&{\d}_{\bf w}(A-\frac{3}{4}\phi)-ae^{-4C}{\d}_{\bf u}f_3+
\frac{b}{4}e^{\phi/2-2B}{\d}_{\bf v}f_2=0,\\
\label{AppEqnAv}
&&{\d}_{\bf v}(A-\frac{3}{4}\phi)-\frac{b}{4}e^{\phi/2-2B}e_w^{-1}{\d}_wf_2=0,\\
\label{AppEqnAf}
&&e^{-4C}{\d}_{\bf v}f_3+\frac{ab}{4}e^{\phi/2-2B}{\d}_{\bf u}f_2=0.
\eea
Finally we look at the dilatino equation:
\bea
\frac{1}{2}{\not\d}\phi\eps+\frac{1}{4}e^{-\phi/2-A}{\not\omega}_2\Gamma_t\eps^*-
\frac{1}{4}e^{\phi/2-2B}{\not\d}f_2 b\Gamma_v\Gamma_w\eps=0.
\eea
Acting by various projectors, we find:
\bea
&&2e^{-\phi/2-A}\Gamma^{uv}{\omega}_{\bf uv}\Gamma_t\eps^*-
be^{\phi/2-2B}\Gamma^u{\d}_{\bf u}f_2 \Gamma_v\Gamma_w\eps=0,\nonumber\\
&&2\Gamma^u\d_{\bf u}\phi\eps+2e^{-\phi/2-A}\Gamma^{uw}{\omega}_{\bf uw}\Gamma_t\eps^*=0,
\nonumber\\
&&2\Gamma^v\d_{\bf v}\phi\eps+2e^{-\phi/2-A}\Gamma^{vw}{\omega}_{\bf vw}\Gamma_t\eps^*+
be^{\phi/2-2B}e_w^{-1}\d_w f_2\Gamma_v\eps=0,\nonumber\\
&&2\Gamma^w\d_{\bf w}\phi\eps-be^{\phi/2-2B}\Gamma^v{\d}_{\bf v}f_2 \Gamma_v\Gamma_w\eps=0.
\nonumber
\eea
Assuming that $\omega$ does not vanish, we find the last projector:
\bea
\Gamma_w\Gamma_t\eps^*=c\eps,\quad \Gamma_w\Gamma_t\eps=-c\eps^*, c^2=1,
\eea
and the equations become
\bea\label{Dltno1}
&&2ce^{-\phi/2-A}{\omega}_{\bf uv}-be^{\phi/2-2B}{\d}_{\bf u}f_2=0,\\
\label{Dltno2}
&&\d_{\bf u}\phi+ce_w^{-1}e^{-\phi/2-A}{\omega}_{{\bf u}w}=0,\\
\label{Dltno3}
&&\d_{\bf v}\phi+ce_w^{-1}e^{-\phi/2-A}{\omega}_{{\bf v}w}+\frac{b}{2}e^{\phi/2-2B}e_w^{-1}\d_w f_2
=0,\\
\label{Dltno4}
&&2e_w^{-1}\d_w\phi-be^{\phi/2-2B}{\d}_{\bf v}f_2 =0.
\eea

At this point 
we already accounted for all projections which should be imposed on the spinor, let us summarize these
projections:
\bea\label{AppAproj}
&&(\Gamma_{11}+1)\eps=(1-i\Gamma_v P_S)\eps=(1-i\Gamma_u P_\Omega)\eps=
(\Gamma_u\Gamma_w\Gamma_\Omega-ia)\eps=0,\nonumber\\
&&\Gamma_w\Gamma_v\Gamma_S\eps^*=b\eps,\quad \Gamma_w\Gamma_t\eps^*=c\eps.
\eea
Notice that only five of these projectors are independent since
\bea
\Gamma_{11}=i\Gamma_u\Gamma_w\Gamma_\Omega\Gamma_t\Gamma_v\Gamma_S:\qquad
\Gamma_{11}\eps=acb~\eps.
\eea
This reproduces the chirality projection in ten dimensions once we require that
\bea\label{10DChiral}
abc=-1.
\eea
To count the number of supersymmetries one should recall that we encountered only eight different matrices 
in the spinor equations (\ref{App1eqnDltn})--(\ref{App1eqnDiff}). These objects can be realized as 
$16\times 16$ matrices and just for illustration we write a particular explicit representation (although 
it is only existence of such representation which will be used):
\bea
\Gamma_a=\gamma^{(4)}_a\otimes {\bf 1}_4,&\ &
P_S=\gamma_5\otimes \sigma_3\otimes {\bf 1}_2,\quad 
P_\Omega=\gamma_5\otimes \sigma_2\otimes \sigma_2,
\nonumber\\
&&\Gamma_S={\bf 1}_4\otimes \sigma_2\otimes {\bf 1}_2,\quad 
\Gamma_\Omega={\bf 1}_8\otimes \sigma_3.
\eea
In this representation $\eps$ is a 16--component complex spinor and five independent projections reduce it 
to one--component real object, so as expected no additional projection can be imposed. To count the number of supersymmetries in ten dimensions we recall that the gamma matrices on $S^2$ and $S^4$ were 
suppressed in this discussion, and once they are re--introduced the size of spinor grows by a factor of 
$2\times 4=8$. So in type IIB we end up with a spinor with eight real independent components, this corresponds to a $1/4$ BPS state. This is consistent with a brane probe analysis. 

To summarize, we have analyzed the dilatino equation as well as three components of gravitino equation (\ref{App1eqnDltn})--(\ref{App1eqnC}) and we showed that these four projectors lead to the restrictions on the Killing spinor (\ref{AppAproj}) and to the bosonic relations (\ref{DefUV}), 
(\ref{SemiFrame2}), (\ref{AppEqnAu})--(\ref{AppEqnAf}), (\ref{Dltno1})--(\ref{Dltno4}). 
We can still use the differential equations (\ref{App1eqnDiff}) to extract some additional information about bosonic fields, and we will do this in the next subsection.

\subsection{Analysis of bilinears}

Let us now look at the differential equation (\ref{App1eqnDiff}) along with its conjugate:
\bea
&&\nabla_\mu\eps+i\frac{e^{-4C}}{2}{\not\d}f_3\gamma_\mu\Gamma_\Omega\eps+
\frac{1}{96}(\gamma_\mu
{\not G}-2\{{\not G},\gamma_\mu\})\eps^*=0,\\
&&\nabla_\mu\eps^\dagger-i\frac{e^{-4C}}{2}\eps^\dagger\gamma_\mu {\not\d}f_3\Gamma_\Omega+
\frac{1}{96}\eps^T(
{\not G}\gamma_\mu-2\{{\not G},\gamma_\mu\})=0.
\eea
These two equations can be combined to evaluate a derivative of the bilinear $\eps^\dagger\eps$:
\bea\label{DerEdE}
&&\nabla_\mu(\eps^\dagger\eps)+i\frac{e^{-4C}}{2}\eps^\dagger[{\not\d}f_3,\gamma_\mu]\Gamma_\Omega\eps+
\frac{1}{96}\left[-2\eps^\dagger {\not G}\gamma_\mu\eps^*-\eps^\dagger\gamma_\mu{\not G}\eps^*+cc\right]=0.
\eea
To eliminate ${\not G}$ from this equation we use (\ref{App1eqnC}) and a hermitean conjugate of  
(\ref{App1eqnDltn}):
\bea
-\frac{1}{48}{\not G}\eps^*=(-ie^{-C}P_\Omega+{\not\d}C)\eps+ie^{-4C}{\not\d}f_3\Gamma_{\Omega}\eps,
\quad
\frac{1}{24}\eps^\dagger{\not G}=\frac{1}{2}\eps^T{\not\d}\phi.\nonumber
\eea
Substituting these relations into (\ref{DerEdE}), one finds:
\bea\label{InterBil21}
&&\nabla_\mu(\eps^\dagger\eps)+i\frac{e^{-4C}}{2}\eps^\dagger[{\not\d}f_3,\gamma_\mu]\Gamma_\Omega\eps-
\frac{1}{2}\d_\mu\phi\eps^\dagger\eps+\\
&&\qquad +
\left[-e^{\bf u}_\mu e^{-C}+\d_\mu C\right]\eps^\dagger\eps+ie^{-4C}\eps^\dagger\gamma_\mu{\not\d}f_3\Gamma_\Omega\eps=0.
\nonumber
\eea
Next we notice that the projection $\Gamma_\Omega\eps=-i\Gamma_u\eps$ implies that 
$\eps^\dagger\Gamma_\Omega\eps=0$, so the terms with $f_3$ cancel out in the above equation. We also 
recall that according to (\ref{SemiFrame1}), (\ref{SemiFrame2}), 
$e^{\bf u}_\mu dx^\mu=e^{-A-\phi/2}du$, then equation (\ref{InterBil21}) simplifies:
\bea
e^{\phi/2-C}\nabla_\mu(e^{-\phi/2+C}\eps^\dagger\eps)-e^{-C-A-\phi/2}\eps^\dagger\eps~\d_\mu u=0.
\nonumber
\eea
Finally recalling the definition (\ref{DefUV}), we eliminate $C$ from the last relation and solve the resulting equation for the bilinear:
\bea\label{NormKill}
&&\nabla_\mu(e^{-A}\eps^\dagger\eps)=0:\qquad \eps^\dagger\eps=e^A.
\eea
At the last stage we fixed a constant in normalization of $\eps$. 

Let us now consider a vector $\eps^\dagger\Gamma_\Omega\gamma_\mu P_\Omega\eps$, which 
has a 
very simple form due to various projectors:
\bea\label{AppAVecForm}
\eps^\dagger\Gamma_\Omega\gamma_\nu P_\Omega\eps~dx^\nu=-a\eps^\dagger\gamma_\nu\Gamma_w\eps~dx^\nu=
-ae^{\bf w}_\mu\eps^\dagger\eps~dx^\mu.
\eea
This vector obeys a differential equation:
\bea\label{AppAVec1}
&&\nabla_\mu(\eps^\dagger\Gamma_\Omega\gamma_\nu P_\Omega\eps)-
ie^{-4C}\eps^\dagger\gamma_{(\nu} {\not\d}f_3\gamma_{\mu)} P_\Omega\eps+
\frac{1}{96}\left[U_{\mu\nu}+V_{\mu\nu}\right]=0,\\
&&U_{\mu\nu}=\eps^\dagger\Gamma_\Omega\gamma_\nu P_\Omega(\gamma_\mu
{\not G}-2\{{\not G},\gamma_\mu\})\eps^*,\quad
V_{\mu\nu}=\eps^T(
{\not G}\gamma_\mu-2\{{\not G},\gamma_\mu\})\Gamma_\Omega\gamma_\nu 
P_\Omega\eps.\nonumber
\eea
In particular, we will be interested in the antisymmetric part of this relation since it will give an exterior derivative of the one--form (\ref{AppAVecForm}). Let us consider various terms separately. 
\bea
&&U_{[\mu\nu]}=-a\eps^\dagger\gamma_{[\nu} (-\gamma_{\mu]}
{\not G}-2{\not G}\gamma_{\mu]})\Gamma_w\eps^*:\nonumber\\
\nonumber\\
&&\quad-4e^{\phi/2+A}\eps^\dagger\gamma_{[\mu}G_+\gamma_{\nu]}\Gamma_w\eps^*=
\eps^\dagger\gamma_{[\mu}{\not\omega}\Gamma_t\gamma_{\nu]}\Gamma_w\eps^*=
c\eps^\dagger\gamma_{[\mu}{\not\omega}\gamma_{\nu]}\eps=2c\omega_{\mu\nu}\eps^\dagger\eps
\nonumber\\
&&\quad
-4e^{\phi/2+A}\eps^\dagger\gamma_{\mu\nu}G_+\Gamma_w\eps^*=
\eps^\dagger\gamma_{\mu\nu}{\not\omega}\Gamma_t\Gamma_w\eps^*=
-c\eps^\dagger\gamma_{\mu\nu}{\not\omega}\eps=2c\omega_{\mu\nu}\eps^\dagger\eps
\nonumber\\
&&\quad-4e^{2B-\phi/2}
\eps^\dagger\gamma_{[\mu}G_-\gamma_{\nu]}\Gamma_w\eps^*=
\eps^\dagger\gamma_{[\mu}{\not\d}f_2\Gamma_S\gamma_{\nu]}\Gamma_w\eps^*=
-b\eps^\dagger\gamma_{[\mu}{\not\d}f_2\gamma_{\nu]}\Gamma_v\eps=0\nonumber\\
&&\quad-4e^{2B-\phi/2}\eps^\dagger\gamma_{\mu\nu}G_-\Gamma_w\eps^*
=\eps^\dagger\gamma_{\mu\nu}{\not\d}f_2\Gamma_S\Gamma_w\eps^*=
-b\eps^\dagger\gamma_{\mu\nu}{\not\d}f_2\Gamma_v\eps\nonumber\\
\nonumber\\
&&U_{[\mu\nu]}=6a(6ce^{-A-\phi/2}\omega_{\mu\nu}\eps^\dagger\eps
-e^{-2B+\phi/2}b\eps^\dagger\gamma_{\mu\nu}{\not\d}f_2\Gamma_v\eps),
\eea
\bea
V_{\mu\nu}&=&
-a\eps^T(-2\gamma_\mu{\not G}\gamma_\nu-
{\not G}\gamma_\mu\gamma_\nu) \Gamma_w\eps:\nonumber\\
\label{Vstar}
\frac{1}{6}V^*_{[\mu\nu]}&=&
-4a\eps^\dagger(-2\gamma_{[\mu}(G_+-G_-)\gamma_{\nu]}-
(G_+-G_-)\gamma_{\mu\nu}) \Gamma_w\eps^*\nonumber\\
&=&a(-6ce^{-A-\phi/2}\omega_{\mu\nu}\eps^\dagger\eps+
e^{-2B+\phi/2}b\eps^\dagger\gamma_{\mu\nu}{\not\d}f_2\Gamma_v\eps).
\eea
To evaluate the above expressions it was useful to construct a combination of projectors:
\bea
\Gamma_\Omega P_\Omega\eps=-a\Gamma_w\eps,\quad
\Gamma_\Omega P_\Omega\eps^*=a\Gamma_w\eps^*.\nonumber
\eea
Notice that the right--hand side of (\ref{Vstar}) is real: it is obvious for the first term, while for the second one 
one needs to use projectors to evaluate
\bea
\eps^\dagger\gamma_{\mu\nu}{\not\d}f_2\Gamma_v\eps=e^A
\left(\d_\nu f_2e^{\bf v}_\mu-\d_\nu f_2e^{\bf v}_\mu\right).
\eea
Then we conclude that $V_{[\mu\nu]}=V^*_{[\mu\nu]}=U_{[\mu\nu]}$ and equation (\ref{AppAVec1}) 
becomes
\bea
-d(e^Ae^{\bf w}_\mu dx^\mu)
-\frac{1}{8}\left[6ce^{-\phi/2}\omega_{\mu\nu}-
2be^{A+\phi/2-2B}{\d}_\nu f_2 e^{\bf v}_\mu \right]dx^{\mu\nu}=0.
\eea
Let us look at various components of this two--form. We begin with
a coefficient in front of $du\wedge dv$:
\bea\label{VecEqnUV}
&&d(ae^Ae^{\bf w}_\mu dx^\mu)_{uv}
-\frac{1}{4}\left[6ce^{-\phi/2}\omega_{uv}+
be^{2A-\phi/2-2B}{\d}_u f_2 \right]=0.
\eea
Similarly we evaluate the two remaining components:
\bea\label{VecEqnUW}
&&d(ae^Ae^{\bf w}_\mu dx^\mu)_{uw}-\frac{3}{2}e^{-\phi/2}c\omega_{uw}=0,\\
\label{VecEqnVW}
&&d(ae^Ae^{\bf w}_\mu dx^\mu)_{vw}+
\left[-\frac{3c}{2}e^{-\phi/2}\omega_{vw}+\frac{1}{4}e^{2A-\phi/2-2B}b{\d}_w f_2\right]=0.
\eea
To summarize, in this subsection we analyzed the equations for the scalar and vector bilinears, this 
led to normalization of the Killing spinor (\ref{NormKill}) and to three differential equations 
(\ref{VecEqnUV})--(\ref{VecEqnVW}) for the bosonic fields. These equations along with relations discussed in subsection \ref{AppSubsProj} give a system which is equivalent to the equations for the Killing spinors, and now we will try to simplify this system.

\subsection{Solving the equations.}

In the previous two subsections we have reduced the dilatino and gravitino equations to a set of relations for the bosonic fields. The relevant differential equations are 
(\ref{AppEqnAu})--(\ref{AppEqnAf}), (\ref{Dltno1})--(\ref{Dltno4})
and (\ref{VecEqnUV})--(\ref{VecEqnVW}). Let us simplify this set of eleven equations. 

We begin with recalling the expression (\ref{SemiFrame1}) for $e^{\bf w}$ and the gauge 
condition $A_v=0$. Introducing $h=e^A e_w$, we can write
\bea
e^Ae^{\bf w}_\mu dx^\mu=h(dw+A_u du),\quad \omega=d(B_w dw+B_u du).
\eea
Here we parameterized the exact two--form $\omega$ in terms of two functions $B_w$, $B_u$ and one can still perform a $v$--independent gauge transformation of a one--form 
$\eta_1\equiv B_w dw+B_u du$. 

Equation (\ref{VecEqnUW}) takes the form
\bea
-\d_u h+\d_w(hA_u)-\frac{3c}{2}e^{-\phi/2}(\d_u B_w-\d_w B_u)=0.
\eea
Comparing this with equation (\ref{Dltno2}):
\bea
\d_u\phi-A_u\d_w \phi+ch^{-1}e^{-\phi/2}(\d_u B_w-\d_w B_u)=0,
\eea
we arrive at the relation which does not contain fluxes:
\bea\label{AppFllEqn1}
&&\d_u h-\d_w(hA_u)-\frac{3}{2}h(\d_u\phi-A_u\d_w \phi)=0:\nonumber\\
&&\d_u(he^{-3\phi/2})-\d_w(e^{-3\phi/2}hA_u)=0.
\eea
It is useful to eliminate fluxes from the equations (\ref{VecEqnUV}), (\ref{VecEqnVW}) as well.
To do so one can use (\ref{Dltno3}), (\ref{Dltno2}):
\bea
&&\frac{b}{2}e^{A-2B-\phi/2}\d_w f_2=-\left[e_w\d_v \phi+ce^{-\phi/2-A}\omega_{vw}\right],\\
&&be^{\phi/2-2B}(\d_u-A_u\d_w)f_2=2ce^{-\phi/2-A}e^{-A+\phi}(\omega_{uv}-A_u\omega_{wv}):
\nonumber\\
&&be^{2A-2B}\d_u f_2=2(c\omega_{uv}-A_ue_w e^{\phi/2+A}\d_v\phi).
\eea
Substituting this into the equation (\ref{VecEqnVW}), we find
\bea\label{AppFllEqn2}
&&d(ae^Ae^{\bf w}_\mu dx^\mu)_{vw}-2ce^{-\phi/2}\omega_{vw}-\frac{h}{2}\d_v\phi
=0\nonumber\\
&&-\d_v h-2ce^{-\phi/2}\omega_{vw}-\frac{h}{2}\d_v\phi=0\nonumber\\
&&\d_v(he^{\phi/2}+2cB_w)=0,
\eea
while equation (\ref{VecEqnUV}) gives
\bea\label{AppFllEqn3}
&&d(ae^Ae^{\bf w}_\mu dx^\mu)_{uv}+\frac{h}{2}A_u\d_{v}\phi
-2ce^{-\phi/2}\omega_{uv}=0\nonumber\\
&&\d_v (hA_u)+\frac{h}{2}A_u\d_v \phi+2ce^{-\phi/2}\d_v B_u=0\nonumber\\
&&\d_v (he^{\phi/2}A_u+2c B_u)=0.
\eea
The last equation implies that we can use the gauge transformation of $\eta_1$ to set
\bea
B_u=-\frac{c}{2}he^{\phi/2}A_u,
\eea
and equation (\ref{AppFllEqn2}) will still hold. We still have a freedom 
$\eta_1\rightarrow \eta_1+dW(w)$ and to fix it we rewrite the equation (\ref{Dltno2}):
\bea\label{AppFllEqn4}
&&\d_u\phi-A_u\d_w \phi+ch^{-1}e^{-\phi/2}(\d_u B_w-\d_w B_u)=0\nonumber\\
&&\d_u\phi+\frac{e^{3\phi/2}}{2h}\d_w (he^{-3\phi/2}A_u)+ch^{-1}e^{-\phi/2}\d_u B_w=0\nonumber\\
&&\d_u(e^{2\phi}he^{-3\phi/2}+2cB_w)=0.
\eea
At the last step we used the relation (\ref{AppFllEqn1}). Comparing (\ref{AppFllEqn2}) and 
(\ref{AppFllEqn4}), we observe that $e^{\phi/2}h+2cB_w$ can only depend on $w$, so the remaining gauge freedom can be fixed by requiring that
\bea
B_w=-\frac{c}{2}he^{\phi/2}.
\eea

At this point the equations (\ref{VecEqnUV})--(\ref{VecEqnVW}), (\ref{Dltno2}) were used to show that 
\bea\label{AppADblEqn}
\omega=-\frac{c}{2}d\left[he^{\phi/2}(dw+A_u du)\right],\quad 
\d_u(he^{-3\phi/2})-\d_w(e^{-3\phi/2}hA_u)=0.
\eea
The remaining dilatino equations (\ref{Dltno1}), (\ref{Dltno3}), (\ref{Dltno4}) can be rewritten as expressions for derivatives of the flux $f_2$:
\bea\label{FlEqnW}
&&\frac{b}{2}e^{2A-2B}\d_w f_2=-he^{\phi/2}\d_v\phi+\frac{1}{2}\d_v(he^{\phi/2})=\frac{e^{2\phi}}{2}
\d_v(he^{-3\phi/2}),\\
\label{FlEqnU}
&&be^{2A-2B}\d_u f_2=2(c\omega_{uv}-A_ue_w e^{\phi/2+A}\d_v\phi)=
e^{2\phi}\d_v(he^{-3\phi/2}A_u),\\
\label{FlEqnV}
&&be^{3\phi/2-2B-A}\d_v f_2=2h^{-1}e^A\d_w \phi.
\eea
This accounts for seven equations, and the remaining four equations
(\ref{AppEqnAu})--(\ref{AppEqnAf}) will be analyzed later. 

The second equation in (\ref{AppADblEqn}) implies local relations
\bea\label{DefineFunc}
he^{-3\phi/2}=\d_w F,\qquad e^{-3\phi/2}hA_u=\d_uF
\eea
with some function $F$. Then equations (\ref{FlEqnW}), (\ref{FlEqnU}) can be simplified and the resulting relations can be easily integrated:
\bea\label{FlSolut}
&&{b}\d_w f_2=v^2\d_v\d_w F,\quad {b}\d_u f_2=v^2\d_v\d_u F:\qquad bf_2=v^2 \d_v F.
\eea
We also used the fact that the relations (\ref{DefineFunc}) define $F$ up to an additive 
$v$--dependent function, and this freedom was fixed in the last equation. Finally (\ref{FlEqnV}) leads to a differential equation relating $F$ and the dilaton:
\bea\label{FlEqnDiff}
\d_w e^{-2\phi}=-\Delta_v F.
\eea

Let us now go back to the equations (\ref{AppEqnAu})--(\ref{AppEqnAf}). We begin with simplifying 
(\ref{AppEqnAv}):
\bea
&&\d_v(A-\frac{3}{4}\phi)-\frac{b}{4}e^{2A-2B-\phi/2}h^{-1}\d_w f_2=0\nonumber\\
&&\d_v(4A-3\phi)-v^{-2}e^{3\phi/2}h^{-1}\d_w (v^2\d_v F)=0\nonumber\\
&&\d_v(4A-3\phi)-e^{3\phi/2}h^{-1}\d_v (he^{-3\phi/2})=0\nonumber\\
&&\d_v(e^{-4A+3\phi/2}h)=0:\qquad h=e^{4A-3\phi/2}{\tilde h}(u,w).
\eea 
Let us recall that $h$ originally appeared in the veilbein as
\bea
e^{\bf w}=he^{-A}(dw+A_u du)=e^{3A-3\phi/2}\left[{\tilde h}(u,w)(dw+A_u du)\right].
\eea
We still have a freedom in reparameterizing $w$ as $w\rightarrow w'(w,u)$, and it can be used to simplify the expression in the square bracket by setting ${\tilde h}=1$. This fixes the choice of coordinates up to a shift $w\rightarrow w+W(u)$.  We conclude that equation (\ref{AppEqnAv}) 
combined with gauge fixing leads to the relation
\bea
h=e^{4A-3\phi/2}.
\eea
Next we simplify the equation (\ref{AppEqnAu}):
\bea\label{AppTrash1}
&&({\d}_{u}-A_u\d_w)(A-\frac{3}{4}\phi)+ae^{-4C+\phi/2}e^{-4A+3\phi/2}{\d}_wf_3=0\nonumber\\
&&{\d}_{u}(4A-3\phi)+\d_u F\d_w e^{3\phi-4A}+4au^{-4}{\d}_wf_3=0\nonumber\\
&&{\d}_{u}(4A-3\phi)-e^{3\phi-4A}\d_u e^{4A-3\phi}+{\d}_w(4au^{-4}f_3+e^{3\phi-4A}\d_u F)=0\nonumber\\
&&{\d}_w(4au^{-4}f_3+e^{3\phi-4A}\d_u F)=0.
\eea
Similar manipulations with (\ref{AppEqnAv}) give
\bea\label{AppTrash2}
&&e^{-4C}{\d}_{v}f_3+\frac{a}{4}v^2e^{2A-2B-\phi}({\d}_{u}-A_u\d_w)\d_v F=0\nonumber\\
&&e^{-4C}{\d}_{v}f_3+\frac{a}{4}e^{\phi}({\d}_{v}(e^{-3\phi/2}hA_u)-A_u\d_v(he^{-3\phi/2}))=0\nonumber\\
&&e^{-4C}{\d}_{v}f_3+\frac{a}{4}e^{4A-2\phi}{\d}_{v}A_u=0:\qquad
\d_v(4au^{-4}f_3+e^{3\phi-4A}\d_u F)=0.
\eea
Equations (\ref{AppTrash1}), (\ref{AppTrash2}) imply that 
\bea
f_3=-\frac{a}{4}u^4 e^{3\phi-4A}\d_u F+{\tilde f}_3(u)=-\frac{a}{4}u^4 A_u+{\tilde f}_3(u).
\eea
This relation can be used to express for the one--form $e^{\bf w}$ in terms of $f_3$:
\bea
e^{\bf w}=e^{3A-3\phi/2}(dw-4au^{-4}f_3 du+4au^{-4}{\tilde f}_3(u) du).
\eea
We recall that at this point the diffeomorphism invariance is fixed up to a shift $w\rightarrow w+W(u)$,
and this remaining transformation can be used to set ${\tilde f}_3(u)=0$. 

To summarize, we have solved all equations for the Killing spinor except for (\ref{AppEqnAw}) and we
also uniquely specified the choice of coordinates and veilbeins. This led to the following relations for the metric and the fluxes:
\bea
&&e^{\bf u}=e^{-A+\phi/2}du,\quad e^{\bf v}=e^{A-\phi}dv,\quad
e^{\bf w}=e^{3A-3\phi/2}(dw+A_u du),\\
&&e^{4A-3\phi}=\d_w F,\quad A_u=\frac{\d_u F}{\d_w F},\nonumber\\
&&\omega=-\frac{c}{2}d\left[e^{4A-\phi}(dw+A_u du)\right],\quad
f_2=bv^2 \d_v F,\quad 
f_3=-\frac{a}{4}u^4 A_u.
\eea
and to a differential equation (\ref{FlEqnDiff}) relating $F$ and the dilaton. 

Finally we simplify the equation (\ref{AppEqnAw}):
\bea\label{LastDiffEqn}
&&e^{-3A+3\phi/2}{\d}_{w}(A-\frac{3}{4}\phi)-ae^{A-4C-\phi/2}({\d}_{u}-A_u\d_w)f_3+
\frac{b}{4}e^{3\phi/2-2B-A}{\d}_{v}f_2=0\nonumber\\
&&{\d}_{w}(4A-3\phi)+e^{4A-4C-2\phi}({\d}_{u}-A_u\d_w)(u^4 A_u)+
e^{2\phi}\Delta_{v}F=0\nonumber\\
&&u^4 e^{-8A+4\phi}{\d}_{w}(4A-\phi)+({\d}_{u}-A_u\d_w)(u^4 A_u)=0\nonumber\\
&&-u^4 e^{-4A+3\phi}\d_w e^{-4A+\phi}+(\d_u-A_u\d_w)(u^4 A_u)=0.
\eea
The last equation can be simplified even further if we use $(u,v,F)$ rather than $(u,v,w)$ 
as a set of independent variables. This set of variables turns out to be useful for analyzing the 
regularity of the solution,
so we give the map here. First, it is convenient to introduce a new function $H$:
\bea
e^{2H}\equiv e^{4A-3\phi}=\d_w F.
\eea
Then we can relate various derivatives:
\bea
\d_u|_{F,v}=\d_u|_{w,v}-A_u\d_w|_{u,v},\quad \d_w|_{u,v}=e^{2H}\d_F|_{u,v},\quad
\d_v|_{F,u}=\d_v|_{w,u}+\frac{\d_v w}{\d_F w}\d_F|_{v,u}.\nonumber
\eea
Going back to the equation (\ref{LastDiffEqn}), we can rewrite it in terms of $(u,v,F)$ coordinates:
\bea
-u^4\d_F e^{-2H-2\phi}+\d_u(-u^4 \d_u w)|_{v,F}=0,
\eea
and one more derivative eliminates $w$ from this equation:
\bea
u^4\d^2_F e^{-2H-2\phi}+\d_u(u^4 \d_u e^{-2H})=0.
\eea
At this point we have solved all equations for the Killing spinors. We showed that the solution is completely parameterized by two functions $F$, $e^\phi$ and we found two equations 
(\ref{FlEqnDiff}), (\ref{LastDiffEqn}) which relate them. Unfortunately these two equations are not sufficient for 
finding the complete solution, and they should be supplemented by an equation of motion 
for the Kalb--Ramond field. We will analyze that equation in the next subsection.


\subsection{Equation of motion for $B_{\mu\nu}$.}
\label{AppSbEOM}


As mentioned in section \ref{SectSpike}, even to construct a geometry describing a 
fundamental string, one needs 
to supplement the equations for Killing spinors by the equation for the NS--NS B 
field. In the present context, this equation becomes
\bea
d*(e^{-\phi}H_3)=-4F_5\wedge F_3.
\eea 
Substituting the information which has been accumulated so far, we arrive at the relation
\bea\label{May29Eqn1}
d\left[e^{-\phi}v^2 u^4 e^{-3H/2}e^{-9\phi/4}(*_3d\left[e^{2\phi}(dF+e^{2H}\d_v w dv)\right]\right]=
abc~
d(u^4A_u)\wedge d(v^2\d_v F).
\eea
Here the Hodge dual is taken with respect to a three--dimensional metric:
\bea
ds_3^2=e^{H-\phi/2}dv^2+e^{-H-\phi/2}du^2+e^{-H+3\phi/2}(dF+e^{2H}\d_v w dv)^2.
\eea
It appears that we need only $u,v$ component of the equation (\ref{May29Eqn1}). 
We begin with simplifying the left hand side :
\bea
&&d\left[v^2 u^4 e^{-3H/2}e^{-13\phi/4}(*_3d\left[e^{2\phi}(dF+e^{2H}\d_v w dv)\right]\right]_{uv}
\nonumber\\
&&=
d\left[v^2 u^4 e^{-H}e^{-4\phi}*_2de^{2\phi}\right]_{uv}
-d\left[v^2 u^4 e^{-2H}e^{-4\phi}\d_F e^{2\phi}e^{2H}\d_v w du\right]_{uv}\nonumber\\
&&\qquad-d\left[v^2 u^4 e^{-3H/2}e^{-5\phi/4}\d_F(e^{2H}\d_v w)e^{H/2-3\phi/4}*_2 dv \right]_{uv}
\nonumber\\
&&\qquad+
d\left[v^2 u^4 e^{-3H/2}e^{-5\phi/4}\d_u(e^{2H}\d_v w)(*_3d\left[du dv\right])\right]_{uv}\nonumber\\
&&=d\left[v^2 u^4 e^{-H}e^{-4\phi}*_2de^{2\phi}\right]_{uv}+
\d_v\left[v^2 u^4 e^{-2H}e^{-2\phi}\d_F(e^{2H}\d_v w)\right]
\nonumber\\
&&\qquad+
\d_u\left[v^2 u^4 \d_u(e^{2H}\d_v w)\d_v w\right]-\d_v(v^2 u^4 \d_Fe^{-2\phi}\d_v w)\nonumber\\
&&=\d_u\left[v^2 u^4(\d_u e^{-2\phi}+\d_u(e^{2H}\d_v w)\d_v w)\right]-\d_v(v^2 u^4 \d_Fe^{-2\phi}\d_v w)
\nonumber\\
&&\qquad-\d_v\left[v^2 u^4e^{-2H}(-\d_v e^{-2\phi}-e^{-2\phi}\d_F(e^{2H}\d_v w))\right]\nonumber\\
&&=v^2\d_u\left[u^4\d_u e^{-2\phi}\right]+\d_v\left[v^2 u^4\d_ve^{-2\phi-2H}\right]+
\d_u\left[v^2u^4\d_u(e^{2H}\d_v w)\d_v w\right]\nonumber\\
&&\qquad+\d_v\left[v^2 u^4e^{-2H-2\phi}\d_Fe^{2H}\d_v w\right]-\d_v(v^2 u^4 \d_Fe^{-2\phi}\d_v w)
\nonumber\\
&&=v^2\d_u\left[u^4\d_u e^{-2\phi}\right]+\d_v\left[v^2 u^4\d_ve^{-2\phi-2H}\right]+
\d_u\left[v^2u^4\d_u(e^{2H}\d_v w)\d_v w\right]\nonumber\\
&&\qquad+\d_v\left[v^2 u^4 e^{2H}\Delta_u w\d_v w\right].
\eea
We used the duality convention $^*_2 du=-e^H dv$ as well as relation
\bea
&&d\left[v^2 u^4 e^{-3H/2}e^{-13\phi/4}(*_3\left[\d_F e^{2\phi}
dF\wedge (dF+e^{2H}\d_v w dv)\right]\right]_{uv}\nonumber\\
&&\qquad=-d\left[v^2 u^4 e^{-2H}e^{-4\phi}\d_F e^{2\phi}e^{2H}\d_v w du\right]_{uv}=
-\d_v(v^2 u^4 \d_Fe^{-2\phi}\d_v w).\nonumber
\eea 
The right hand side of (\ref{May29Eqn1}) can also be simplified (we use (\ref{10DChiral}) to eliminate $abc$):
\bea
&&abc[d(-u^4\d_u w)\wedge d(-v^2e^{2H}\d_v w)]_{uv}=-u^4 v^2
[\Delta_u w{\tilde\Delta}_v w-\d_u\d_v w\d_u(e^{2H}\d_v w)]\nonumber\\
&&=-\left\{
\d_u\left[v^2u^4\d_u w{\tilde\Delta}_v w\right]-\d_v\left[v^2 u^4\d_u w\d_u(e^{2H}\d_v w)\right]\right\}
\nonumber\\
&&=-\left\{
\d_v\left[v^2u^4e^{2H}\Delta_u w\d_v w\right]-\d_u\left[v^2 u^4\d_u\d_v w e^{2H}\d_v w\right]\right\}
\nonumber\\
&&=-\left\{
-\d_v\left[v^2u^4e^{2H}\d_F(e^{-2\phi}\d_F w)\d_v w\right]-\d_u\left[v^2 u^4\d_u\d_v w e^{2H}\d_v w\right]\right\}.
\nonumber
\eea
This leads to the final form of (\ref{May29Eqn1}):
\bea\label{FieldEqnApp}
0&=&v^2\d_u\left[u^4\d_u e^{-2\phi}\right]+\d_v\left[v^2 u^4\d_ve^{-2\phi-2H}\right]+
\d_u\left[v^2u^4\d_u(e^{2H}\d_v w)\d_v w\right]\nonumber\\
&&\qquad+\d_v\left[v^2 u^4 e^{2H}\Delta_u w\d_v w\right]\nonumber\\
&&-\left\{
\d_v\left[v^2u^4e^{2H}\Delta_u w\d_v w\right]-\d_u\left[v^2 u^4\d_u\d_v w e^{2H}\d_v w\right]\right\}
\nonumber\\
&=&v^2\d_u\left[u^4\d_u e^{-2\phi}\right]+\d_v\left[v^2 u^4\d_ve^{-2\phi-2H}\right]+
u^4 v^4\Delta_u(e^{2H}\d_v w\d_v w).
\eea
Notice that while this equation does not follow from the relations which we have extracted from Killing spinor, it is consistent with those relations. For example, acting on (\ref{FieldEqnApp}) by 
$\d_F$ and using the relation
\bea
&&v^2\d_u\left[u^4\d_u (e^{-2H}{\tilde \Delta}_v w-\d_v w\d_F(e^{2H}\d_v w))\right]
-\d_v\left[v^2 u^4\d_v\Delta_u w\right]
\nonumber\\
&&=v^2\d_u u^4\d_u \left[-e^{2H}\d_F \d_v w \d_v e^{2H}-\d_v w\d_F(e^{2H}\d_v w)\right]=
-u^4 v^2\d_F\Delta_u(e^{2H}\d_v w\d_v w),\nonumber
\eea
we arrive at an identity. 

As already mentioned, it is only $(u,v)$ component of (\ref{May29Eqn1}) which gives a new relation. For example, looking at $(v,F)$ component of that equation and evaluating lhs and rhs, we find:
\bea
&&d\left[v^2 u^4 e^{-3H/2}e^{-13\phi/4}(*_3d\left[e^{2\phi}(dF+e^{2H}\d_v w dv)\right]\right]_{vF}
=2u^4\left[\d_u\d_{[v}w \d_{F]}(e^{2H}v^2\d_v w)\right]\nonumber\\
&&abc[d(-u^4\d_u w)\wedge d(-v^2e^{2H}\d_v w)]_{vF}=
2[\d_{[v}(u^4\d_u w)\d_{F]}(v^2e^{2H}\d_v w)],
\nonumber
\eea
so the $(v,F)$ component of (\ref{May29Eqn1}) becomes an identity. 
The $(u,F)$ component works in the same way.

\subsection{Summary of the solution}

In this appendix we have looked at solutions of type IIB supergravity with $SO(3)\times SO(5)$ 
rotational symmetry. We solved the equations for the Killing spinors and found the following expressions for the metric and fluxes:
\bea
ds^2=e^H\left[-e^{3\phi/2}dt^2+e^{-\phi/2}(dv^2+v^2 d\Omega_2^2)\right]+
e^{-H-\phi/2}(du^2+u^2 d\Omega^2_4)+e^{3H+3\phi/2}(dw+{\cal A})^2\nonumber
\eea
\bea
&&{\cal A}=A_u du=\frac{\d_u F}{\d_w F},\quad e^{2H}=\d_w F,\quad
F_5=-\frac{a}{4}d(u^4A_u)\wedge d\Omega_4+dual,\nonumber\\
&&H_3=-c~d\left[e^{2H+2\phi}(dw+{\cal A})\right]dt,\quad 
F_3=d(bv^2\d_v F)\wedge d\Omega_2.
\eea
The geometries are parameterized by two functions $F$, $e^\phi$ and these two functions obey differential equations:
\bea
&&\d_w e^{-2\phi}+\Delta_v F=0,\\
&&u^4 e^{-2H}\d_w e^{-2H-2\phi}-(\d_u-A_u\d_w)(u^4 A_u)=0.
\eea
The last equation can also be rewritten in terms of the coordinates $(u,v,F)$:
\bea\label{ScwdDifEqn}
u^4\d_F e^{-2H-2\phi}+\d_u(u^4 \d_u w)|_{v,F}=0.
\eea
It turns out that the equations for the Killing spinors are not sufficient to determine $F$ and dilaton completely and we also had to look at the equations of motion for the Kalb--Ramond two--form. This supplied an extra equation (\ref{FieldEqnApp}):
\bea
v^2\d_u\left[u^4\d_u e^{-2\phi}\right]+\d_v\left[v^2 u^4\d_ve^{-2\phi-2H}\right]+
u^4 v^4\Delta_u(e^{2H}\d_v w\d_v w)=0.
\eea
The Killing spinor $\eps$ satisfies five independent projections (\ref{AppAproj})
and chirality condition translates into the relation (\ref{10DChiral}).

Suppose we started with solutions which has $a=-1$. The flipping the orientation of $S^4$ and reversing the direction of time (this procedure keeps the relation $\Gamma_{11}\eps=-\eps$ 
untouched), we arrive at the solution with $a=1$. Similarly, starting with $b=-1$, we can recover solution with
$b=1$ by reversing orientation of $S^2$ and $t$. Thus without loss of generality, we can set
\bea
a=b=1,\qquad c=-1.
\eea
These conventions are used in the main part of the paper.


\section{Generalization: geometries without $S^2$}

\label{AppGenSln}

\renewcommand{\theequation}{C.\arabic{equation}}
\setcounter{equation}{0}


In the Appendix \ref{AppSpike} we {\it derived} supersymmetric solutions of type IIB supergravity assuming $U(1)\times SO(3)\times SO(5)$ symmetry. Once that solution is constructed, it is fairly 
easy to make a guess and generalize it to the case without non--abelian isometries. 
Of course, in this case we do not claim to construct the most general solution, but rather we 
use analogy to 
make a guess and then check that the geometry indeed preserved $1/4$ of supersymmetries. In this section we outline this procedure for the solutions which have $SO(5)\times U(1)$ isometries, and the more general case works in the same way. Notice that keeping $SO(5)$ seems natural if we want to consider a superposition of D3 spikes, since in this case the branes are located at $x_5=x_6=x_7=x_8=x_9=0$. More general solutions without $SO(5)$ are discussed in section 
\ref{SecGnrSln} and they could correspond to either separate stacks of D3 branes or to spikes on D5 branes.

\subsection{Analysis of the projectors}

Motivated by the solutions with $SO(3)\times SO(5)\times U(1)$ symmetry, we require the metric and the fluxes to take the form:
\bea\label{MetrGssAppB}
ds^2&=&e^H\left[-e^{3\phi/2}dt^2+e^{-\phi/2}d{\bf x}_3^2\right]+e^{-H-\phi/2}\left[(du^2+u^2 d\Omega^2_4)+
e^{2\phi+4H}(dw+A_u du)^2\right]\nonumber\\
&&H_3=-\frac{c}{2}d\left[e^{2\phi+2H}(dw+A_u du)\right]dt,\quad
F_5=-\frac{a}{4}d(u^4 A_u)\wedge d\Omega_4+dual,\nonumber\\
&&e^{2H}=\d_w F,\quad A_u=\frac{\d_u F}{\d_w F},\quad f_3=-\frac{a}{4}u^4 A_u.
\eea
The flux $F_3$ is still undetermined.  To check this guess and to find 
$F_3$, we go back to the original set of equations for the Killing spinor\footnote{To 
connect with discussion in the Appendix \ref{AppSpike}, we defined the warp factors 
$e^{2A}=e^{H+3\phi/2}$ and $e^{2C}=u^2e^{-H-\phi/2}$.}:
\bea
&&\frac{1}{2}{\not\d}\phi\eps^*-(G_++G_-)\eps=0,\qquad 
\frac{1}{2}{\not\d}\phi\eps-(G_+-G_-)\eps^*=0,\nonumber\\
&&{\not\d}A\eps-ie^{-4C}{\not\d}f_3\Gamma_{\Omega}\eps+\frac{1}{2}
(-3G_++G_-)\eps^*=0,\nonumber\\
\label{App2eqnA}
&&(-ie^{-C}P_\Omega+{\not\d}C)\eps+ie^{-4C}{\not\d}f_3\Gamma_{\Omega}\eps+
\frac{1}{2}(G_++G_-)\eps^*=0,\\
&&\nabla_\mu\eps+i\frac{e^{-4C}}{2}{\not\d}f_3\gamma_\mu\Gamma_\Omega\eps+
\frac{1}{96}(\gamma_\mu
{\not G}-2\{{\not G},\gamma_\mu\})\eps^*=0.\nonumber
\eea
As before, we find one geometric projector:
\bea
&&\left[-ie^{-C}P_\Omega+{\not\d}(A+C-\frac{\phi}{2})\right]\eps=0,
\eea
which is consistent with ansatz provided that 
\bea\label{AppBExtraProj}
\left[-iP_\Omega+\Gamma_u\right]\eps=0.
\eea
This is one of the projectors (\ref{AppAproj}) which easily generalizes to the present setup. Three other independent projectors can be generalized as well:
\bea\label{AppBProj}
&&\Gamma_u\Gamma_w\Gamma_\Omega\eps=ia\eps,\quad 
-i\Gamma_w\Gamma_{123}\eps^*=b\eps,\quad \Gamma_{11}\eps=-\eps.
\eea
Notice that in (\ref{AppAproj}) we also had a projector containing $P_S$ and it is lost in the present setup since we do not assume an existence of $S^2$. Similarly, had we not assumed the $SO(5)$ symmetry, the projector (\ref{AppBExtraProj}) would disappear and the first relation in 
(\ref{AppBProj}) would become $-\Gamma_w\Gamma_{45678}\eps=ia\eps$. This would lead to complex $32$--component spinor restricted by three projectors, i.e. as expected, we would preserve $1/4$ of SUSY.  

Let us now go back to the solutions with $SO(5)$ symmetry. They are also $1/4$-supersymmetric due 
to the projections (\ref{AppBExtraProj}), (\ref{AppBProj}). It is useful to take a combination of 
(\ref{AppBProj}) to produce another projector
\bea\label{TimeProj}
\Gamma_w\Gamma_{t}\eps^*=c\eps,\quad c=-ab.
\eea
We already checked one linear combination of the projectors appearing in 
(\ref{App2eqnA}) and now we look at a different combination which does not contain $G_+$:
\bea\label{GenEqnA0}
&&{\not\d}(A-\frac{3}{4}\phi)\eps-ie^{-4C}{\not\d}f_3\Gamma_{\Omega}\eps-
G_-\eps^*=0.
\eea
Acting on this relation by $(1-i\Gamma_u P_\Omega)$, we find
\bea\label{GenEqnA}
2\Gamma_u\left[{\d}_{\bf u}(A-\frac{3}{4}\phi)+ae^{-4C}{\d}_{\bf w}f_3\right]\eps-
2ie^{-4C}\Gamma_{\bf i}{\d}_{\bf i}f_3\Gamma_{\Omega}\eps-
(1-i\Gamma_u P_\Omega)G_-\eps^*=0.
\eea
Notice that our ansatz implies that the expression is square brackets vanishes:
\bea
&&{\d}_{\bf u}(A-\frac{3}{4}\phi)+ae^{-4C}{\d}_{\bf w}f_3=
\frac{1}{2}e^{H/2+\phi/4}(\d_u-A_u\d_w) H+au^{-4}e^{H/2+\phi/4}\d_w f_3\nonumber\\
&&\qquad=\frac{e^{\phi/4-3H/2}}{4}\left[(\d_u-A_u\d_w) e^{2H}-e^{2H}\d_w A_u\right]=
\frac{e^{\phi/4-3H/2}}{4}\left[\d_u e^{2H}-\d_w \d_u F\right]=0,\nonumber
\eea
then equation (\ref{GenEqnA}) can be simplified further:
\bea\label{GenEqnB}
&&-ae^{-4C}\Gamma_{uw}\Gamma_{\bf i}{\d}_{\bf i}f_3\eps-
(G_-)_{\bf u}\Gamma_u\eps^*=0\nonumber\\
&&-ae^{-4C}\Gamma_{\bf i}{\d}_{\bf i}f_3\eps-
bi(G_-)_{\bf u}\Gamma_{123}\eps=0.
\eea
Acting on this relation by the $(\Gamma_u\Gamma_w\Gamma_\Omega-ia)$, we derive a restriction on $G_-$:
\bea
(G_-)_{\bf iwu}\Gamma_{i}\Gamma_{123}\eps=0:\quad (G_-)_{\bf iwu}=0.
\eea
This implies that one can choose a gauge
\bea\label{GMinus}
&&G_-=\frac{i}{2}e^{\phi/2}{d}\left[\eps_{ijk}h_k dx^{ij}\right],
\eea
and various components of $G_-$ become
\bea\label{GMinusComp}
({\not G}_-)_{\bf u}=3ie^{-H/2+5\phi/4}\eps_{ijk}(\d_u-A_u \d_w) h_k \Gamma^{ij},
\quad (G_-)_{\bf 123}=ie^{5\phi/4-3H/2}\d_i h_i.
\eea
Plugging this into the equation (\ref{GenEqnB}), we find and equation for $h_i$:
\bea
\left[au^{-4}e^{2H}{\d}_{i}f_3+6b(\d_u-A_u \d_w) h_i \right]\Gamma_i\eps=0:
\quad 6b(\d_u-A_u \d_w) h_i=\frac{1}{4}e^{2H}\d_i A_u.
\eea
Let us now look at the remaining piece in (\ref{GenEqnA0}):
\bea
&&(1+i\Gamma_u P_\Omega)\left[{\not\d}(A-\frac{3}{4}\phi)\eps-ie^{-4C}{\not\d}f_3\Gamma_{\Omega}\eps-
G_-\eps^*\right]=0.
\eea
Multiplying this by $(1-ia\Gamma_{uw}\Gamma_\Omega)$ and making simplifications, we find
\bea
&&\left[\Gamma_x{\d}_{\bf x}(A-\frac{3}{4}\phi)\eps-(G_-)_{\bf w}\Gamma_w\eps^*\right]=0\nonumber\\
&&\left[\frac{1}{2}\Gamma_x{\d}_{\bf x}H-ib(G_-)_{\bf w}\Gamma_{123}\right]\eps=0\nonumber\\
&&\left[\frac{1}{2}e^{-H/2+\phi/4}\Gamma_i{\d}_iH+
3be^{H/2-\phi/4-2H}e^{-H+\phi/2}\eps_{ijk}\d_w h_k \Gamma^{ij}\Gamma_{123}\right]\eps=0\nonumber\\
&&\left[\frac{1}{2}e^{2H}\Gamma_i{\d}_iH-
6b\d_w h_k \Gamma_{k}\right]\eps=0:\qquad 24b\d_w h_k=\d_k e^{2H}.
\eea
To summarize, at this point we have two equations for $h_i$:
\bea
24b\d_w h_k=\d_k e^{2H},\quad 24b(\d_u-A_u \d_w) h_i=e^{2H}\d_i A_u,
\eea 
and they can be combined to produce a relation for $\d_u h_i$:
\bea
24b\d_u h_i=\d_i(e^{2H} A_u)=\d_i\d_u F.
\eea
In this form the equations for $h_i$ can be easily integrated and we find a solution up to $x$--dependent functions:
\bea
24bh_i=\d_i F+24b{\tilde h}_i(x).\nonumber
\eea
Notice that due to the definition (\ref{GMinus}) the functions ${\tilde h}_i(x)$ appear in the flux only through 
${\tilde g}=\d_i{\tilde h}_i$. Suppose that ${\tilde g}\ne 0$, then making a shift 
$F\rightarrow F-24b\Delta_{\bf x}^{-1}{\tilde g}$ we absorb ${\tilde g}$ in $F$, so without loss of generality we can set ${\tilde g}=0$ and ${\tilde h}_i=0$:
\bea\label{ExprHi}
24bh_i=\d_i F.
\eea

We have analyzed various projections of the equation (\ref{GenEqnA0}) to derive 
(\ref{GMinus}) and 
(\ref{ExprHi}), now we use these relations to simplify (\ref{GenEqnA0}):
\bea
&&\left[\Gamma_w{\d}_{\bf w}(A-\frac{3}{4}\phi)\eps-
iu^{-4}e^{2H+\phi}\Gamma_u{\d}_{\bf u}f_3\Gamma_{\Omega}\eps-
6(G_-)_{\bf 123}\Gamma_{123}\eps^*\right]=0\nonumber\\
&&\left[\frac{1}{2}e^{-3\phi/4-3H/2}\Gamma_w{\d}_w H\eps-
iu^{-4}e^{5H/2+5\phi/4}\Gamma_u({\d}_{u}-A_u\d_w)f_3\Gamma_{\Omega}\eps-
6ib(G_-)_{\bf 123}\Gamma_{w}\eps\right]=0\nonumber\\
&&\left[\frac{1}{2}{\d}_w H-
au^{-4}e^{4H+2\phi}({\d}_{u}-A_u\d_w)f_3+
6be^{2\phi}\d_i h_i\right]=0.\nonumber
\eea
Here we used the projector $\Gamma_\Omega\eps=-ia\Gamma_u\Gamma_w\eps$ as well as the 
expression (\ref{GMinusComp}) for $(G_-)_{\bf 123}$. Writing $h_i$ in terms of $F$ and 
$f_3$ in terms of $A_u$, we arrive at the equation
\bea\label{ApGnrSemiLast}
2\d_w H+u^{-4}e^{4H+2\phi}({\d}_{u}-A_u\d_w)(u^4A_u)+e^{2\phi}\d_i \d_i F=0,
\eea
which is equivalent to (\ref{GenEqnA0}). 

At this point we have confirmed two out of three projectors appearing in (\ref{App2eqnA}) and now we look at the dilatino projector:
\bea
\frac{1}{2}{\not\d}\phi\eps+\frac{1}{4}e^{-\phi/2-A}{\not\omega}_2\Gamma_t\eps^*+G_-\eps^*=0.
\eea
Acting by various projectors, we find:
\bea
(1-i\Gamma_uP_\Omega)(1-ia\Gamma_{uw}\Gamma_\Omega):
&&e^{-\phi/2-A}\Gamma^{uk}{\omega}_{\bf uk}\Gamma_t\eps^*+
2(G_-)_{\bf u}\Gamma_u\eps^*=0,\nonumber\\
(1-i\Gamma_uP_\Omega)(1+ia\Gamma_{uw}\Gamma_\Omega):
&&\Gamma^u\d_{\bf u}\phi\eps+e^{-\phi/2-A}\Gamma^{uw}{\omega}_{\bf uw}\Gamma_t\eps^*=0,
\nonumber\\
(1+i\Gamma_uP_\Omega)(1-ia\Gamma_{uw}\Gamma_\Omega):
&&\Gamma^k\d_{\bf k}\phi\eps+e^{-\phi/2-A}\Gamma^{kw}{\omega}_{\bf kw}\Gamma_t\eps^*+
2(G_-)_{\bf w}\Gamma_{w}\eps^*=0,\nonumber\\
(1+i\Gamma_uP_\Omega)(1+ia\Gamma_{uw}\Gamma_\Omega):
&&\Gamma^w\d_{\bf w}\phi\eps+12(G_-)_{\bf 123}\Gamma_{123}\eps^*=0.\nonumber
\eea
Although we already have a reasonable guess for $\omega$ (which we have not used so far!), 
it might be useful to relax it by setting
\bea
H_3=2\omega\wedge dt
\eea
and to {\it derive} the expression appearing in (\ref{MetrGssAppB}). 

Using relation $\eps^*=ib\Gamma_w\Gamma_{123}\eps$, we eliminate $\eps^*$ from projectors:
\bea\label{Tmp14a}
&&\left[-e^{-5\phi/4-H/2}\Gamma^{k}{\omega}_{\bf uk}\Gamma_t+
2(G_-)_{\bf u}\right]\eps=0,\\
\label{Tmp14b}
&&\left[\d_{\bf u}\phi\eps-ibe^{-5\phi/4-H/2}{\omega}_{\bf uw}\Gamma_t\Gamma_{123}\right]\eps=0,
\\
\label{Tmp14c}
&&\left[-ib\Gamma_{321}\Gamma^k\d_{\bf k}\phi\eps+e^{-5\phi/4-H/2}\Gamma^{k}{\omega}_{\bf kw}\Gamma_t+
2(G_-)_{\bf w}\right]\eps=0,\\
\label{Tmp14d}
&&\left[\d_{\bf w}\phi+12ib(G_-)_{\bf 123}\right]\eps=0.
\eea
Substituting the value of $(G_-)_{\bf 123}$ from (\ref{GMinusComp}), we can simplify the last equation:
\bea\label{Tmp13d}
12b\d_i h_i=e^{-2\phi}\d_{w}\phi.
\eea
Recalling the projection $\Gamma_t\Gamma_{123}\eps=ibc\eps$, we rewrite the equation 
(\ref{Tmp14b}) as an expression for $\omega_{uw}$:
\bea\label{Tmp13b}
\omega_{uw}=-ce^{2\phi+2H}(\d_{u}-A_u\d_w)\phi.
\eea
The same projection can be used to eliminate $\Gamma_t$ from (\ref{Tmp14a}) and (\ref{Tmp14c}).
To analyze these two equations we use the expression (\ref{GMinus}) to evaluate
\bea
({G}_-)_{\bf u}\Gamma_{123}=-6ie^{-H/2+5\phi/4}(\d_u-A_u \d_w) h_k \Gamma^{k},\quad
({G}_-)_{\bf w}\Gamma_{123}=-6ie^{-5H/2+\phi/4}\d_w h_k \Gamma^{k}.\nonumber
\eea
Simplifications in the equation (\ref{Tmp14c}) give
\bea\label{Tmp13c}
&&\left[c\d_{\bf k}\phi\eps+e^{-5\phi/4-H/2}{\omega}_{\bf kw}+
12bce^{-5H/2+\phi/4}\d_w h_k\right]\Gamma_k\eps=0\nonumber\\
&&c\d_{k}\phi+e^{-2\phi-2H}{\omega}_{kw}+
12bce^{-2H}\d_w h_k=0\nonumber\\
&&\qquad \qquad {\omega}_{kw}=-\frac{c}{2}\d_k e^{2H+2\phi}.
\eea
To arrive at the second equation one should notice that $A_k\Gamma_k\eps=0$ implies $A_k=0$ if all $A_k$ are real. 

Finally we simplify equation (\ref{Tmp14a}):
\bea\label{Tmp13a}
&&\left[-e^{-5\phi/4-H/2}{\omega}_{\bf uk}+
12bce^{-H/2+5\phi/4}(\d_u-A_u\d_w) h_k \right]\Gamma^{k}\eps=0\nonumber\\
&&-e^{-2\phi}({\omega}_{uk}-A_u\omega_{wk})+
12bc(\d_u-A_u\d_w) h_k=0\nonumber\\
&&-e^{-2\phi}{\omega}_{uk}+12bc\d_u h_k+cA_ue^{2H}\d_k\phi=0\nonumber\\
&&\qquad \qquad\quad {\omega}_{uk}=\frac{c}{2}\d_k(e^{2\phi}\d_u F).
\eea
Notice that equations (\ref{Tmp13b})--(\ref{Tmp13a}) imply that 
\bea
\omega=-\frac{c}{2}d(e^{2\phi+2H} dw+e^{2\phi}\d_u F du)+{\tilde\omega},
\eea
where two--form ${\tilde\omega}$ can only have legs in the directions $x^i$ and it has no 
$u$-- or $w$--dependence (from now on we 
will {\it assume} that such contribution is absent). This clearly reproduces (\ref{Tmp13c}) and (\ref{Tmp13a}), to check 
(\ref{Tmp13b}) we compute
\bea
&&\left[-\frac{c}{2}d(e^{2\phi+2H} dw+e^{2\phi}\d_u F du)\right]_{uw}=-\frac{c}{2}\left\{
\d_u (e^{2\phi}\d_w F)-\d_w(e^{2\phi}\d_u F)\right\}\nonumber\\
&&\qquad=-ce^{2\phi}\left(\d_w F\d_u-\d_u F\d_w\right)\phi=
-ce^{2\phi+2H}\left(\d_u-A_u\d_w\right)\phi.
\eea
This confirms the expression for $\omega$ which can also be rewritten as
\bea\label{Tmp13Om}
\omega=-\frac{c}{2}d(e^{2\phi}\d_w F dw+e^{2\phi}\d_u F du).
\eea

To summarize, we analyzed the projectors appearing in (\ref{App2eqnA}) and showed that 
they reduce to 
(\ref{GMinus}), (\ref{ExprHi}), (\ref{ApGnrSemiLast}), (\ref{Tmp13d}), (\ref{Tmp13Om}). 
Three of these relations give expressions for the fluxes, while (\ref{ApGnrSemiLast}) and 
(\ref{Tmp13d}) lead to differential equations which should be satisfied by $F$ and $e^\phi$:
\bea\label{DifEqn14a}
&&\d_i\d_i F=-\d_w e^{-2\phi},\\
\label{DifEqn14b}
&&u^4 e^{-2H}\d_w e^{-2H-2\phi}-({\d}_{u}-A_u\d_w)(u^4A_u)=0.
\eea
These relations are obvious generalizations of (\ref{IIBeqn1}), (\ref{IIBeqn2}). 

Thus we see that the gravitino and dilatino projectors confirm the ansatz 
(\ref{MetrGssAppB}), moreover they lead to the unique expression for the $F_3$:
\bea
G_-=\frac{ib}{48}e^{\phi/2}d\left[\eps_{ijk}\d_k F dx^{ij}\right]
\eea
and to differential equations (\ref{DifEqn14a}), (\ref{DifEqn14b}).
In the next subsection we check that the differential equations for $\eps$ are also consistent with this
solution.

\subsection{Checking differential equations.}

Let us check the ${\bf x}$, $u$ and $w$  components of the gravitino equation. Starting with the 
general expression for the spin--connection
\bea
\omega_\mu=\left[e^{\nu A}(\d_\mu e^B_\nu-\d_\nu e^B_\mu)-e^{\rho A}e^{\sigma B}e^C_\mu\d_\rho 
e_{\sigma C}\right]\Gamma_{AB},
\eea
we compute its component along $x_k$:
\bea
\omega_k&=&(e^{\nu {\bf u}}\d_k e^{\bf w}_\nu-e^{\nu {\bf w}}\d_k e^{\bf u}_\nu)\Gamma_{uw}
-\frac{1}{2}{\gamma^\nu}_k\d_\nu(H-\frac{\phi}{2})-
\frac{1}{2}e^{\rho A}e^{B}_k \d_\rho (H-\frac{\phi}{2})\Gamma_{AB}
\nonumber\\
&=&e^{2H+\phi}\d_k A_u\Gamma_{uw}-{\gamma^\nu}_k\d_\nu(H-\frac{\phi}{2}).
\eea
Then the $k$ projection of the spinor equation becomes
\bea
&&\left[\d_k+\frac{1}{4}e^{2H+\phi}\d_k(e^{-2H}\d_u F)\Gamma_{uw}\right]\eps-
\frac{1}{4}{\gamma^\nu}_k\d_\nu(H-\frac{\phi}{2})\eps+
i\frac{e^{-4C}}{2}{\not\d}f_3\gamma_k\Gamma_\Omega\eps\nonumber\\
&&\qquad+\frac{1}{96}(\gamma_k
{\not G}-2\{{\not G},\gamma_k\})\eps^*=0.
\eea
Projector (\ref{GenEqnA0}) and equation for the dilatino can be used to compute
\bea\label{GepsStr}
\frac{1}{24}{\not G}\eps^*=(G_++G_-)\eps^*=\frac{1}{2}{\not\d}\phi\eps+2G_-\eps^*=
{\not\d}(2A-\phi)\eps-2ie^{-4C}{\not\d}f_3\Gamma_\Omega\eps.
\eea
This leads to simplification in the differential equation:
\bea\label{DifTmp14a}
&&\left[\d_k+\frac{1}{4}e^{2H+\phi}\d_k(e^{-2H}\d_u F)\Gamma_{uw}+\frac{1}{4}\d_k(H+\frac{\phi}{2})
\right]\eps-
\frac{1}{2}{\gamma^\nu}_k\d_\nu H\eps+
ie^{-4C}{\gamma^\nu}_k{\d}_\nu f_3\Gamma_\Omega\eps\nonumber\\
&&\qquad
-\frac{1}{8}\gamma^{\mu\nu}G_{\mu\nu k}\eps^*=0.
\eea
It is convenient to decompose a three--form $G$ as
\bea
G=G^{(+)}+G^{(-)}:\qquad G^{(+)}=e^{-\phi/2}H_3,\quad G^{(-)}=ie^{\phi/2}F_3,
\eea
and evaluate the contributions of $G^{(+)}$ and $G^{(-)}$ separately:
\bea
\frac{1}{2}G^{(+)}_{\mu\nu k}dx^{\mu\nu}&=&-ce^{-\phi/2}
\d_k\left[e^{2\phi+2H}(dw+A_u du)\right]\wedge dt:\nonumber\\
G^{(+)}_{\mu\nu k}\gamma^{\mu\nu}&=&
-2ce^{-\phi/2}\d_k e^{2\phi+2H} e^{-2H-3\phi/2}\Gamma_w\Gamma^t-
2ce^{2H+\phi}\d_k A_u\Gamma_u\Gamma^t\nonumber\\
&=&-2ce^{2H+2\phi}\left[\d_k e^{-2\phi-2H}-
e^{-\phi}\d_k A_u \Gamma_{uw}\right]\Gamma_w\Gamma_t,\nonumber
\eea
\bea
&&G^{(-)}=24\frac{i}{2}e^{\phi/2}\eps_{ijk}\d_\mu h_k dx^\mu dx^{ij}=
24ie^{\phi/2}\left[\d_i h_i dx^{123}+\frac{\eps_{ijk}}{2}\d_\sigma h_k dy^\sigma dx^{ij}\right]:\nonumber\\
&&G^{(-)}_{\mu\nu k}\gamma^{\mu\nu}=
24ie^{\phi/2}\left[2e^{-H+\phi/2}\d_i h_i \Gamma_{123}\Gamma_k-2\eps_{ijk}\d_\sigma h_j 
\gamma^\sigma e^{-H/2+\phi/4}\Gamma_i\right]
\nonumber\\
&&\qquad\qquad =
bie^{\phi/2}\left[2e^{-H+\phi/2}\d_i \d_i F\Gamma_{123}\Gamma_k-2\eps_{ijk}\d_\sigma \d_j F 
\gamma^\sigma e^{-H/2+\phi/4}\Gamma_i\right].
\nonumber
\eea
In the last two equations index $\sigma$ goes over coordinates $u$ and $w$. Substituting these results into (\ref{DifTmp14a}) and using projectors (\ref{AppBProj}), (\ref{TimeProj}), we find
\bea
&&\left[\d_k-\frac{1}{4}\d_k(H+\frac{3\phi}{2})
\right]\eps-
\frac{1}{2}{\gamma^\nu}_k\d_\nu H\eps+
ie^{-4C}{\gamma^\nu}_k{\d}_\nu f_3\Gamma_\Omega\eps\nonumber\\
&&\qquad
-\frac{ib}{4}e^{\phi/2}\left[e^{-H+\phi/2}\Delta_x F\Gamma_k (ib\Gamma_w)\eps-
\eps_{ijk}\d_\sigma \d_j F 
\gamma^\sigma e^{-H/2+\phi/4}\Gamma_i\eps^*\right]=0.\nonumber
\eea
To eliminate $\eps^*$ we use the relation
\bea
\eps_{ijk}\Gamma_i\eps^*=ib\eps_{ijk}\Gamma_i\Gamma_w\Gamma_{123}\eps=
-ib\Gamma_w\Gamma_{jk}\eps,
\eea
and we also recall that
\bea
ie^{-4C}{\gamma^\nu}_k{\d}_\nu f_3\Gamma_\Omega\eps=-\frac{1}{4}e^{2H+\phi}u^{-4}
{\gamma^\nu}_k{\d}_\nu (u^4 A_u)\Gamma_{uw}\eps.
\eea
Using this information, one can rewrite the differential equation as
\bea\label{CheckDifUr}
&&\left[\d_k-\frac{1}{4}\d_k(H+\frac{3\phi}{2})
\right]\eps-
\frac{1}{2}{\gamma^\nu}_k\d_\nu H\eps-\frac{1}{4}e^{2H+\phi}u^{-4}
{\gamma^\nu}_k{\d}_\nu (u^4 A_u)\Gamma_{uw}\eps\nonumber\\
&&\quad
+\frac{1}{4}e^{\phi/2}\left[-e^{-H+\phi/2}\d_w e^{-2\phi}\Gamma_k\Gamma_w+
\d_\sigma \d_j F 
\gamma^\sigma e^{-H/2+\phi/4}\Gamma_w\Gamma_{jk}\right]\eps=0.
\eea
It is convenient to do a separate analysis of different terms appearing in this equation.  
We begin with contributions proportional to $\Gamma_{jk}$:
\bea
&&-\frac{1}{2}\d_j H\eps-\frac{1}{4}e^{2H+\phi}u^{-4}
{\d}_j (u^4 A_u)\Gamma_{uw}\eps\nonumber\\
&&\qquad+\frac{1}{4}e^{\phi/2}\left[
(\Gamma_u e^{H/2+\phi/4}(\d_u-A_u\d_w)+\Gamma_we^{-3H/2-3\phi/4}\d_w) \d_j F\right]  
e^{-H/2+\phi/4}
\Gamma_w\eps\nonumber\\
&&=(-\frac{1}{2}\d_j H+\frac{1}{4}e^{-2H}\d_j\d_w F)\eps-\frac{1}{4}\left[e^{2H+\phi}u^{-4}
{\d}_j (u^4 A_u)-e^{\phi}(\d_u-A_u\d_w)\d_j F\right]\Gamma_{uw}\eps\nonumber\\
&&=\frac{1}{4}e^{-2H}(-\d_j e^{2H}+\d_j\d_w F)\eps-\frac{1}{4}\left[e^{2H+\phi}
{\d}_j (u^4 A_ue^{2H})-e^{\phi}\d_u\d_j F\right]\Gamma_{uw}\eps.\nonumber
\eea 
Both terms in the right hand side vanish due to the definitions (\ref{MetrGssAppB}). Now we collect 
the remaining contributions
to (\ref{CheckDifUr}) which contain $\Gamma_k$:
\bea
&&\left[\frac{1}{2}{\gamma^\sigma}\d_\sigma H+\frac{e^{2H+\phi}}{4u^{4}}
{\gamma^\sigma}{\d}_\sigma (u^4 A_u)\Gamma_{uw}\right]\eps
-\frac{1}{4}e^{-H+\phi}\d_w e^{-2\phi}\Gamma_w\eps\nonumber\\
&&\quad=\frac{1}{4}\Gamma_u\left[2e^{H/2+\phi/4}(\d_u-A_u\d_w)H-
e^{2H+\phi}e^{-3H/2-3\phi/4}\d_w A_u\right]\eps\nonumber\\
&&\qquad+\frac{1}{4}\Gamma_w\left[2e^{-3H/2-3\phi/4}\d_w H+
\frac{e^{2H+\phi}}{u^{4}}e^{H/2+\phi/4}(\d_u-A_u\d_w)(u^4A_u)-
e^{-H+\phi}\d_w e^{-2\phi}
\right]\eps\nonumber\\
&&\quad=\frac{e^{-3H/2+\phi/4}}{4}\Gamma_u\left[\d_u e^{2H}-\d_w(A_u e^{2H})\right]\eps
\nonumber\\
&&\qquad+\frac{e^{-3H/2-3\phi/4}}{4}\Gamma_w\left[2\d_w H+
\frac{e^{4H+2\phi}}{u^{4}}(\d_u-A_u\d_w)(uA_u)+e^{2\phi}\d_i\d_i F
\right]\eps.\nonumber
\eea
The right hand side of this expression vanishes due to (\ref{MetrGssAppB}) and 
(\ref{ApGnrSemiLast}). We conclude that equation (\ref{CheckDifUr}) reduces to
\bea
\left[\d_k-\frac{1}{4}\d_k(H+\frac{3\phi}{2})
\right]\eps=0,
\eea
which implies that
\bea\label{GenEpsSol}
\eps=\exp\left[\frac{1}{4}(H+\frac{3\phi}{2})\right]\eps_0(u,w).
\eea

Next we check $w$ components of the differential equation:
\bea\label{DifUrW}
&&\nabla_w\eps+i\frac{e^{-4C}}{2}{\not\d}f_3\gamma_w\Gamma_\Omega\eps+
\frac{1}{96}(\gamma_w
{\not G}-2\{{\not G},\gamma_w\})\eps^*=0\nonumber\\
&&\nabla_w\eps+iu^{-4}e^{2H+\phi}{\gamma^\nu}_w{\d}_\nu f_3\Gamma_\Omega\eps+
\frac{1}{4}\gamma_w{\not\d}(2A-\phi)\eps-\frac{1}{8}\gamma^{\mu\nu}G_{\mu\nu w}\eps^*=0.
\eea
The spin connection along $w$ direction is
\bea
\omega_w=e^{2H+\phi}\d_w A_u \Gamma_{uw}-
\frac{1}{4}[{\not\d}(3H+\frac{3\phi}{2}),\gamma_w]-
e^{3H/2+3\phi/4}\gamma^{\mu\nu}\d_\mu e^{\bf w}_\nu.
\eea
We also need the expression for $\gamma^{\mu\nu}G_{\mu\nu w}\eps^*$:
\bea
\frac{1}{2}G^{(+)}_{\mu\nu w}dx^{\mu\nu}&=&-ce^{-\phi/2}
\d_w\left[e^{2\phi+2H}(dw+A_u du)\right]\wedge dt+
ce^{-\phi/2}\d_\mu\left[e^{2\phi+2H}\right]dx^\mu\wedge dt:\nonumber\\
G^{(+)}_{\mu\nu k}\gamma^{\mu\nu}&=&
-2c\left[e^{-2H-2\phi}\d_w e^{2\phi+2H} \Gamma_w+
e^{2H+\phi}\d_w A_u\Gamma_u-e^{-H/2-5\phi/4}{\not\d}e^{2\phi+2H}\right]\Gamma^t\nonumber\\
&=&-2ce^{2H+2\phi}\left[\d_w e^{-2\phi-2H}-
e^{-\phi}\d_w A_u \Gamma_{uw}\right]\Gamma_w\Gamma_t\nonumber\\
&&-2ce^{-H/2-5\phi/4}{\not\d}e^{2\phi+2H}\Gamma_t,\nonumber\\
G^{(-)}&=&12ie^{\phi/2}\eps_{ijk}\d_\mu h_k dx^\mu dx^{ij}:\nonumber\\
G^{(-)}_{\mu\nu w}\gamma^{\mu\nu}&=&
24ie^{\phi/2}\eps_{ijk}e^{-H+\phi/2}\d_w h_k \Gamma_{ij}=
48ie^{-H+\phi}\d_w h_k \Gamma_{k}\Gamma_{123},
\nonumber
\eea
\bea
\frac{1}{8}\gamma^{\mu\nu}G_{\mu\nu w}\eps^*&=&-\frac{1}{4}e^{2H+2\phi}\left[\d_w e^{-2\phi-2H}-
e^{-\phi}\d_w A_u \Gamma_{uw}\right]\eps\nonumber\\
&&-\frac{1}{4}e^{-H/2-5\phi/4}{\not\d}e^{2\phi+2H}\Gamma_w\eps-
6be^{-H+\phi}\d_w h_k \Gamma_{k}\Gamma_{w}\eps.
\eea
Substitution of these expressions into (\ref{DifUrW}) gives a complicated equation. We begin with analyzing the coefficient in front of $\Gamma_k$:
\bea
&&\left[-\frac{3}{4}\frac{e^{\phi+H}}{2}\d_k(H+\frac{\phi}{2})\Gamma_w-
\frac{1}{4}e^{H+\phi}\gamma^\nu\d_k e^{\bf w}_\nu 
+ie^{3H+2\phi}\Gamma_w{\d}_k\frac{f_3}{u^4}\Gamma_\Omega-
\frac{e^{\phi+H}}{4}\Gamma_w{\d}_k(H+\frac{\phi}{2})\right]\eps\nonumber\\
&&\qquad+\frac{1}{4}e^{-H-\phi}{\d}_ke^{2\phi+2H}\Gamma_w\eps+
\frac{1}{4}e^{-H+\phi}\d_w \d_k F\Gamma_{w}\eps\nonumber\\
&&=e^{\phi+H}\left[\left(-\frac{3}{8}-\frac{3}{8}-\frac{1}{4}\right)\d_k(H+\frac{\phi}{2})+
\frac{1}{2}\d_k(\phi+H)+\frac{1}{2}\d_k H\right]\Gamma_w\eps\nonumber\\
&&\qquad+\left[-\frac{1}{4}e^{H+\phi}\gamma^u e^{3H/2+3\phi/4}\d_k A_u 
-e^{3H+2\phi}{\d}_k\left(-\frac{A_u}{4}\right)\Gamma_u\right]\eps.
\eea
One can see that the right hand side is zero, so equation  (\ref{DifUrW}) does not contain terms with 
$\Gamma_k$. Next we look at the contributions proportional to $\Gamma_{uw}\eps$:
\bea\label{Trash16}
&&\frac{1}{4}\left[e^{2H+\phi}\d_w A_u -
\frac{1}{2}{\eps^\sigma}_w\d_\sigma(3H+\frac{3\phi}{2})-
e^{3H/2+3\phi/4}\eps^{\sigma\tau}\d_\sigma e^{\bf w}_\tau\right]\Gamma_{uw}\eps\nonumber\\
&&\quad-
\frac{1}{4}{\eps^\sigma}_w\d_\sigma(H+\frac{\phi}{2})\Gamma_{uw}\eps+
\frac{1}{4}\left[-e^{2H+\phi}\d_w A_u 
+e^{-2H-2\phi}{\eps^\sigma}_w{\d}_\sigma e^{2\phi+2H}
\right]\Gamma_{uw}\eps\nonumber\\
&&=\frac{1}{4}\left[
\frac{1}{2}{\eps^\sigma}_w\d_\sigma(-H+\frac{3\phi}{2})-
e^{3H/2+3\phi/4}\eps^{\sigma\tau}\d_\sigma e^{\bf w}_\tau\right]\Gamma_{uw}\eps.
\eea
Using relations
\bea
&&e^{3H/2+3\phi/4}\eps^{\sigma\tau}\d_\sigma e^{\bf w}_\tau=
\frac{3}{2}{\eps^\sigma}_w\d_\sigma(H+\frac{\phi}{2})+e^{3H+3\phi/2}\eps^{\sigma u}\d_\sigma A_u,\nonumber\\
&&e^{3H+3\phi/2}\eps^{\sigma u}\d_\sigma A_u=-e^{2H+\phi}\d_w A_u,\nonumber\\
&&{\eps^\sigma}_w\d_\sigma H=e^{\bf w}_we^\sigma_{\bf u}\d_\sigma H=
e^{2H+\phi}(\d_u-A_u\d_w)H\nonumber\\
&&\quad\qquad\quad=\frac{e^{\phi}}{2}\left[\d_u\d_w F-\d_w(e^{2H}A_u)+
e^{2H}\d_w A_u\right]=\frac{1}{2}e^{2H+\phi}\d_w A_u,\nonumber
\eea
we conclude that the right hand side of (\ref{Trash16}) vanishes. 

Thus the left hand side of the equation (\ref{DifUrW}) reduces to expression which does not contain gamma matrices:
\bea
&&\left[\d_w-au^{-4}e^{2H+\phi}{\eps^\sigma}_w\d_\sigma f_3+\frac{1}{4}\d_w(H+\frac{\phi}{2})-
\frac{1-1}{2}\d_w(\phi+H)\right]\eps\nonumber\\
&&=\left[\d_w+\frac{e^{4H+2\phi}}{4u^4}(\d_u-A_u\d_w) (u^4 A_u)+\frac{1}{4}\d_w(H+\frac{\phi}{2})
\right]\eps\nonumber\\
&&=\left[\d_w-\frac{1}{2}\d_w (H+\phi)+\frac{1}{4}\d_w(H+\frac{\phi}{2})
\right]\eps=e^{H/4+3\phi/8}\d_we^{-H/4-3\phi/8}\eps,
\eea
and it vanishes if $\eps_0$ in (\ref{GenEpsSol}) does not depend on $w$. This completes 
the check of the equation (\ref{DifUrW}).

At this point we have shown that the geometry (\ref{MetrGssAppB}) satisfies all equations for the Killing spinor, except the $u$ projection of the gravitino equation. Rather that checking this last relation explicitly, we multiply the differential equation appearing in (\ref{App2eqnA}) by $\gamma^\mu$ 
and sum over six indices corresponding to $(t,u,w,x_i)$\footnote{We also used (\ref{GepsStr}) to eliminate $\frac{1}{96}\gamma^\mu\gamma_\mu{\not G}\eps^*$.}:
\bea\label{ContrcDifUr}
\gamma^\mu \nabla_\mu\eps-5iu^{-4}e^{2H+\phi}{\not\d} f_3
\Gamma_\Omega\eps+
\frac{6}{4}{\not\d}(2A-\phi)\eps-\frac{1}{8}
\gamma^\mu\gamma^{\alpha\beta}G_{\alpha\beta \mu}\eps^*=0.
\eea
Notice that this relation is equivalent to the $u$ component of the gravitino equation 
since all other components were already shown to vanish. Let us simplify the left hand side of the last equation:
\bea\label{Trash16b}
S&\equiv&\left[\gamma^\mu \nabla_\mu-5iu^{-4}e^{2H+\phi}{\not\d} f_3
\Gamma_\Omega+
\frac{3}{2}{\not\d}(H+\frac{\phi}{2})\right]\eps\nonumber\\
&&-3\left[{\not\d}(H+\frac{\phi}{2})-
2iu^{-4}e^{2H+\phi}{\not\d}f_3\Gamma_\Omega\right]\eps\nonumber\\
&=&\gamma^\mu \nabla_\mu\eps-\frac{1}{4}u^{-4}e^{2H+\phi}{\not\d} (u^4 A_u)
\Gamma_{uw}\eps-
\frac{3}{2}{\not\d}(H+\frac{\phi}{2})\eps.
\eea
We have used first projector in (\ref{AppBProj}) as well as expression for $f_3$ from 
(\ref{MetrGssAppB}).
To proceed we evaluate
\bea
&&\gamma^\mu\omega_\mu=e^{\nu A}{\not\d} e^B_\nu\Gamma_{AB}+
\d_\nu \gamma^\mu {\gamma^\nu}_\mu-\d_\rho \gamma_{\sigma}
\gamma^{\rho\sigma}=e^{2H+\phi}{\not\d} A_u\Gamma_{uw}+
\d_\nu g^{\mu\sigma}\gamma_\sigma{\gamma^\nu}_\mu,\nonumber\\
&&\d_\nu g^{\mu\sigma}\gamma_\sigma{\gamma^\nu}_\mu=
\frac{1}{2}\d_\nu g^{\mu\sigma}(\gamma_\sigma\gamma^\nu\gamma_\mu-
g_{\sigma\mu}\gamma^\nu)=\gamma_\mu\d_\nu g^{\mu\nu}-
{\not\d} g^{\mu\sigma}g_{\sigma\mu}.
\eea
Substituting this into the right hand side of (\ref{Trash16b}), we find
\bea
S&=&{\not\d}\eps+\frac{1}{4}(\gamma_\mu\d_\nu g^{\mu\nu}-
{\not\d} g^{\mu\sigma}g_{\sigma\mu})\eps-\frac{e^{2H+\phi}}{u}A_u \gamma^u
\Gamma_{uw}\eps-
\frac{3}{2}{\not\d}(H+\frac{\phi}{2})\eps\nonumber\\
&=&{\not\d}\eps+\frac{1}{4}\gamma_\mu\eps~ \d_\nu g^{\mu\nu}+
\frac{1}{4}{\not\d} (6H+\phi)\eps-\frac{e^{5(H+\phi/2)/2}}{u}A_u \Gamma_{w}\eps-
\frac{3}{2}{\not\d}(H+\frac{\phi}{2})\eps\nonumber\\
&=&e^{H/4+3\phi/8}{\not\d}(e^{-H/4-3\phi/8}\eps)+\frac{e^{-H+\phi/2}}{4}\gamma_\mu\eps~ 
\d_\nu (e^{H-\phi/2}g^{\mu\nu})-\frac{e^{5(H+\phi/2)/2}}{u}A_u \Gamma_{w}\eps.\nonumber
\eea
The first term in the right hand side vanishes due to relation (\ref{GenEpsSol}) and the fact that 
$\eps_0$ is a constant spinor. Recalling the relevant six--dimensional metric
\bea
ds_6^2&=&e^{H-\phi/2}\left[-e^{2\phi}dt^2+d{\bf x}_3^2+e^{-2H}du^2+
e^{2\phi+2H}(dw+A_u du)^2\right]\equiv e^{H-\phi/2}{\tilde g}_{\mu\nu}dx^\mu dx^\nu,\nonumber
\eea
we can simplify $S$ even further:
\bea
S&=&\frac{e^{-H/2+\phi/4}}{4}{\tilde\gamma}_\mu\eps~ 
\d_\nu ({\tilde g}^{\mu\nu})-\frac{e^{5(H+\phi/2)/2}}{u}A_u \Gamma_{w}\eps\nonumber\\
&=&\frac{e^{-H/2+\phi/4}}{4}\left[{\tilde\gamma}_u\eps~
\{-\d_w(A_ue^{2H})+\d_u e^{2H}\}-\frac{4}{u}e^{3H+\phi}A_u \Gamma_{w}\eps\right.\nonumber\\
&&\left.+{\tilde\gamma}_w\eps~\{\d_w(e^{-2H-2\phi}+A_u^2 e^{2H})-\d_u(A_ue^{2H})\}
\right]\nonumber\\
&=&\frac{e^{-H/2+\phi/4}}{4}\left[(e^{-H}\Gamma_u+e^{\phi+H}A_u\Gamma_w)\eps~
\{-\d_w(A_ue^{2H})+\d_u e^{2H}\}\right.\nonumber\\
&&\left.+e^{\phi+H}{\Gamma}_w\eps~\left\{\d_w(e^{-2H-2\phi}+A_u^2 e^{2H})-\d_u(A_ue^{2H})
-\frac{4}{u}e^{2H}A_u\right\}
\right].\nonumber
\eea
Expressing the terms proportional to $\Gamma_u\eps$ in terms of $F$ (see (\ref{MetrGssAppB})), 
we conclude that they cancel out and one only needs to analyze the last line of the equation written above. Substituting the expression for $\d_w e^{-2H-2\phi}$ from (\ref{DifEqn14b}), we arrive at a relation
\bea
S=\frac{e^{H/2+5\phi/4}}{4}{\Gamma}_w\eps~
\left[A_u\d_w(A_u e^{2H})-A_u\d_u e^{2H})\right],
\eea
and the right hand side is clearly equal to zero. Then we conclude that equation (\ref{ContrcDifUr}) is satisfied by the system (\ref{MetrGssAppB}), (\ref{GMinus}), (\ref{DifEqn14a}), (\ref{DifEqn14b}).

\bigskip

To summarize, in this appendix we generalized the metric (\ref{SymSltnSum}) to the situation without $SO(3)$ symmetry and we explicitly checked that all dilatino and gravitino equations are satisfied for the resulting solution. The $SO(5)$ symmetry can be lifted in the same way, and introducing minor modifications to the procedure outlined in this appendix, one can check that a more general geometry 
(\ref{GenSltnSum}) is also supersymmetric.


\section{1/2--BPS geometries and near--horizon limit}
\label{AppCmpr}

\renewcommand{\theequation}{D.\arabic{equation}}
\setcounter{equation}{0}


In section \ref{SectSpike} we constructed geometries preserving eight supercharges along with 
$SO(5)\times SO(3)$ isometries. It is natural to ask whether some special solutions can have an enhanced supersymmetry. In the case of asymptotically--flat space the answer is well--known: to preserve 16 supercharges, one should set two out of three field--strengths $(H_3,F_3,F_5)$ to zero. 
The resulting solutions describe sets of parallel branes with flat worldvolumes. In the spaces with
$AdS_5\times S^5$ asymptotics the situation is different, and it is possible to find 1/2--BPS geometries 
with all fluxes being turned on \cite{yama,myWils,Gutprl}. It is interesting to find a relation between 
these solutions and the metrics constructed in this paper. We outline the procedure for embedding 
the solutions of \cite{myWils} into the ansatz (\ref{SymSltnSum}) in section \ref{SecSbWils} and in this appendix we 
provide some computational details. 

In section \ref{SectSbNrhrz} we showed that starting with an 
asymptotically--flat solution one can construct a geometry with $AdS_5\times S^5$ asymptotics by taking a near--horizon limit (\ref{BSymmNearHor}). If this limit is applied to a metric produced by a stack of flat D3 branes, one arrives at 
$AdS_5\times S^5$ which preserves twice as much supersymmetry as the original solution, and it is very natural to ask whether similar enhancement happens for a more general 1/4--BPS state. This problem is analyzed below and we find that a D3 brane metric is the only solution which has an 
enhanced symmetry in the limit (\ref{BSymmNearHor}). 

We begin by embedding the solutions with $SO(5)\times SO(3)\times SO(2,1)$ symmetry 
into a more general class of geometries described by (\ref{SymSltnSum}). To do this we need to 
recall the metrics found in 
\cite{myWils}:
\bea\label{WilsMetric}
ds^2=ye^{S-\phi/2}dH_2^2+ye^{G-\phi/2}d\Omega_2^2+ye^{-G-\phi/2}d\Omega_4^2+
\frac{e^{-\phi}}{2y\cosh G}(dx^2+dy^2).
\eea
The warp factors entering this expressions are specified in terms of one harmonic function and
we refer to \cite{myWils} for details. Here we will need only 
one of the equations satisfied by the warp factors:
\bea\label{WilsDifEqn}
&&d(S-G-2\phi)=-\frac{1}{y\cosh G}(e^{-G}dy+{\cal F}
(ye^{S-\phi/2})^{-1/2}dx),\\
&&{\cal F}\equiv \sqrt{ye^{-\phi/2}(e^S-e^G-e^{-G})}.\nonumber
\eea
To find the map between coordinates $(x,y,z)$ used in (\ref{WilsMetric}) and $(u,v,w)$ describing (\ref{SymSltnSum}), we begin with relations (\ref{BWlsToNew}):
\bea\label{WlsToNew}
e^{H}=yz^2 e^{S-2\phi},\quad
u^2=ye^{H-G}=y^2 z^2 e^{S-G-2\phi}\quad
v^2=ye^{G-H}=z^{-2} e^{-S+G+2\phi}.
\eea
This leaves only one undetermined coordinate $w$. Unfortunately we will only be able to find an expression for its differential. Let us introduce a scalar $\omega_z$ and a one--form $\omega$ which has legs in two dimensional space spanned by $x,y$:
\bea\label{IntermOmega}
e^{2H}(dw+{\cal A})=(\d_w Fdw+\d_u Fdu)\equiv \omega_z dz+\omega.
\eea
Substituting this into (\ref{SymSltnSum}) and matching the resulting metric with (\ref{WilsMetric}),
we find a relation
\bea
ye^S\frac{dz^2}{z^2}+\frac{1}{2y\cosh G}(dx^2+dy^2)
=e^H dv^2+e^{-H}du^2+e^{-H+2\phi}(\omega_z dz+\omega)^2.
\eea
Extracting the coefficient in front of $z^{-2}dz^2$, we determine $\omega_z$:
\bea
\omega_z=y^{1/2}e^{(S-4\phi)/2+\phi/4}{\cal F}.
\eea
To find $\omega$ we look at the coefficient in front of $dz$:
\bea
-z^{-3}e^H de^{-S+G+2\phi} +
e^{-H}z d(y^2 e^{S-G-2\phi})+2e^{-H+2\phi}\omega_z\omega=0,
\nonumber
\eea
then using equation (\ref{WilsDifEqn}) we compute
\bea
\omega&=&-z\frac{e^{-(S/2+\phi/4)}}{\sqrt{y}{\cal F}}\left[
y^2 e^{S-2\phi}\cosh Gd(S-G-2\phi)+ye^{S-G-2\phi}dy
\right]\nonumber\\
&=&-z\frac{e^{-(S/2+\phi/4)}}{\sqrt{y}{\cal F}}\left[
-ye^{S-2\phi}{\cal F}(ye^{S-\phi/2})^{-1/2}dx\right]
=ze^{-2\phi}dx.
\eea
To summarize, we found an expression for the differential of $w$:
\bea\label{WlsToNew1}
dw+A_udu=e^{-2H}\left[y^{1/2}e^{(S-4\phi)/2+\phi/4}{\cal F}dz+ze^{-2\phi}dx\right],
\eea  
and one can use this relation along with (\ref{WlsToNew}) to recover a unique set of coordinates 
$(u,v,w)$ starting from any solution with $SO(5)\times SO(3)\times SO(2,1)$ symmetry. As a consistency
check, we observe that the substitution of (\ref{WlsToNew1}) into (\ref{SymSltnSum}) gives an expression for the NS--NS 3--form
$H_3=df(x,y)\wedge \mbox{vol}_{AdS}$ which is expected for the solutions with $SO(2,1)$ symmetry. 

So far we started with an assumption that geometry (\ref{SymSltnSum}) has a hidden 
$SO(2,1)$ symmetry and 
showed that such solutions can be matched into the construction of \cite{myWils}. Unfortunately, a metric containing $AdS_2\times S^2\times S^4$ factors cannot be asymptotically flat, so this match is 
useful only for the geometries with $AdS_5\times S^5$ asymptotics. However, one can still start from a metric produced by branes in flat space and hope that a 
symmetry gets enhanced in a certain limit (for example, starting with a geometry produced by 
D3 brane and going close to the source, one finds $AdS_5\times S^5$), it would be interesting to see whether a region with $SO(2,1)\times SO(3)\times SO(5)$ isometry can be recovered from a generic asymptotically--flat solution (\ref{SymSltnSum}). A natural generalization of the near--horizon limit for 
this case has been proposed in section \ref{SectSbNrhrz} and now we will analyze whether it is possible to enhance the symmetry from $SO(3)\times SO(5)$ to 
$SO(3)\times SO(5)\times SO(2,1)$ in this limit.

We begin by recalling the "near--horizon map" (\ref{BSymmNearHor}) which transforms any
asymptotically--flat solution into a geometry with $AdS^5\times S^5$ asymptotics:
\bea\label{SymmNearHor}
u=\eps{\tilde u},\quad  e^H=\eps^2 e^{\tilde H},\quad 
v=\eps^{-1}{\tilde v},\quad  w=\eps^{-3}{\tilde w},\quad t=\eps^{-1}{\tilde t} ,\quad
F=\eps{\tilde F}.
\eea
Let us look at various fields in the limit $\eps\rightarrow 0$:
\bea\label{ScalingBefHat}
e^\phi&=&f_0(uv,u^3 w,u)\rightarrow f_0({\tilde u}{\tilde v},{\tilde u}^3 {\tilde w},0)\equiv
e^{\hat \phi},\nonumber\\
e^{\tilde H}&=&\eps^{-2}u^2f_1(uv,u^3 w,u)\rightarrow 
{\tilde u}^2f_1({\tilde u}{\tilde v},{\tilde u}^3 {\tilde w},0)\equiv {\tilde u}^2 e^{\hat H},\\
{\tilde A}_u&=&\eps^{4}u^{-4}f_2(uv,u^3 w,u)\rightarrow 
{\tilde u}^{-4}f_2({\tilde u}{\tilde v},{\tilde u}^3 {\tilde w},0)\equiv {\tilde u}^{-4} {\hat A}_u,\nonumber\\
{\tilde F}&=&\eps^{-1}uf_3(uv,u^3 w,u)\rightarrow 
{\tilde u}f_3({\tilde u}{\tilde v},{\tilde u}^3 {\tilde w},0)\equiv {\tilde u} {\hat F}.\nonumber
\eea
Assuming that the translational invariance in $t$ gets enhanced to $SO(2,1)$, 
we take the metric of the resulting AdS space to be 
$-z^{2}dt^2+\frac{dz^2}{z^2}$. Then it is convenient to use $z$ as one of the coordinates, and we will call the two remaining coordinates $x={\tilde u}{\tilde v}$ and $y={\tilde u}^3{\tilde w}$, so that 
$(u,v,w)$ are functions of $(z,x,y)$. An assumption of the enhanced symmetry requires the dilaton 
and warp factors of the spheres to be $z$--independent and it also imposes a relation 
$g_{tt}\sim z^2$. The 
only consistent way to satisfy these requirements is to make the following rescalings:
\bea
{\tilde u}=z{\hat u},\quad {\tilde v}=z^{-1}{\hat v},\quad {\tilde w}=z^{-3}{\hat w}
\eea
and to assume that all variables with hats are functions of $x$ and $y$ only. Then we can  
rewrite (\ref{ScalingBefHat}) in terms of "hatted functions" which depend on $x$ and $y$:
\bea\label{ScalingHat}
e^\phi=e^{\hat \phi},\quad
e^{\tilde H}=z^2 e^{\hat H},\quad 
{\tilde A}_u=z^{-4} {\hat A}_u,\quad {\tilde F}=z{\hat F}.
\eea
Substituting this into (\ref{SymSltnSum}) we observe that $z$ disappears from RR fluxes\footnote{For example, to see that $F_3$ has no $z$ dependence, we compute ${\tilde v}^2\d_{\tilde v}{\tilde F}\rightarrow 
z^{-2}{\hat v}^2{\tilde u}^2\d_x{\hat F}={\hat v}^2{\hat u}^2\d_x{\hat F}$.} and from the warp factors 
in front of $S^2$ and $S^4$. The remaining part of the metric and NS--NS three--form become
\bea\label{NearHorH3}
ds_4^2&=&-e^{{\hat H}+3{\hat\phi}/2}z^2d{\tilde t}^2+
e^{{\hat H}-{\hat\phi}/2}z^2d(z^{-1}{\hat v})^2+
e^{-{\hat H}-{\hat\phi}/2}z^{-2}d(z{\hat u})^2\nonumber\\
&&+z^{-2}e^{-{\hat H}+3{\hat\phi}/2}\left[d(z{\hat F})-
\d_{\tilde v}(z{\hat F})d(z^{-1}{\hat v})\right]^2,\nonumber\\
H_3&=&-\frac{1}{2}d\left\{e^{2{\hat\phi}}\left[
d(z{\hat F})-\d_{\tilde v}(z{\hat F})d(z^{-1}{\hat v})\right]\right\}\wedge dt.
\eea
A necessary condition for the $SO(2,1)$ invariance is a particular $z$--dependence of $H_3$:
\bea
H_3=d{\hat f}\wedge dz\wedge dt
\eea 
and comparison with (\ref{NearHorH3}) implies that 
\bea
\Omega\equiv d\left\{e^{2{\hat\phi}}\left[d{\hat F}-z^{-2}\d_{\tilde v}(z{\hat F})d{\hat v}
\right]\right\}.
\eea
should vanish. Since we are dealing with three functions $({\hat u},{\hat v},{\hat w})$ which depend on two variables $(x,y)$, we can impose a gauge ${\hat u}=1$, then 
$\d_{\tilde v}=z\d_x$ and we find a simpler expression for $\Omega$:
\bea\label{OmegaForm}
\Omega= d\left\{e^{2{\hat\phi}}\left[d{\hat F}-\d_x {\hat F}dx
\right]\right\}=d\left\{e^{2{\hat\phi}}\d_y {\hat F}dy\right\}=\d_x(e^{2{\hat\phi}}\d_y {\hat F})
dx\wedge dy.
\eea
Thus the assumption of $SO(2,1)$ invariance translates into the relation
\bea
\d_x(e^{2{\hat\phi}}\d_y {\hat F})=0,
\eea
which can be reformulated as $x$--independence of the following function:
\bea\label{defFxind}
f\equiv \frac{e^{-2\hat\phi}}{\d_y{\hat F}},\qquad \d_x f=0.
\eea
Let us write the equations (\ref{IIBeqn1}), (\ref{IIBeqn2a}) in the ${\hat u}=1$ gauge:
\bea
\label{NearHorAds1}
&&\d_ye^{-2{\hat\phi}}+x^{-2}\d_x(x^2\d_x{\hat F})=0,\\
\label{NearHorAds}
&&\d_{\hat F}(e^{-2{\hat\phi}}\d_{\hat F}{\hat w})+
({\hat F}\d_{\hat F}-x\d_x)(3{\hat w}+({\hat F}\d_{\hat F}-x\d_x){\hat w})=0,\\
&&e^{2{\hat H}}=\d_y{\hat F},\quad A_u=\frac{\hat F}{\d_y{\hat F}}.\nonumber
\eea
To arrive at (\ref{NearHorAds}) we used the following manipulations:
\bea
&&\d_F|_{u,v}=z\d_{\hat F},\quad 
\d_u w|_{v,F}=\d_u\left(z^{-3}{\hat w}(\frac{F}{u},uv)\right)=
-z^{-4}(3{\hat w}+({\hat F}\d_{\hat F}-x\d_x){\hat w}),\nonumber\\
&&\d_u(u^4\d_u w)|_{v,F}=z^{-1}({\hat F}\d_{\hat F}-x\d_x)
(3{\hat w}+({\hat F}\d_{\hat F}-x\d_x){\hat w}).\nonumber
\eea
It is convenient to eliminate dilaton from equations (\ref{NearHorAds1}), (\ref{NearHorAds}) and rewrite them in terms of $f(y)$, ${\hat F}(x,y)$. Then we arrive at the system
\bea
\label{NearHorAds1a}
&&\d_y(f\d_y{\hat F})+\Delta_x {\hat F}=0,\\
\label{NearHorAds2a}
&&\d_y f+\left[({\hat F}+x\d_x{\hat F})\d_y-x\d_y{\hat F}\d_x\right]\left[
3y+\frac{1}{\d_y{\hat F}}({\hat F}+x\d_x{\hat F})\right]=0.
\eea
To simplify the last equation we used the relations
\bea
{\hat F}\d_{\hat F}-x\d_x|_{\hat F}=\frac{1}{\d_y{\hat F}}({\hat F}+x\d_x{\hat F})\d_y-x\d_x|_y,\quad
\d_{\hat F}=\frac{1}{\d_y {\hat F}}\d_y.\nonumber
\eea
Thus we arrived at a system of two PDEs for one function of two variables ${\hat F}(x,y)$ and one function $f(y)$. Since these differential equations are independent and $f$ does not depend on $x$, one can easily show that there are no solutions with nontrivial $x$--dependence. 

To summarize, we started with an assumption that {\it a near--horizon limit} of an\\
{\it asymptotically--flat solution} (\ref{SymSltnSum}) has an enhanced $SO(2,1)$ symmetry, this led 
to the requirement (\ref{defFxind}), which in turn implied the trivial $x$--dependence in the solution. 
Of course, there are many interesting geometries with $SO(2,1)\times SO(3)\times SO(5)$ isometries
\cite{myWils}, however they cannot be constructed as a near horizon limit (\ref{BSymmNearHor}) of 
solutions with flat asymptotics. The exceptions are $x$--independent solutions of (\ref{NearHorAds1a}), 
(\ref{NearHorAds2a})\footnote{Notice that in this case we have two ODEs for two functions of $y$.} 
and now we will show that the only such solution is $AdS_5\times S^5$.

{\bf Example: $AdS_5\times S^5$.}
Let us assume that function ${\hat F}$ does not depend on $x$. Then equation (\ref{NearHorAds1a}) implies that the dilaton is constant ($e^{-2\phi}=f\d_y{\hat F}$). One can proceed by solving equation 
(\ref{NearHorAds2a}) to find ${\hat F}(y)$, but it turns out that there is an alternative route which leads 
to a simpler equation. Once we know that the dilaton is constant and there is no $x$--dependence in the system, the equation (\ref{NearHorAds}) can be rewritten in terms of 
$e^{-2{\hat H}}=\d_{\hat F}{\hat w}$:
\bea
\d_{\hat F}e^{-2{\hat H}-2\phi}+
({\hat F}^2\d_{\hat F}+4{\hat F})e^{-2\hat H}=0.
\eea
This equation can be easily solved:
\bea\label{May16Soln}
e^{-2\hat H}=\frac{Q}{(1+e^{2\phi}{\hat F}^2)^2}=\frac{Qz^4}{(z^2+e^{2\phi}F^2)^2},
\eea
and the metric takes the standard form (for simplicity we set $e^{2\phi}=1$):
\bea
ds^2&=&z^2 e^{\hat H}\left[-dt^2+dv^2+v^2d\Omega_2^2\right]+
z^{-2}e^{-\hat H}\left[du^2+u^2 d\Omega_4^2+(dF-\d_v F dv)^2\right]\nonumber\\
&=&z^2 e^{\hat H}\left[-dt^2+dv^2+v^2d\Omega_2^2\right]+
z^{-2}e^{-\hat H}\left[dz^2+z^2 d\Omega_4^2+dF^2\right].
\eea


\begin{thebibliography}{99}
%
\bibitem{LeighPolch}
J.~Dai, R.~G.~Leigh and J.~Polchinski,
  Mod.\ Phys.\ Lett.\  A {\bf 4}, 2073 (1989);\\
R.~G.~Leigh,
  Mod.\ Phys.\ Lett.\  A {\bf 4}, 2767 (1989);\\
P.~Horava,
  Nucl.\ Phys.\  B {\bf 327}, 461 (1989);
  Phys.\ Lett.\  B {\bf 231}, 251 (1989).
%
\bibitem{HorowStrom}
G.~T.~Horowitz and A.~Strominger,
  Nucl.\ Phys.\  B {\bf 360}, 197 (1991).
%
\bibitem{Polch}
J.~Polchinski,
  Phys.\ Rev.\ Lett.\  {\bf 75}, 4724 (1995), hep-th/9510017.
%
\bibitem{malda}
J.~M.~Maldacena,
  Adv.\ Theor.\ Math.\ Phys.\  {\bf 2}, 231 (1998),
  Int.\ J.\ Theor.\ Phys.\  {\bf 38}, 1113 (1999), hep-th/9711200.
%
\bibitem{StromVafa}
A.~Strominger and C.~Vafa,
  Phys.\ Lett.\  B {\bf 379}, 99 (1996), hep-th/9601029.
%
\bibitem{giant}
J.~McGreevy, L.~Susskind and N.~Toumbas,
  JHEP {\bf 0006}, 008 (2000), hep-th/0003075;\\
M.~T.~Grisaru, R.~C.~Myers and O.~Tafjord,
  JHEP {\bf 0008}, 040 (2000), hep-th/0008015.
%
\bibitem{rey}
S.~J.~Rey and J.~T.~Yee,
  Eur.\ Phys.\ J.\  C {\bf 22}, 379 (2001), hep-th/9803001;\\
J.~Pawelczyk and S.~J.~Rey,
  Phys.\ Lett.\  B {\bf 493}, 395 (2000), hep-th/0007154].
%
\bibitem{Drukker}
N.~Drukker and B.~Fiol,
  JHEP {\bf 0502}, 010 (2005), hep-th/0501109;\\
S.~Yamaguchi,
  JHEP {\bf 0605}, 037 (2006), hep-th/0603208;\\
J.~Gomis and F.~Passerini,
  JHEP {\bf 0608}, 074 (2006), hep-th/0604007.
%
\bibitem{LLM}
H.~Lin, O.~Lunin and J.~M.~Maldacena,
  JHEP {\bf 0410}, 025 (2004), hep-th/0409174.
%
\bibitem{LinMaldGom}
H.~Lin and J.~M.~Maldacena,
  Phys.\ Rev.\  D {\bf 74}, 084014 (2006), hep-th/0509235;\\
J.~Gomis and S.~Matsuura,
  JHEP {\bf 0706}, 025 (2007), arXiv:0704.1657 [hep-th].
\bibitem{yama}
S.~Yamaguchi,
  Int.\ J.\ Mod.\ Phys.\  A {\bf 22}, 1353 (2007), hep-th/0601089.
%
\bibitem{myWils}
O.~Lunin,
  JHEP {\bf 0606}, 026 (2006), hep-th/0604133.
%
\bibitem{GomRom}
J.~Gomis and C.~Romelsberger,
  JHEP {\bf 0608}, 050 (2006), hep-th/0604155.
%
\bibitem{Gutprl}
E.~D'Hoker, J.~Estes and M.~Gutperle,
  arXiv:0705.1004 [hep-th].
%
\bibitem{CalMal}
C.~G.~.~Callan and J.~M.~Maldacena,
  Nucl.\ Phys.\  B {\bf 513}, 198 (1998), hep-th/9708147.
%
\bibitem{gauntRev}
J.~P.~Gauntlett,
  arXiv:hep-th/9705011.
%
\bibitem{smithRev}
D.~J.~Smith,
  Class.\ Quant.\ Grav.\  {\bf 20}, R233 (2003), hep-th/0210157.
%
\bibitem{HanWitten}
A.~Hanany and E.~Witten,
  Nucl.\ Phys.\  B {\bf 492}, 152 (1997), hep-th/9611230;\\
E.~Witten,
  Nucl.\ Phys.\  B {\bf 500}, 3 (1997), hep-th/9703166;
  Nucl.\ Phys.\  B {\bf 507}, 658 (1997), hep-th/9706109;\\
S.~Elitzur, A.~Giveon, D.~Kutasov, E.~Rabinovici and A.~Schwimmer,
  Nucl.\ Phys.\  B {\bf 505}, 202 (1997), hep-th/9704104;\\
A.~Giveon and D.~Kutasov,
  Rev.\ Mod.\ Phys.\  {\bf 71}, 983 (1999), hep-th/9802067.
%
\bibitem{StrmTwns}
A.~Strominger,
  Phys.\ Lett.\  B {\bf 383}, 44 (1996), hep-th/9512059;\\
P.~K.~Townsend,
  Nucl.\ Phys.\ Proc.\ Suppl.\  {\bf 58}, 163 (1997), hep-th/9609217.
%
\bibitem{ScanRef}
M.~Cederwall, A.~von Gussich, B.~E.~W.~Nilsson and A.~Westerberg,
  Nucl.\ Phys.\  B {\bf 490}, 163 (1997), hep-th/9610148;\\
M.~Aganagic, C.~Popescu and J.~H.~Schwarz,
  Phys.\ Lett.\  B {\bf 393}, 311 (1997), hep-th/9610249;\\
M.~Cederwall, A.~von Gussich, B.~E.~W.~Nilsson, P.~Sundell and A.~Westerberg,
  Nucl.\ Phys.\  B {\bf 490}, 179 (1997), hep-th/9611159;\\
E.~Bergshoeff and P.~K.~Townsend,
  Nucl.\ Phys.\  B {\bf 490}, 145 (1997), hep-th/9611173.
%
\bibitem{HorTseytl}
G.~T.~Horowitz and A.~A.~Tseytlin,
  Phys.\ Rev.\  D {\bf 51}, 2896 (1995), hep-th/9409021;\\
A.~A.~Tseytlin,
  Nucl.\ Phys.\  B {\bf 475}, 149 (1996), hep-th/9604035;\\
A.~A.~Tseytlin,
  Class.\ Quant.\ Grav.\  {\bf 14}, 2085 (1997), hep-th/9702163.
%
\bibitem{11DScan}
E.~Bergshoeff, E.~Sezgin and P.~K.~Townsend,
  Phys.\ Lett.\  B {\bf 189}, 75 (1987).
%
\bibitem{gueven}
R.~Gueven,
  Phys.\ Lett.\  B {\bf 276}, 49 (1992).
%
\bibitem{MDBI}
M.~Perry and J.~H.~Schwarz,
  Nucl.\ Phys.\  B {\bf 489}, 47 (1997), hep-th/9611065;\\
  M.~Aganagic, J.~Park, C.~Popescu and J.~H.~Schwarz,
  Nucl.\ Phys.\  B {\bf 496}, 191 (1997), hep-th/9701166.
%
\bibitem{PST}
I.~A.~Bandos, K.~Lechner, A.~Nurmagambetov, P.~Pasti, D.~P.~Sorokin and M.~Tonin,
  Phys.\ Lett.\  B {\bf 408}, 135 (1997), hep-th/9703127;\\
P.~Pasti, D.~P.~Sorokin and M.~Tonin,
  Phys.\ Lett.\  B {\bf 398}, 41 (1997), hep-th/9701037;\\
I.~A.~Bandos, K.~Lechner, A.~Nurmagambetov, P.~Pasti, D.~P.~Sorokin and M.~Tonin,
  Phys.\ Rev.\ Lett.\  {\bf 78}, 4332 (1997), hep-th/9701149.
%
\bibitem{bachas}
C.~Bachas, M.~R.~Douglas and C.~Schweigert,
  JHEP {\bf 0005}, 048 (2000), hep-th/0003037.
%
\bibitem{GKMTZ}
J.~P.~Gauntlett, C.~Koehl, D.~Mateos, P.~K.~Townsend and M.~Zamaklar,
  Phys.\ Rev.\  D {\bf 60}, 045004 (1999), hep-th/9903156.
\bibitem{WittBar}
E.~Witten,
  JHEP {\bf 9807}, 006 (1998), hep-th/9805112.
\bibitem{CalGui}
Y.~Imamura,
  Nucl.\ Phys.\  B {\bf 537}, 184 (1999), hep-th/9807179;\\
C.~G.~Callan, A.~Guijosa and K.~G.~Savvidy,
  Nucl.\ Phys.\  B {\bf 547}, 127 (1999), hep-th/9810092;\\
J.~M.~Camino, A.~V.~Ramallo and J.~M.~Sanchez de Santos,
  Nucl.\ Phys.\  B {\bf 562}, 103 (1999), hep-th/9905118;\\
J.~Gomis, A.~V.~Ramallo, J.~Simon and P.~K.~Townsend,
  JHEP {\bf 9911}, 019 (1999), hep-th/9907022.
%
\bibitem{granPolDil}
M.~Grana and J.~Polchinski,
  Phys.\ Rev.\  D {\bf 65}, 126005 (2002), hep-th/0106014.
%
\bibitem{tseytlin}
A.~A.~Tseytlin,
  Nucl.\ Phys.\  B {\bf 469}, 51 (1996), hep-th/9602064.
%
\bibitem{SeibWitNC}
N.~Seiberg and E.~Witten,
  JHEP {\bf 9909}, 032 (1999), hep-th/9908142.
%
\bibitem{HasItz}
A.~Hashimoto and N.~Itzhaki,
  Phys.\ Lett.\  B {\bf 465}, 142 (1999), hep-th/9907166.
%
\bibitem{MalRus}
J.~M.~Maldacena and J.~G.~Russo,
  JHEP {\bf 9909}, 025 (1999), hep-th/9908134.
%
\bibitem{KrLarTriv}
P.~Kraus, F.~Larsen and S.~P.~Trivedi,
  JHEP {\bf 9903}, 003 (1999), hep-th/9811120.
%
\bibitem{johnson}
C.~V.~Johnson,
  hep-th/0007170.
%
\bibitem{schwarz}
J.~H.~Schwarz,
  Nucl.\ Phys.\  B {\bf 226}, 269 (1983);\\
J.~H.~Schwarz and P.~C.~West,
  Phys.\ Lett.\  B {\bf 126}, 301 (1983);\\
P.~S.~Howe and P.~C.~West,
  Nucl.\ Phys.\  B {\bf 238}, 181 (1984).
%
\bibitem{MarlfPeet}
D.~Marolf and A.~W.~Peet,
  Phys.\ Rev.\  D {\bf 60}, 105007 (1999), hep-th/9903213.
%
\bibitem{CherkHash}
S.~A.~Cherkis and A.~Hashimoto,
  JHEP {\bf 0211}, 036 (2002), hep-th/0210105.
%
\bibitem{lmmm}
O.~Lunin and S.~D.~Mathur,
  Nucl.\ Phys.\  B {\bf 623}, 342 (2002), hep-th/0109154;\\
O.~Lunin, J.~M.~Maldacena and L.~Maoz,
  arXiv:hep-th/0212210.
%
\bibitem{GMR}
J.~B.~Gutowski, D.~Martelli and H.~S.~Reall,
  Class.\ Quant.\ Grav.\  {\bf 20}, 5049 (2003), hep-th/0306235.
%
\bibitem{BW}
I.~Bena and N.~P.~Warner,
  Adv.\ Theor.\ Math.\ Phys.\  {\bf 9}, 667 (2005), hep-th/0408106.
%
\bibitem{AddMom}
O.~Lunin,
  JHEP {\bf 0404}, 054 (2004), hep-th/0404006;\\
S.~Giusto, S.~D.~Mathur and A.~Saxena,
  Nucl.\ Phys.\  B {\bf 701}, 357 (2004), hep-th/0405017;
  Nucl.\ Phys.\  B {\bf 710}, 425 (2005), hep-th/0406103;\\
S.~Giusto and S.~D.~Mathur,
  Nucl.\ Phys.\  B {\bf 729}, 203 (2005), hep-th/0409067\\
P.~Berglund, E.~G.~Gimon and T.~S.~Levi,
  JHEP {\bf 0606}, 007 (2006), hep-th/0505167.
%
\bibitem{myDefect}
O.~Lunin,
  arXiv:0704.3442 [hep-th].
%
\bibitem{granaPolch}
M.~Grana and J.~Polchinski,
  Phys.\ Rev.\  D {\bf 63}, 026001 (2001), hep-th/0009211.

\end{thebibliography}
\end{document}